\documentclass{article}
\usepackage[top=1.5in, bottom=1.5in, left=1.2in, right=1.2in]{geometry}
\usepackage{authblk}
\usepackage{amssymb, amsmath,framed}
\usepackage{graphicx}
\usepackage[round]{natbib}

\usepackage{bm}
\usepackage[colorlinks=true,linkcolor=blue,urlcolor=blue,citecolor=blue,anchorcolor=blue]{hyperref}
\expandafter\ifx\csname package@font\endcsname\relax\else
 \expandafter\expandafter
 \expandafter\usepackage
 \expandafter\expandafter
 \expandafter{\csname package@font\endcsname}
\fi
\def\bq{\begin{equation}}
\def\eq{\end{equation}}
\def\bqy{\begin{eqnarray}}
\def\eqy{\end{eqnarray}}
\def\calc{\mathcal{C}}
\def\cald{\mathcal{D}}

\def\calh{\mathcal{H}}

\def\calk{\mathcal{K}}
\def\call{\mathcal{L}}

\def\calr{\mathcal{R}}

\def\calu{\mathcal{U}}

\makeatother

\begin{document}

\title{Theoretical constraints imposed by gradient detection and dispersal on microbial size in astrobiological environments}

\author{Manasvi Lingam\thanks{Electronic address: \texttt{mlingam@fit.edu}}}

\affil{Department of Aerospace, Physics and Space Science, Florida Institute of Technology, Melbourne, FL 32901, USA}
\affil{Institute for Theory and Computation, Harvard University, Cambridge, MA 02138, USA}

\date{}

\maketitle

\begin{abstract}
The capacity to sense gradients efficiently and acquire information about the ambient environment confers many advantages like facilitating movement toward nutrient sources or away from toxic chemicals. The amplified dispersal evinced by organisms endowed with motility is possibly beneficial in related contexts. Hence, the connections between information acquisition, motility, and microbial size are explored from an explicitly astrobiological standpoint. By using prior theoretical models, the constraints on organism size imposed by gradient detection and motility are elucidated in the form of simple heuristic scaling relations. It is argued that environments such as alkaline hydrothermal vents, which are distinguished by the presence of steep gradients, might be conducive to the existence of ``small'' microbes (with radii of $\gtrsim 0.1$ $\mu$m) in principle, when only the above two factors are considered; other biological functions (e.g., metabolism and genetic exchange) could, however, regulate the lower bound on microbial size and elevate it. The derived expressions are potentially applicable to a diverse array of settings, including those entailing solvents other than water; for example, the lakes and seas of Titan. The paper concludes with a brief exposition of how this formalism may be of practical and theoretical value to astrobiology.
\end{abstract}

\section{Introduction}
The ability to precisely sense physical and chemical gradients is a ubiquitous feature of the microbial world, which is not surprising in light of the many desirable outcomes that are documented to result from this ability \citep{AGZ,ML06,HT14,BS18}. For instance, the ability to sense gradients in the chemical concentration enables organisms to move either toward or away from these compounds \citep[pp. 159-161]{PKTG}; see also \citet{WA04}. In the former case, one of the chief advantages is the ability to reach nutrient sources and thereby benefit from enhanced uptake \citep{LMT,Sto12,Kir18}, while other positives include range expansion, biofilm formation and the onset of symbiosis \citep{PWA11,CHY,WCV18,RFL19}. In the latter (i.e., movement away from the sources), microbes can endeavor to avoid toxic chemicals by travelling in the direction with lower concentration, or they could come up ways to mitigate the damage \citep{Kir18}. 

This mechanism, which constitutes a response to chemical stimuli and gradients, is known as chemotaxis. It has a long and distinguished history spanning myriad disciplines of science dating back to the nineteenth century \citep{Eng81,Pfe84,Bon47,Ad66,KS71,BB72,BP77,GLH12,Tu13,Cam18,Pa19}. Chemotaxis in bacteria, with the well-known model organism \emph{Escherichia coli} being one example, exhibits many remarkable features including high sensitivity, flexibility and robustness vis-\`a-vis chemical stimuli \citep{ASBL,BS05,SW12,ME16,WFE,FSJ20}.

Aside from chemotaxis, the likes of phototaxis and thermotaxis---which are reliant on identifying light and thermal gradients, respectively---have also been widely investigated in model organisms and via theoretical modelling \citep{Ar99,Mo99,Jek09,DS12,HT14,LTL16,WW17,YRC}. In each of these mechanisms, the key common point that deserves to be appreciated is: organisms sense \emph{information} about their environments. While this information can be put to many uses, one of the key aspects that we shall focus on herein is targeted locomotion \citep{WW09}---that is to say, movement along some preferred direction. Thus, if microbes are not efficient at acquiring information about their habitats, their capacity for meaningful motility could be stymied, although there are latent subtleties at play \citep{CV10,TSZ13}. This process of detecting gradients and gaining information is reliant on accurately measuring the ``intensity'' of the stimulant at multiple points in space or time \citep{BP77,Berg}. 

As the preceding sentence indicates, there exist two basic modes of sensing, namely, the spatial and the temporal \citep{Dus09}. In the former, microbes use the data garnered from receptors at different spatial locations to determine the direction of locomotion. In the latter, measurements from the receptors are collected at various moments in time to regulate the direction. The upshot of the prior discussion is that gradient sensitivity is generally expected to play a crucial role in modulating the efficacy of motility \citep{Fen02}. This statement yields a vital corollary: if the organism is too small to sense gradients efficiently, this bottleneck would pose difficulties for meaningful locomotion. In other words, the constraints imposed by information specify a minimum cell radius ($R_\mathrm{min}$) for microbes that are efficient at gradient sensing and dispersal. Naturally, this does not represent the only limiting factor, and we will touch upon other controls later in the paper.

It is worth highlighting that motility is regarded a viable biosignature candidate for \emph{in situ} life-detection missions \citep{NLD16,NBL18,NHV18,NAD20,RSS21}. In fact, the spatial resolution required to identify motile organisms is lower when compared to their non-motile counterparts \citep[pg. 755]{NLD16}. The ability to pin down the minimum cell size, and the cell density of motile lifeforms thence \citep{MaLi21}, is valuable therefore from the practical goal of gauging and designing suitable instruments for future missions. However, we emphasize that the \emph{raison d'{\^e}tre} of this paper is not to solely assess the feasibility of motility as such. Our major objective is, instead, to unveil the constraints on gradient sensing imposed by size; to put it differently, the chief purpose is to explore how size conditions the efficiency and efficacy of organisms to acquire information about their environment, which in turn permits them to act in a number of ways delineated above. It is essential to recognize that this is \emph{not} the only bottleneck on size because factors like energy, nutrients, cell structure and physiology also play a role \citep{LA15,LRL15,Kemp19,LC20}. As stated earlier, our goal is to explicate some of the inherent connections between gradient detection and organism dispersal on the one hand and microbial size on the other.

For the purposes of this paper, we will adopt the heuristic framework explicated in \citet{Dus97} due to its comparative simplicity and broad scope. To reiterate, the objective is to develop uncomplicated scaling relations that may pave the way toward understanding how the sizes of microbes in variegated astrobiological environments might be constrained by the capacity to promote dispersal as well as resolve gradients and obtain information about the neighborhood. The importance of studying the physical, chemical and biological constraints on size and its attendant evolutionary and ecological consequences has a rich history, dating back to at least the pioneering essay by \citet{Hal26}. While this topic has been commonly explored at the level of macroscopic organisms \citep{Went,Den93,WB00,Ang09,GFK15}, size constraints at the microscopic level have also been investigated \citep{Mor67,Koch,KOB99,ABG,Co18}.

There are certain aspects, however, whereby our treatment diverges from that of \citet{Dus97}. First, one of the principal purposes of the aforementioned publication is to estimate the signal-to-noise ratio (S/N), the metric used for measuring the efficacy of gradient detection, as a function of the cell size. Here, we tackle the converse problem, in which we employ the condition $\mathrm{S/N} = 1$ to deduce the corresponding $R_\mathrm{min}$. Second, we modify some of the fiducial values employed in \citet{Dus97} in light of current developments; the pertinent references are cited when we deviate from the canonical estimates. Third, and most importantly, we frame our analysis in an astrobiological context by focusing on domains and worlds that are perceived as promising from this standpoint. 

The outline of the paper is as follows. We begin by presenting the salient equations and model parameters in Section \ref{SecMod}. We explore the predictions of this framework for two distinct environments in Section \ref{SSecApp}---to wit, submarine hydrothermal vents on Earth (and elsewhere), and generic lakes and seas of Titan. Finally, we conclude with a summary of our findings along with a brief exposition of their practical and theoretical implications in Section \ref{SecConc}.

\section{Model description and results}\label{SecMod}
Given that our analysis mirrors that of \citet{Dus97}, we begin with a brief summary of the caveats and simplifications involved in constructing the heuristic model for surface and subsurface habitats. For starters, the organisms are taken to have spherical symmetry; changing the shape to ellipsoidal or cylindrical is anticipated to yield noteworthy benefits but also incur concomitant costs, as elucidated in \citet[pp. 31-32]{Kio08} and \citet{SHS19}. Second, the power per unit volume accessible by an organism for swimming is held roughly constant \citep[cf.][]{MGL08,DOM10}. Third, the effect of noise (the S/N to be more precise) is taken to directly impact the capability of organisms to obtain information about stimuli and it is consequently pressed into service as a proxy. Fourth, the temporal duration over which an organism can undertake rectilinear motion or ascertain the direction of stimuli is governed by rotational Brownian motion. 

Fifth, the scaling relations expounded herein are applicable at low Reynolds number \citep{Pur77,Lau16}. Turbulence could lead to sizable quantitative changes, such as through the effects of turbulent diffusion \citep{Wei00,OL01}. Sixth, the methodology is apropos only for single microbes and is therefore not applicable to collective behavior, which is characterized by much richer dynamics \citep{BLB98,VZ12,HS17,FKL18}. When microbes form consortia, which is particularly pertinent in harsh environmental conditions, it might be feasible for these aggregates to perform gradient sensing beyond the limits of single-cell chemotaxis analyzed herein \citep{VHM16,Cam18}. Lastly, it is assumed that the gradients in question are maintained continuously (with spatiotemporal variations) by virtue of geological, chemical, and physical activity and are consequently not dissipated by metabolism and other biological processes. Certain environments on Earth (e.g., submarine hydrothermal vents) are known to sustain long-lived chemical and thermal gradients \citep{RBB14,MASD}.

In spite of these limitations, we note that the results have proven to be fairly accurate for prokaryotes \citep[e.g.,][]{MWJ15,Kir18,BFT20}; similar considerations could apply to eukaryotes in differentiating between motile and non-motile species \citep{WJ20}.

\subsection{Expressions for the minimal cell size}\label{SSecExpCS}
The first point to note is that motility, by definition, can facilitate faster dispersal. In the absence of motility, the behavior of microbes can be described by the classical diffusion coefficient $D_0$, whereas its inclusion leads to an ``effective'' diffusion coefficient $D_M$ \citep{Berg}. The ratio of these two diffusion coefficients is $\zeta \equiv D_M/D_0$. We can solve for $R_\mathrm{min}$ by demanding that this ratio must exceed the minimum threshold of $\zeta = 1$ in order for dispersal to start becoming effective. By implementing this procedure, we end up with
\begin{equation}\label{Rdisp}
  R_\mathrm{min} \approx 0.36\,\mathrm{\mu m}\, \left(\frac{\calu}{\calu_\oplus}\right)^{-1/3} \left(\frac{\eta}{\eta_\oplus}\right)^{-1/3} \left(\frac{T}{T_\oplus}\right)^{1/3},
\end{equation}
where the subscript `$\oplus$' hereafter is taken to signify representative values on Earth, chosen based on \citet[Table 1]{Dus97} and other sources, with the explicit \emph{proviso} understanding that Earth-based organisms and habitats display a substantial degree of heterogeneity. In the above equation, $\eta$ denotes the dynamic viscosity of the environment and $\calu$ embodies the speed of swimming relative to the organism's size. The reason behind introducing $\calu$ (units of s$^{-1}$) has to do with the empirical linear scaling discerned between the speeds and sizes of organisms \citep[Figure 34]{Bon06}, which is supported by theoretical arguments \citep{Vog08,Dus09,MR16}. We have adopted $T_\oplus = 293$ K ($20$ $^\circ$C), $\eta_\oplus = 10^{-3}$ N s m$^{-2}$, and $\calu_\oplus = 10$ s$^{-1}$; note that the latter parameter is close to the median value for swimming microbes \citep{MR16} and to \emph{Escherichia coli} in particular \citep[pg. 270]{MP16}. With that said, some species of \emph{Archaea} evince fast swimming speeds that are more than one order of magnitude higher than \emph{E. coli} \citep{HW12}. Our understanding of archaeal motility remains incomplete despite the available data \citep{AJ18}, partly due to the variety of propulsion methods accessible in theory \citep{BDL16}.

Next, we turn our attention to the various constraints imposed by garnering information from gradients in stimuli via spatial and temporal methods. The rest of the formulae are derived by taking the corresponding equations from \citet[Table 2]{Dus97} for the signal-to-noise ratio (S/N), and deploying the criterion $\mathrm{S/N} = 1$ to solve for $R_\mathrm{min}$. The first example from this category are chemical gradients. For the spatial mode, the relevant scaling is given by
\begin{equation}\label{RchemS}
  R_\mathrm{min} \approx 0.32\,\mathrm{\mu m}\,\left(\frac{\cald_c}{\cald_\oplus}\right)^{-1/6} \left(\frac{\calc}{\calc_\oplus}\right)^{-1/6} \left(\frac{\call_c}{\call_{c\,\oplus}}\right)^{1/3} \left(\frac{\eta}{\eta_\oplus}\right)^{-1/6} \left(\frac{T}{T_\oplus}\right)^{1/6}
\end{equation}
whereas the equivalent expression for the temporal mode is
\begin{equation}\label{RchemT}
  R_\mathrm{min} \approx 0.38\,\mathrm{\mu m}\,\left(\frac{\calu}{\calu_\oplus}\right)^{-1/6} \left(\frac{\cald_c}{\cald_\oplus}\right)^{-1/12} \left(\frac{\calc}{\calc_\oplus}\right)^{-1/12} \left(\frac{\call_c}{\call_{c,\,\oplus}}\right)^{1/6} \left(\frac{\eta}{\eta_\oplus}\right)^{-1/4} \left(\frac{T}{T_\oplus}\right)^{1/4},
\end{equation}
where $\calc$ represents the average concentration of the appropriate chemical in the environment, $\cald_c$ denotes the diffusion coefficient corresponding to that chemical in the given solvent, and $\call_c \approx \calc \left(d \calc/dz\right)^{-1}$ embodies the characteristic length scale associated with the chemical gradients. The fiducial values chosen are $\cald_\oplus = 10^{-9}$ m$^2$ s$^{-1}$, $\call_{c,\,\oplus} = 10^{-3}$ m, and $\calc_\oplus \approx 10$ $\mu$M; note that the latter is commensurate with the concentrations of nutrients and other chemicals observed in marine environments \citep{SG06,SB13}.

In the same vein, we tackle the second category---to wit, thermal gradients. For the spatial mode of obtaining information, the minimum cell size is
\begin{equation}\label{RtempS}
  R_\mathrm{min} \approx 0.46\,\mathrm{\mu m}\,\left(\frac{\kappa}{\kappa_\oplus}\right)^{-3/13} \left(\frac{\calh}{\calh_\oplus}\right)^{1/13} \left(\frac{\call_t}{\call_{t,\,\oplus}}\right)^{4/13} \left(\frac{\eta}{\eta_\oplus}\right)^{-3/13} \left(\frac{T}{T_\oplus}\right)^{3/13},
\end{equation}
while this quantity under the temporal mode is estimated to be
\begin{equation}\label{RtempT}
  R_\mathrm{min} \approx 0.43\,\mathrm{\mu m}\,\left(\frac{\calu}{\calu_\oplus}\right)^{-4/25}\left(\frac{\kappa}{\kappa_\oplus}\right)^{-3/25} \left(\frac{\calh}{\calh_\oplus}\right)^{1/25} \left(\frac{\call_t}{\call_{t,\,\oplus}}\right)^{4/25} \left(\frac{\eta}{\eta_\oplus}\right)^{-7/25} \left(\frac{T}{T_\oplus}\right)^{7/25},
\end{equation}
where $\kappa$ constitutes the thermal conductivity of the organism and the ambient solvent, $\calh$ represents the volumetric heat capacity of this system, and $\call_t \approx T \left(d T/dz\right)^{-1}$ encapsulates the characteristic length scale linked with the thermal gradient. The normalization factors for these parameters are $\kappa_\oplus = 0.6$ W m$^{-1}$ K$^{-1}$, $\calh_\oplus = 4.2 \times 10^6$ J K$^{-1}$ m$^{-3}$, and $\call_{t,\,\oplus} = 10^4$ m; the last relation follows from the above definition of $\call_t$ along with an average thermal gradient of $30$ K/km and temperature of $288$ K for Earth \citep[pg. 65]{Chi16}.

The third, and last, category of interest is gradients in photon flux (i.e., intensity). As before, one can proceed to determine the minimum cell sizes for the spatial and temporal pathways. After simplification, they are respectively given by
\begin{equation}\label{RradS}
  R_\mathrm{min} \approx 0.64\,\mathrm{\mu m}\,\left(\frac{\delta}{\delta_\oplus}\right)^{-1/7} \left(\frac{\Phi}{\Phi_\oplus}\right)^{-1/7} \left(\frac{\call_\ell}{\call_{\ell,\,\oplus}}\right)^{2/7} \left(\frac{\eta}{\eta_\oplus}\right)^{-1/7} \left(\frac{T}{T_\oplus}\right)^{1/7},
\end{equation}
\begin{equation}\label{RradT}
  R_\mathrm{min} \approx 0.49\,\mathrm{\mu m}\,\left(\frac{\calu}{\calu_\oplus}\right)^{-2/13}\left(\frac{\delta}{\delta_\oplus}\right)^{-1/13} \left(\frac{\Phi}{\Phi_\oplus}\right)^{-1/13} \left(\frac{\call_\ell}{\call_{\ell,\,\oplus}}\right)^{2/13} \left(\frac{\eta}{\eta_\oplus}\right)^{-3/13} \left(\frac{T}{T_\oplus}\right)^{3/13},
\end{equation}
where $\delta$ quantifies the fraction of photons that are absorbed by a single layer composed of photoreceptors (taken to be rhodopsins), $\Phi$ is the flux of photons in a suitable wavelength range, and $\call_{\ell} \approx \Phi \left(d \Phi/dz\right)^{-1}$ represents the characteristic length scale associated with light gradients. The fiducial value of $\delta_\oplus = 3 \times 10^{-4}$ is constructed from the absorption coefficient of photoreceptors \citep[Table 1]{WN98} and the thickness of an individual rhodopsin molecule \citep{Har01}. As the longest dimension of rhodopsin is $\sim 7.5$ nm \citep[pg. 748]{Pal06}, it should not be an issue in principle to fit these molecules into a cell of radius $\gtrsim 0.1$ $\mu$m; the exact number of such molecules will depend on the undetermined packing fraction. We choose the normalization $\Phi_\oplus = 10^{21}$ photons m$^{-2}$ s$^{-1}$, as it corresponds to the maximal flux of $400$-$700$ nm photons incident on the Earth's surface.\footnote{\url{https://www.nrel.gov/grid/solar-resource/spectra.html}} Lastly, in order to maintain consistency with \citet[Table 1]{Dus97}, we have chosen $\call_{\ell,\,\oplus} = 10^{-3}$ m. 

Apart from the spatial and temporal modes of information sensing, microbes are also capable of detecting the orientation of the light source, i.e., discerning the direction in which light is propagating. This process is, however, feasible only when the radiation passing across the organism is subject to substantial attenuation. The minimum cell size required in order to determine the positioning of the light source is expressible as
\begin{equation}\label{RradD}
  R_\mathrm{min} \approx 0.45\,\mathrm{\mu m}\,\left(\frac{\delta}{\delta_\oplus}\right)^{-1/7} \left(\frac{\Phi}{\Phi_\oplus}\right)^{-1/7} \left(\frac{\calk}{\calk_\oplus}\right)^{-2/7} \left(\frac{\eta}{\eta_\oplus}\right)^{-1/7} \left(\frac{T}{T_\oplus}\right)^{1/7},
\end{equation}
where $\calk$ denotes the attenuation coefficient for photoreceptors, and the nominal value of $\calk_\oplus = 10^4$ m$^{-1}$ is taken from \citet[Table 1]{WN98}. 

\begin{figure}
\includegraphics[width=10.5cm]{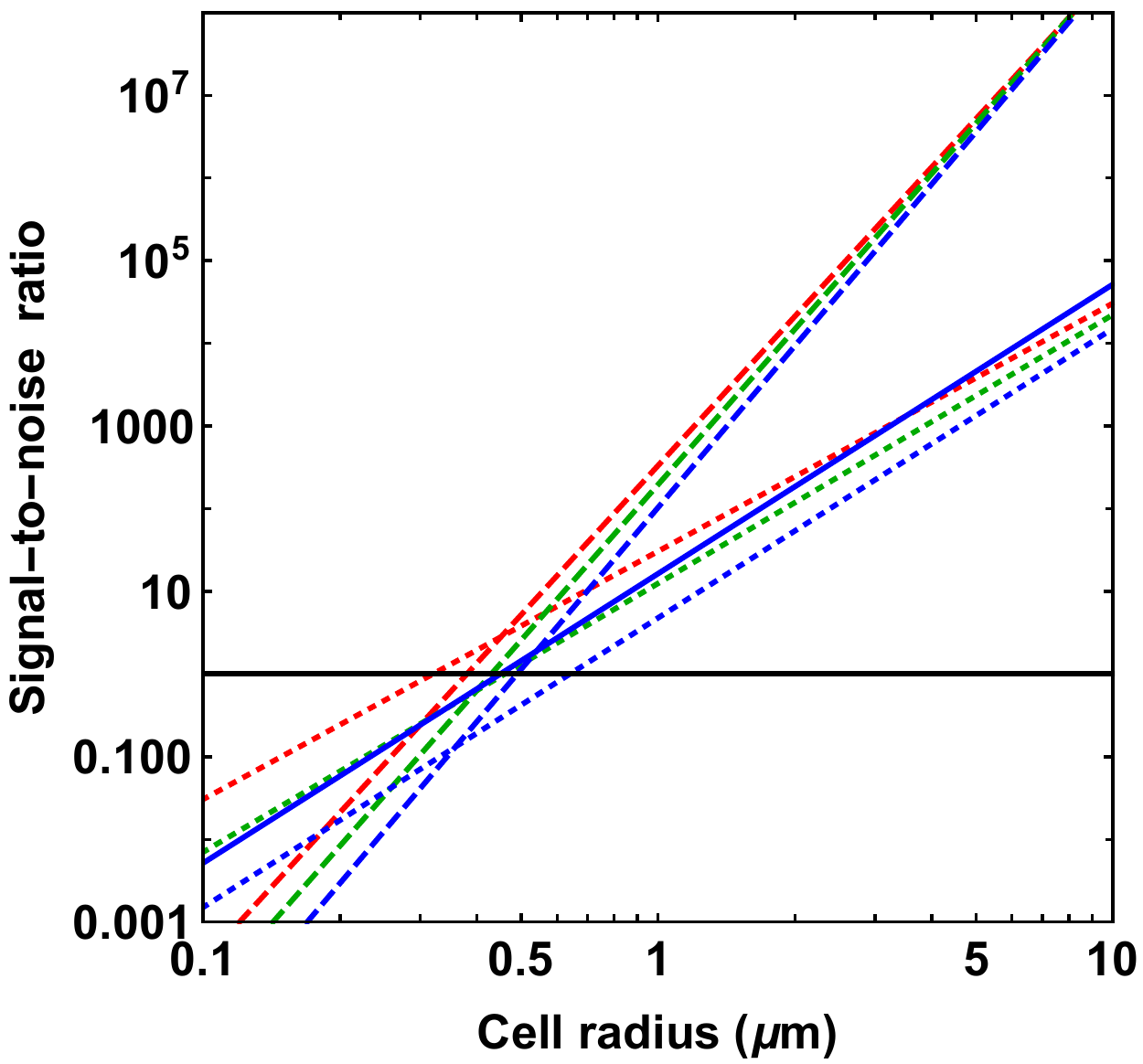} \\
\caption{The signal-to-noise ratio (S/N) as a function of the cell size $R$ (in $\mu$m) for Earth-analogs. The curves reflect the various ambient gradients and attendant pathways for perceiving them. Red, green and blue are used to demarcate chemical, thermal and light gradients, respectively. The dotted and dashed lines (for all colours) indicate the spatial and temporal means for identifying gradients. The solid blue line represents the S/N for detecting the direction of the light source via gradient sensing. All results were obtained by combining (\ref{SNR}) with (\ref{RchemS})-(\ref{RradD}).}
\label{FigSNR}
\end{figure}

For the sake of completeness, the consolidated expression for the signal-to-noise ratio for arbitrary cell radius $R$ is furnished below:
\begin{equation}\label{SNR}
    \mathrm{S/N} = \left(\frac{R}{R_\mathrm{min}}\right)^\beta,
\end{equation}
where the values of $R_\mathrm{min}$ and $\beta$ are dependent on the type and modality of gradient detection. 
\begin{itemize}
    \item Chemical gradients: for the spatial mode, $R_\mathrm{min}$ is calculated using (\ref{RchemS}) and $\beta = 3$, whereas for the temporal mode $R_\mathrm{min}$ is determined by (\ref{RchemT}) and $\beta = 6$.
    \item Thermal gradients: for the spatial mode, $R_\mathrm{min}$ is computed from (\ref{RtempS}) and $\beta = 13/4$, whereas for the temporal mode $R_\mathrm{min}$ is estimated using (\ref{RtempT}) and $\beta = 25/4$.
    \item Light gradients: for the spatial mode, $R_\mathrm{min}$ is evaluated via (\ref{RradS}) and $\beta = 7/2$, whereas for the temporal mode $R_\mathrm{min}$ is ascertained using (\ref{RradT}) and $\beta = 13/2$. When it comes to sensing the direction of the light source, $R_\mathrm{min}$ is set by (\ref{RradD}) and $\beta = 7/2$.
\end{itemize}
From this list, we see that even a modest increase in $R$ would result in a significant gain in the signal-to-noise ratio. To illustrate our point, let us choose $R \approx 3 R_\mathrm{min}$, which translates to $\mathrm{S/N} \approx 30$-$1260$ depending on the modality of gradient sensing. We have plotted S/N as a function of $R$ in Figure \ref{FigSNR}, where the expressions for $R_\mathrm{min}$ in (\ref{SNR}) were held fixed at their normalization factors in (\ref{RchemS})-(\ref{RradD}). In all models, we see that S/N grows rapidly with the size, along expected lines.

Finally, as we had remarked at the beginning of Section \ref{SSecExpCS}, larger cell size permits effective dispersal. The latter is measured in terms of $\zeta$, the ratio of the two motile and non-motile diffusion coefficients, and obeys the relationship:
\begin{equation}
    \zeta = \left(\frac{R}{R_\mathrm{min}}\right)^6,
\end{equation}
where $R_\mathrm{min}$ is given by (\ref{Rdisp}). If we adopt $R \approx 3 R_\mathrm{min}$, as we did in the preceding paragraph, we obtain $\zeta \approx 730$. Thus, even when the cell size is elevated by a modest amount, the accompanying boost in the dispersal is strikingly high. 

\subsection{On the categories of parameters in the equations}
Our final results for the minimal cell size are exemplified by (\ref{Rdisp})-(\ref{RradD}). It is instructive to define $\calr_\mathrm{min}$, which is the \emph{global minimum} of these equations, namely, it refers to the $R_\mathrm{min}$ that is smaller than all the other expressions; it goes without saying, the exact form of $\calr_\mathrm{min}$ will vary based on the parameters of the system under consideration. The mathematical significance of $\calr_\mathrm{min}$ can be understood as follows. If $R < \calr_\mathrm{min}$, it would imply that the organism is incapable of efficient dispersal and gradient sensing. Hence, at least insofar as these issues are concerned, $\calr_\mathrm{min}$ would comprise a viable lower bound for the radius of microbes. 

At this stage, it is helpful to carry out a meticulous scrutiny of the parameters involved in our formulae prior to applying them to specific astrobiological settings. The first group is composed of variables that are chiefly dictated by the nature of the environment(s), which includes the solvent(s). The likes of $\eta$, $\calh$, $\kappa$, and $T$ are straightforward examples, since they can be predicted from basic properties, e.g., solvent, temperature and pressure. On a more subtle level, $\Phi$, $\cald_c$, $\calc$, and the gradient length scales (collectively denoted by $\call$) also fall under this umbrella. It is harder to quantify them precisely, but one could at least draw upon either physical principles or analogues on Earth (see Section \ref{SSecApp}) to estimate them. For instance, calculating the photon flux ($\Phi$) in underwater environments is quite straightforward if the depth, temperature and solvent are provided \citep{LL20}.

The second group is composed of $\delta$ and $\calk$, both of which are intimately connected to the molecular properties of photoreceptors. It is certainly conceivable that extraterrestrial life may use macromolecules other than rhodopsins. Yet, the ubiquity and diversity of rhodopsins warrants further explication. Rhodopsins are widespread in photoreceptors in plants, animals and unicellular eukaryotes \citep{FSP84,Heg08,MYA10}. Looking beyond eukaryotes, the so-called microbial rhodopsins are known to play myriad roles in phototaxis, intracellular signalling and harvesting electromagnetic energy \citep{GLM02,EL14}. Recent studies indicate that microbial rhodopsins constitute the dominant source of light harvesting in Earth's oceans \citep{GCR19}. Despite their multifarious functions, rhodopsins across taxa share many core structural features in common \citep{Bir90,BF06}. It has been conjectured that rhodopsins might have been widespread on the young Earth and exoplanets \citep{DSS18}. In view of these details, it does not seem unreasonable to utilize fiducial values derived from Earth as proxies for $\delta$ and $\calk$.

The remaining parameter, which is also the one subject to the most ambiguity, is $\calu$. The reason is that $\calu$ is dictated by physiology and not by the properties of the medium or individual macromolecules as in the former two categories. A careful inspection of \citet[pg. 270]{MP16} and \citet[Table 1]{BDL16} reveals that $\calu$ deviates from our choice of $\calu_\oplus$ by roughly an order of magnitude only for certain microbes. Due to the relatively weak dependence of $R_\mathrm{min}$ on $\calu$---as evident from (\ref{Rdisp}), (\ref{RchemT}), (\ref{RtempT}) and (\ref{RradT})---our results are likely to change by a factor of $\lesssim 2$. However, these results pertain solely to Earth-based organisms, which leads us to the question: what about extraterrestrial organisms? Although there is admittedly no clear answer, it has been conjectured by \citet[Section 4.4]{MR16} that the magnitude of $\calu$ has a strong mechanistic basis and is therefore constrained to a somewhat narrow interval.

\section{Applications of the formalism}\label{SSecApp}
Before we tackle a couple of specific environments, a few general trends are discernible from Section \ref{SSecExpCS}, which are adumbrated below.
\begin{enumerate}
    \item $R_\mathrm{min}$ decreases when the swimming velocity measured in units of body size is increased.
    \item In all of the formulae, $R_\mathrm{min}$ is proportional to $(T/\eta)^\gamma$, where $\gamma > 0$ although the exact values varies from one expression to another. 
    \item $R_\mathrm{min}$ displays monotonically increasing behaviour with respect to the gradient length scales ($\call$). To put it differently, sharper gradients (i.e., smaller values of $\call$) are predicted to bring about a reduction in $R_\mathrm{min}$.
    \item Broadly speaking, $R_\mathrm{min}$ decreases monotonically with many environmental parameters such as $\cald_c$, $\calc$, $\kappa$ and $\Phi$, although exceptions can and do exist. For such variables, increasing their magnitude would result in lowered $R_\mathrm{min}$ and \emph{vice versa}.
    \item If $\calu$ is held fixed, we perceive that $R_\mathrm{min}$ evinces a stronger algebraic dependence on the appropriate parameters for the spatial mode as opposed to the temporal mode. 
\end{enumerate}
Even though it is tempting to dismiss the above points because they are qualitative, there are several important trends and consequences that emerge from them. We will illustrate this statement with one notable example by focusing on point \#3. 

In habitats with sharp gradients---whether it be chemical, thermal, or photon flux---we find that the corresponding $R_\mathrm{min}$ can decrease significantly, and so could $\calr_\mathrm{min}$. A microbe with smaller cell size, but equipped with similar gradient detection capabilities, would entail a lower metabolic cost \emph{ceteris paribus}, as per current allometric models and empirical data \citep{BGA04,DOM10}. Moreover, theoretical models indicate that the total energetic costs incurred for protein and nucleic acid repair are reduced for smaller cells because they are posited to have a lower inventory of biomolecules \citep[Figures 2 and 3]{KVB17}. Last, if we make the ostensibly reasonable assumption that the first living organisms were on the smaller side, this premise suggests that environments with sharp gradients might have been conducive to the origin of life in this regard.

It is intriguing, therefore, that both past and recent research has emphasized thermodynamic disequilibria as a \emph{sine qua non} for abiogenesis \citep{Sch44,PN71,RDHS,Fry,SM16,BB17,BBGM,MLi19,Spitz}, and several geological settings have been proposed as viable candidates \citep{WHH18,Sle18,KM18,CLH19,MASD,ASR20}. We will explore one of the leading contenders---to wit, submarine alkaline hydrothermal vents---from this standpoint in more detail shortly hereafter. Geothermal fields in subaerial locations (e.g., hot springs), which have a long history in origins-of-life research \citep{Har24}, exhibit marked gradients in inorganic substances, temperature and redox chemistry \citep{BB68,SDS12,MB12,DDK19,DMW19,DD20,BSS20}. Beaches represent another crucial environment that have been relatively underappreciated in origins-of-life research, despite the presence of strong gradients in salinity, temperature and light intensity \citep{La04,BC05,SAB13,LL18}. In the event that life emerged in one (or more) of these domains, it is conceivable that the concomitant existence of steep gradients may have permitted (proto)cells to efficiently acquire information via chemotaxis, phototaxis or thermotaxis among other avenues.

It is worth taking a brief detour at this juncture, and highlighting that the aforementioned variants of taxis are by no means the only ones that abound on Earth. Magnetotactic bacteria are capable of magnetotaxis, whereby these bacteria orient themselves along Earth's magnetic field \citep{RB75,BF04,Sch07,LB13}. Magnetotactic microbes are not tackled in this study for two principal reasons: (i) important questions pertaining to their evolution, ecological distribution, physiology and mechanistic basis are not yet unambiguously settled \citep{EWO07,FS08,LB13,US16}; and (ii) not all worlds possess strong magnetic fields \citep{URC10,DPL15}; the likes of Titan, Venus and Mars are either weakly magnetized or unmagnetized \citep{DJS10,BBM16,HKM20} and the same might hold true for tidally locked terrestrial exoplanets around M-dwarfs \citep{DLMC,DJL18,SM19}. In spite of these caveats, further research along these lines is clearly warranted.

With this essential qualitative discussion out of the way, we will now outline how our formalism could be harnessed to arrive at quantitative predictions by concentrating on two representative locales of relevance to astrobiology.

\subsection{Submarine alkaline hydrothermal vents}\label{SSSecAHV}
Submarine hydrothermal vents have been recognized as promising sites for abiogenesis ever since the 1980s at the minimum \citep{CBH,BH85}, with alkaline hydrothermal vents (AHVs) garnering the lion's share of attention. Reviews and analyses of this rapidly developing field were expounded in \citet{RDHS,RH97,MB08,RBB14,SHW16,WPX18,CR19,RP20}, whereas dissenting viewpoints and critiques of the underlying principles have been laid out in \citet{Bad04,Org08,CAB,Ja16,Suth17,DD17}. In recent experiments, the availability of mineral and metal catalysts has been shown to facilitate the emerge of protometabolic networks under roughly hydrothermal conditions, with close connections to the reverse tricarboxylic acid cycle and the Wood–Ljungdahl pathway \citep{KNY19,PIM20,MVM20,Hud20}. From a different perspective, the synthesis of prebiotic monomers (e.g., amino acids) and their oligomerization has been documented in AHV-like laboratory conditions \citep{BBT,HL18,BFB}. Lastly, both theoretical modelling and laboratory experiments have revealed that hydrothermal pores are propitious to the synthesis and efficient accumulation of prebiotic compounds, and could initiate their oligomerization in turn \citep{BWD07,KKL15,NAD16,SKH20}.

Several reasons can be identified in favor of our decision to focus on AHVs. First, as explained in the preceding paragraph, there are compelling (although not definitive) grounds to contend that they may represent the sites where the origin of life occurred on Earth. Second, AHVs are viewed as promising candidates for enabling abiogenesis on icy worlds in our Solar system such as Europa and Enceladus \citep{VHK07,RBB14}. Last, and perhaps most importantly, the existence of ongoing hydrothermal processes has been indirectly confirmed on Enceladus by analyzing the data from its plume collected by the Cassini spacecraft \citep{WGP17,PKN18} and there is broad evidence for past hydrothermal activity on Mars \citep{OTB13,WFB15,MOM18,FWS20}. On a related note, Triton exhibits clear signatures of geysers \citep{SKB90,HSHC}, and two independent lines of evidence suggest that Europa has plumes and perhaps hydrothermal activity \citep{SSM17,JKK18}. Hence, when viewed collectively, there are robust reasons to apply the model to hydrothermal vents.

The dynamic environment of AHVs is distinguished by the manifestation of steep gradients in chemical compounds, temperature, pH, and redox chemistry \emph{inter alia} \citep{RM04,SAB13,CR19,MASD,BFV20}. We caution, however, that this very spatial and temporal dynamism makes it challenging to identify average values for the gradient length scales. Bearing this caveat in mind, we remark that the presence of strong gradients ensures that point \#3 comes into play. In other words, due to the prominent role of gradients, it seems likely that the expression for $\calr_\mathrm{min}$ would involve $\call$ in one of its three forms. Next, provided that $\calu$ does not diverge significantly from $\calu_\oplus$, we invoke point \#5 delineated previously, which implies that the spatial mode is likely to have a greater impact when the environmental variables are modified. Thus, by combining these two postulates, $\calr_\mathrm{min}$ is given by one of either (\ref{RchemS}), (\ref{RtempS}) or (\ref{RradS}). 

Let us begin by scrutinizing the last equation of this trio. Field studies have established that bacteria from the family \emph{Chlorobiaceae} are capable of photosynthesis at hydrothermal vents \citep{BOL05,RaD13,LOB}. Furthermore, both empirical evidence and biophysical considerations indicate that photosynthesis at photon fluxes $\sim 5$ orders of magnitude smaller than $\Phi_\oplus$ is feasible \citep{RKB00,BOL05,MGK05}. In fact, \citet{NCV95,NF96} hypothesized that photosynthesis evolved in organisms dwelling near hydrothermal vents as a means of initially detecting infrared radiation via thermotaxis. Thus, there are no compelling \emph{a priori} reasons for ruling out the prospects for phototaxis near hydrothermal vents, all the more so given that non-thermal radiation at these locations dominates its thermal counterpart by more than an order of magnitude in select wavelength bands \citep{VRC96,WCR00}. However, upon carefully inspecting (\ref{RradS}), we notice that $R_\mathrm{min} \propto \Phi^{-1/7}$. Given that $\Phi$ is many orders of magnitude smaller than $\Phi_\oplus$ \citep{WCR02}, it follows that $R_\mathrm{min}$ in (\ref{RradS}) may increase by nearly an order of magnitude unless the light gradients are unusually steep. Hence, we will direct our attention toward (\ref{RchemS}) and (\ref{RtempS}) instead.

The remaining step is to motivate characteristic values for the parameters in the formulae. We begin by adopting $T = 323$ K ($50$ $^\circ$C), which is lower than the temperature of hydrothermal fluids in AHVs by a few tens of K \citep{SHW16}.\footnote{We point out that our ensuing results are only weakly dependent on the temperature, and are thus accurate for temperature changes of $< 10\%$ (measured in K).} This temperature has been employed in hydrothermal pore experiments \citep{MSG13}, and higher temperatures could lead to the swift degradation of biomolecules. At this temperature and selecting a pressure of $\sim 25$ MPa, we use \citet[Table 1]{SZF05} to obtain $\eta \approx 0.55 \eta_\oplus$. Next, we adopt $\kappa \approx \kappa_\oplus$ based on \citet{Cal74},\footnote{\url{https://www.engineeringtoolbox.com/water-liquid-gas-thermal-conductivity-temperature-pressure-d_2012.html}} in conjunction with $\calh \approx \calh_\oplus$.\footnote{\url{https://www.engineeringtoolbox.com/specific-heat-capacity-water-d_660.html}} On-site investigations suggest that the thermal gradients function over length scales of $\sim 0.01$-$1$ m \citep{KBD02,BLT10,PYH17}, which is also consistent with some of the microfluidic experiments simulating hydrothermal pores \citep{BWD07,MSG13,PYH17,WS20}; therefore, we adopt the conservative estimate of $\call_t \approx 1$ m. By substituting these values into (\ref{RtempS}), we end up with $R_\mathrm{min} \approx 30$ nm.

Turning our attention to chemotaxis (chemical gradients), the average concentration $\calc$ is strongly dependent on the compounds under consideration. For instance, there is tentative (albeit equivocal) evidence for the abiotic synthesis of amino acids at nanomolar concentrations in the oceanic lithosphere \citep{MPA18}. If we turn our gaze to simpler molecules, however, the concentrations are much higher. To begin with, we choose CO$_2$ as the chemical of interest, due to its prevalence in oceans and the crucial fact that it constitutes the cornerstone of carbon fixation pathways (autotrophy) on Earth \citep{IB11,Fu11,WS19}. RuBisCO, the primary enzyme in carbon fixation, has a characteristic radius of $\sim 6.5$ nm \citep{ESC15}, which could make it theoretically possible for such molecules to exist within a cell of radius $\gtrsim 100$ nm \citep[cf.][]{Rav94}. The dynamic environments emblematic of AHVs induce extensive fluctuations in the CO$_2$ abundance, which ranges between $\sim 0.1$ $\mu$M to $\gtrsim 1$ mM \citep{GCD97,Mi11,LB20}; to offer a benchmark, the typical concentration of dissolved CO$_2$ in Earth's oceans today is $\sim 10$ $\mu$M \citep[pg. 125]{Gor09}. At the lower end of the spectrum (i.e., when the abundance of bioavailable CO$_2$ is $\lesssim 1$ $\mu$M), it is conceivable that CO$_2$ may function as the limiting resource, which is bolstered by the analyses of the Lost City hydrothermal field \citep{LB20}.

We shall opt for an intermediate value of $\calc \approx 10$ $\mu$M, while the diffusion coefficient is chosen to be $\cald_c \approx 3.7\,\cald_\oplus$ \citep[Table 2]{CMT14}. Lastly, experiments in microfluidic reactors have shown that pH gradients are generated over length scales of $\sim 10^{-4}$ m \citep{MKK17,SOM19,BJP20,Hud20}, and we adopt this estimate for $\call_c$ although it is probably on the optimistic side if AHVs are viewed \emph{in toto}.\footnote{If one excludes these ``sweet spots'', however, it is credible that $\call_c$ may increase by an order of magnitude or more even in the neighbourhood of AHVs.} By substituting the preceding choices into (\ref{RchemS}), we obtain $R_\mathrm{min} \approx 130$ nm. In contrast, if we consider dissolved inorganic carbon (DIC) as our chemical of interest and adopt a fiducial estimate of $\sim 2$ mM for DIC on the basis of data collected from Earth's oceans \citep{HLR}, we arrive at $R_\mathrm{min} \approx 55$ nm after invoking (\ref{RchemS}).

nstead of working with CO$_2$ as the chemical compound of relevance, it is feasible to repeat the analysis for a different species. The preceding emphasis on CO$_2$ was a direct consequence of focusing on carbon fixation (autotrophy), but it is necessary to tackle other chemical substances that are relevant for heterotrophs as well. One natural candidate that springs to mind is phosphorus, specifically in the form of phosphates, because it constitutes the ultimate limiting nutrient for the past and current Earth \citep{Ty99,LS18,Hao20}, and potentially worlds with exclusively (sub)surface oceans \citep{WP13,LiMa18,MLAL,MaLi19,OJA20}. A noteworthy aspect of phosphate is that it represents a vital nutrient for both heterotrophs and autotrophs \citep{Ki94}. When it comes to dissolved phosphate, we adopt $\calc \approx 2$ $\mu$M as per measurements in the vicinity of hydrothermal zones on Earth \citep{WFM96,PM07}. The diffusion coefficient for phosphate is taken to be $\cald_c \approx 0.4\,\cald_\oplus$ \citep{KB80,CZG14}. By repeating the calculation using (\ref{RchemS}) in conjunction with the prior data, we end up with $R_\mathrm{min} \approx 25$ nm.

In light of the results until this stage, we infer that $\calr_\mathrm{min} \lesssim 0.1$ $\mu$m is theoretically feasible in the general proximity of AHVs; recall that $\calr_\mathrm{min}$ is the minimum of all values spanned by $R_\mathrm{min}$. However, this statement should not be misconstrued as implying that microbes with radii $\lesssim 0.1$ $\mu$m would exist in actuality. This limit has been derived via the application of a simple ``high-level'' mechanistic model that does not take into account the accompanying constraints imposed by the intricate molecular machinery associated with cells. For instance, cellular components such as receptor molecules (e.g., rhodopsins) are necessary for gradient sensing, whereas motility could enforce stringent requirements for motor molecules and ATP synthesis. Mechanical action based on the sensed information would call for protein and messenger RNA synthesis (entailing RNA polymerases in turn), which militates against a cell radius of $\lesssim 100$ nm. Moreover, in the case of autotrophs, the necessity of large enzymes---such as RuBisCO, which was briefly introduced and analyzed a few paragraphs earlier---imposes an additional bottleneck on the minimal cell size. This issue is, however, likely to be less significant for heterotrophs because they ought not depend on these macromolecules.

There are several other factors that may collectively restrict the minimum radius of living organisms to $\gtrsim 100$ nm. In a classic study, \citet[pg. 52]{Mor67} calculated that a radius of $\gtrsim 50$ nm is necessary for a minimal self-replicating unit on the basis of the desired genetic and metabolic complexity.\footnote{As a sidenote, the composition of a minimal protocell is much more pared down relative to current microbes \citep{OL84,Dyson}, indicating that smaller sizes are conceivable for these entities.} Subsequent evidence from uncultured bacteria as well as additional theoretical constraints---imposed by cell wall, genome size, and nutrient uptake efficiency---suggests that lifeforms must have a minimum radius of $\sim 70$ nm \citep{Koch,MNP97,Vel01,RBL13,LFW15,ABG}. Lastly, the full panoply of evolutionary processes involving the (ex)change of genetic material, such as harboring prophage (which influences the rate of lateral gene transfer), might be rendered untenable when the cell radius is $\lesssim 100$ nm.

It is instructive to compare the above threshold radius of $\sim 100$ nm against a few of the smallest microbes that have been unearthed to date. For starters, the lower limits for picoplankton are known to approach this threshold \citep{LSH83,SDP91,Jo93}. For instance, members of the ubiquitous free-living SAR11 bacterial clade attain an effective radius of $\sim 130$-$200$ nm \citep{RCV02,MKC04}. The parasitic bacterium \emph{Mycoplasma genitalium} is characterized by a radius of $150$ nm \citep{SFA03} and certain species from the phylum \emph{Nanoarchaeota}, such as the recently discovered symbiont \emph{Nanopusillus acidilobi} \citep{WGB16}, attain cell radii of $\gtrsim 50$ nm. But, it is essential to appreciate that some of these specific examples are either parasites or symbionts and not free-living microbes. Hence, by the same token, if the smallest organisms capable of effective gradient sensing and dispersal are parasites on larger lifeforms (which is not the case \emph{tout court}), discerning the latter would prove to be an easier task for life-detection missions. It goes without saying that the limits for viruses, which are, however, not living entities \emph{sensu stricto}, are much smaller: the porcine circovirus and cowpea mosaic virus have radii of $\sim 10$ nm \citep[pg. 8]{MP16}. 

We will explore the ramifications of our analysis in Section \ref{SecConc}, but there are two points that merit highlighting before proceeding further. First, in order to calculate $R_\mathrm{min}$, a number of variables intimately linked to the ambient environment (e.g., gradient length scales) came into play. Hence, in case other domains evince physico-chemical properties similar to AHVs, \emph{mutatis mutandis}, it is plausible that the above results are broadly applicable to them. Geothermal fields might represent promising candidates in this respect, because they possess some attributes in common with AHVs \citep{DD17,Deam19,LL21}, while diverging from the latter environs in certain notable respects (e.g., wet-dry cycling).

Second, in the event that the Earth-based fiducial values chosen in our analysis of AHVs are applicable to similar environments in (sub)surface ocean worlds, most notably Europa and Enceladus, then the results obtained earlier could also be valid in such settings. In other words, it is conceivable that AHVs in subsurface oceans of certain icy worlds might be conducive to the existence of comparatively small microbes capable of effective gradient sensing and dispersal.

\subsection{The role of the solvent: Titan as case study}
Titan is one of the most intriguing and compelling astrobiological targets in our Solar system \citep{Sag94,McK14,NP16,SMI18,Co20}. Our understanding of this moon---such as its methane cycle, atmospheric composition, sand dunes and transport, lakes and seas---has advanced by leaps and bounds, primarily by virtue of the Cassini-Huygens mission \citep{BLW10,Hay16,Hor17,Lor19}. One of the chief reasons as to why Titan is considered an exciting target stems from the possibility that it may harbour ``weird life'', \emph{sensu} being based on hydrocarbon solvents \citep{BRC04,MS05,SLC15,McK16,SR20,CSB20}; see also \citet{STK92,SG05,RBPC,ISM20}. Moreover, Titan-like objects endowed with seas of methane or ethane might be among the most common type of habitable worlds \citep{Ba04,GMK11,BFM19,ML20}, which increases the significance of Titan and its analogs in the context of gauging the frequency of life in the Universe \citep{Lun09}.

Due to Titan's immanent potentiality for hosting life in non-polar solvents, we are free to take this opportunity to examine how the properties of the solvent influence $R_\mathrm{min}$, that is, in what ways does the solvent alone influence the capacity for putative organisms to garner information and thereby take part in directed movement. Hence, we will proceed to modify only the solvent-related parameters like temperature, viscosity and heat conductivity, while leaving the other variables unaltered. We select $T \approx 90$ K and an atmospheric pressure of roughly $1.5$ atm \citep{JC16,Hay16}; our results remain virtually the same if these magnitudes are slightly altered. Next, we must determine what solvent should be utilized in our calculations. Liquid bodies on Titan are composed of both methane and ethane, but we select the former as the model solvent because the northern lakes are mostly composed of it \citep[pp. 460-461]{Hor17}.

We are now in a position to begin computing $R_\mathrm{min}$ in its diverse forms. For the scenario investigated herein, we have $\eta \approx 0.19 \eta_\oplus$,\footnote{\url{https://www.engineeringtoolbox.com/methane-dynamic-kinematic-viscosity-temperature-pressure-d_2068.html}} and substituting these numbers into (\ref{Rdisp}) yields
\begin{equation}\label{RdispT}
  R_\mathrm{min} \approx 0.42\,\mathrm{\mu m}\, \left(\frac{\calu}{\calu_\oplus}\right)^{-1/3}.
\end{equation}
Next, we consider the prospects for chemotaxis; the respective expressions for the spatial and temporal modes can be simplified to obtain
\begin{equation}\label{RchemST}
  R_\mathrm{min} \approx 0.26\,\mathrm{\mu m}\,\left(\frac{\cald_c}{\cald_\mathrm{CH_4}}\right)^{-1/6} \left(\frac{\calc}{\calc_\oplus}\right)^{-1/6} \left(\frac{\call_c}{\call_{c\,\oplus}}\right)^{1/3},
\end{equation}
\begin{equation}\label{RchemTT}
  R_\mathrm{min} \approx 0.37\,\mathrm{\mu m}\,\left(\frac{\calu}{\calu_\oplus}\right)^{-1/6} \left(\frac{\cald_c}{\cald_\mathrm{CH_4}}\right)^{-1/12} \left(\frac{\calc}{\calc_\oplus}\right)^{-1/12} \left(\frac{\call_c}{\call_{c,\,\oplus}}\right)^{1/6},
\end{equation}
where we have chosen to normalize $\cald_c$ in terms of $\cald_\mathrm{CH_4} \approx 5 \cald_\oplus = 5 \times 10^{-9}$ m$^2$ s$^{-1}$ because it approximates the self-diffusivity of methane \citep{VL78,HT80}. Moving on to thermal gradients, the spatial and temporal paths toward sensing information respectively impose the following constraints:
\begin{equation}\label{RtempST}
  R_\mathrm{min} \approx 0.67\,\mathrm{\mu m}\, \left(\frac{\call_t}{\call_{t,\,T}}\right)^{4/13},
\end{equation}
\begin{equation}\label{RtempTT}
  R_\mathrm{min} \approx 0.56\,\mathrm{\mu m}\,\left(\frac{\calu}{\calu_\oplus}\right)^{-4/25} \left(\frac{\call_t}{\call_{t,\,T}}\right)^{4/25},
\end{equation}
where we have made use of $\kappa \approx \kappa_\oplus/3$ \citep[Table 10.39]{VFTT},\footnote{\url{https://www.engineeringtoolbox.com/methane-thermal-conductivity-temperature-pressure-d_2021.html}} and $\calh \approx 0.2 \calh_\oplus$ \citep[Figure 4]{Yo74}. We have normalized $\call_t$ in terms of $\call_{t,\,T} \approx 1.5 \call_{t,\,\oplus}$, where $\call_{t,\,T}$ is the typical gradient at the surface of Titan; this normalization is chosen to make the results Titan-centric. It is calculated by using the definition of $\call_t$ from Section \ref{SSecExpCS}, the value of $T$ adopted earlier, and the scaling $dT/dz \propto \left(R_P\right)^{1.7}$, with $R_P$ representing the radius of the world \citep[e.g.,][]{ML20}. Finally, we turn our attention to the trio of equations associated with phototaxis. The formulae for the spatial and temporal pathways, respectively, are given by the following:
\begin{equation}\label{RradST}
  R_\mathrm{min} \approx 1.30\,\mathrm{\mu m}\,\left(\frac{\delta}{\delta_\oplus}\right)^{-1/7} \left(\frac{\Phi}{\Phi_T}\right)^{-1/7} \left(\frac{\call_\ell}{\call_{\ell,\,\oplus}}\right)^{2/7},
\end{equation}
\begin{equation}\label{RradTT}
  R_\mathrm{min} \approx 0.77\,\mathrm{\mu m}\,\left(\frac{\calu}{\calu_\oplus}\right)^{-2/13}\left(\frac{\delta}{\delta_\oplus}\right)^{-1/13} \left(\frac{\Phi}{\Phi_T}\right)^{-1/13} \left(\frac{\call_\ell}{\call_{\ell,\,\oplus}}\right)^{2/13},
\end{equation}
where we have normalized $\Phi$ in terms of $\Phi_T \approx 0.01 \Phi_\oplus$, the maximal photon flux at Titan's surface for the same reasons as those underlying $\call_{t,\,T}$ in the previous paragraph. To keep our derivation simple, we have used the inverse square law for the photon flux to determine $\Phi_T$ relative to $\Phi_\oplus$. Lastly, we are left with the analogue of (\ref{RradD}), which duly yields
\begin{equation}\label{RradDT}
  R_\mathrm{min} \approx 0.91\,\mathrm{\mu m}\,\left(\frac{\delta}{\delta_\oplus}\right)^{-1/7} \left(\frac{\Phi}{\Phi_T}\right)^{-1/7} \left(\frac{\calk}{\calk_\oplus}\right)^{-2/7}.
\end{equation}

\begin{figure}
\includegraphics[width=10.5cm]{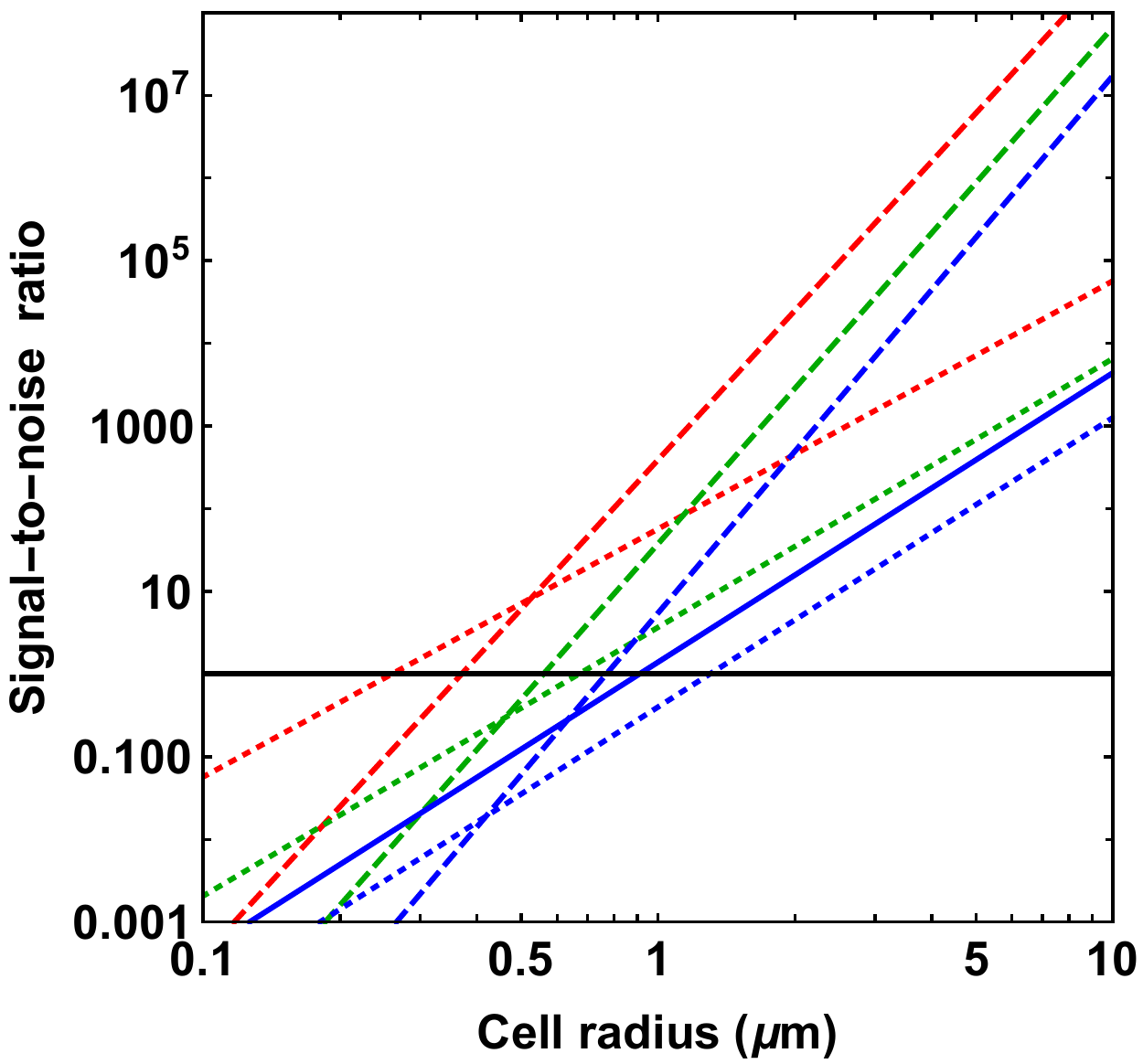} \\
\caption{The signal-to-noise ratio (S/N) as a function of the cell size $R$ (in $\mu$m) for Titan-analogs. The curves reflect the various ambient gradients and attendant pathways for perceiving them. Red, green and blue are used to demarcate chemical, thermal and light gradients, respectively. The dotted and dashed lines (for all colours) indicate the spatial and temporal means for identifying gradients. The solid blue line represents the S/N for detecting the direction of the light source via gradient sensing. All results were obtained by combining (\ref{SNR}) with (\ref{RchemST})-(\ref{RradDT}).}
\label{FigSNRT}
\end{figure}

To recapitulate, (\ref{RdispT})-(\ref{RradDT}) embody the constraints on cell size for Titan and Titan-like worlds, which feature methane as the dominant component of the solvent. By inspecting these equations, we see that, \emph{ceteris paribus}, the smallest magnitude is attained by (\ref{RchemST}). By definition, therefore, we have $\calr_\mathrm{min} \approx 0.26$ $\mu$m, which is nearly identical to Earth's value of $\calr_\mathrm{min} \approx 0.29$ $\mu$m \citep[Table 2]{Dus97}. Thus, at first glimpse, it would appear as though the constraints on cell size on Titan are very close to that on Earth inasmuch as gradient sensing and dispersal is concerned. We caution, however, that these estimates for $\calr_\mathrm{min}$ are applicable to relatively gentle gradients. In contrast, as we have seen in Section \ref{SSSecAHV}, $\calr_\mathrm{min}$ might potentially become lower than the smallest microbes on Earth in settings with prominent chemical or thermal gradients; under similar conditions, these conclusions may apply to Titan as well. 

For the sake of completeness, the signal-to-noise ratio has been plotted as a function of the cell radius in Figure \ref{FigSNRT} for Titan-analogs. In generating this figure, we have made use of (\ref{SNR}) and substituted the fiducial estimates for the expressions for $R_\mathrm{min}$ derived in (\ref{RchemST})-(\ref{RradDT}). In line with our expectations, we notice that S/N is strongly dependent on $R$ and becomes much larger than unity when $R$ exceeds a few $\mu$m, demonstrating that bigger lifeforms are ostensibly advantageous from this perspective, provided that all the other salient factors are held fixed.

There are some general inferences that can be drawn at this juncture. If we compare (\ref{Rdisp})-(\ref{RradD}) and Figure \ref{FigSNR} on the one hand with (\ref{RdispT})-(\ref{RradDT}) and Figure \ref{FigSNRT} on the other, we notice that the two sets are nearly of the same magnitude when the various gradient length scales (e.g., $\call_c$ and $\call_t$) are held fixed. Note that the former set is applicable to Earth as well as other worlds with water as the solvent (e.g., Europa and Enceladus), while the latter pertains to Titan and Titan-like worlds involving methane as the solvent. The similarity is unexpected because these two classes are markedly different in terms of not only their solvents but also other characteristics such as photon flux and temperature.

From a mathematical standpoint, the reason that the two groups yield closely matched results is because the equations for $R_\mathrm{min}$ exhibit a fairly weak dependence on the environmental parameters; to put it differently, there is no single variable that distinctly stands out. The corollary of this statement is that the scalings for $R_\mathrm{min}$ possess a certain degree of universality, provided that the gradient length scales are held invariant. Thus, if we compare $R_\mathrm{min}$ for generic lakes and seas on Titan with those on Earth, the two will be close to each other in terms of their magnitudes. In contrast, if hydrothermal vents on Earth (or Enceladus and Europa) with their accompanying steep gradients are juxtaposed against the lakes and seas of Titan, the former will typically engender a lower value than the latter. To put it more simply, the first scenario is akin to comparing ``apples with apples'', whereas the second is analogous to comparing ``apples with oranges.''

The goal of this section was to demonstrate that the formalism might be suitably modified to tackle worlds with solvents other than water by taking Titan as our case study. Although we shall not delve into this subject further, another intriguing milieu that could merit theoretical analysis in the same vein is the lower and middle cloud decks of Venus, which are known to host liquid sulfuric acid with traces of liquid water \citep{SMI18}. This layer of the Venusian atmosphere has been long regarded as a promising astrobiological target for multiple reasons \citep{MS67,SGA04,GB07,LMS18,GRB20,SPG20,ML21}, but any putative lifeforms are likely to face substantive challenges in this region such as the extremely low water activity, regulation of the osmotic pressure, and high acidity \citep{Co99,CMB21}.

From a more ``exotic'' standpoint, the formalism could be employed to study size constraints in crystalline and amorphous ices. There is growing awareness that glacial and sea ices harbor a diverse range of extremophiles \citep{BAD15,MM18}. There are, however, several challenges that confront putative microbes in these environments insofar as our analysis is concerned. First, the sizes of veins and pores are a few millimeters at the maximum \citep[pg. 2]{MM18}, but at lower temperatures, the typical dimensions are much reduced and may become $\lesssim 1$ $\mu$m for subsurface ocean worlds in the outer Solar system \citep[pg. 218]{PBP07}; this is expected to obstruct motility over extended intervals.

Second, in order for high concentrations of chemicals to exist in the ice---which would be partly responsible for lowering $R_\mathrm{min}$ on the basis of (\ref{RchemS}) and (\ref{RchemT})---they must be incorporated in some fashion from the external environment. However, the diffusion through ice is many orders of magnitude smaller relative to water \citep{MTA13,CWL17}, which is anticipated to impede the accumulation of reactants and nutrients to significant concentrations. Lastly, a number of the parameters inherent in (\ref{Rdisp})-(\ref{RradD}) are poorly constrained in (extra)terrestrial ices, thereby rendering quantitative analyses difficult to undertake. On account of these reasons, we have not attempted to derive explicit estimates for $R_\mathrm{min}$ in the realm of saline and freshwater ices.

\section{Conclusion}\label{SecConc}
The ability of organisms to sense and act upon encountering physical and chemical gradients is of considerable importance. The capacity to garner such information efficiently would, \emph{inter alia}, permit lifeforms to adapt and coexist with their environment, and to undertake locomotion in a directed fashion. Close connections exist between information retrieval and motility on the one hand and cell size on the other, since organisms below a certain size would become inefficient at undertaking the former. By drawing upon results derived by \citet{Dus97,Dus09}, we recast them into formulae for the minimal cell size ($\calr_\mathrm{min}$) that are more amenable to astrobiological analyses; we reiterate that these results should be viewed as heuristic criteria because they are not necessary and sufficient conditions. After deriving these expressions, we applied the framework to a couple of settings.

The first was submarine alkaline hydrothermal vents (AHVs) on Earth and other worlds where they appear to exist (e.g., Enceladus). What we inferred is that, inasmuch as information acquisition via gradient sensing is concerned, AHVs are \emph{theoretically} capable of harboring organisms with radii of $\gtrsim 0.1$ $\mu$m. It is worth reiterating that constraints from other biological functions (e.g., metabolism and replication) could come into play and thereby elevate $\calr_\mathrm{min}$ above the values predicted in this work; some of the relevant caveats and additional factors in this context are delineated in Section \ref{SSSecAHV}. To offer one specific example from the section, autotrophs may require large enzymes for carbon fixation (e.g., RuBisCO) that accordingly elevate the magnitude of $\calr_\mathrm{min}$, whereas this is anticipated to be less significant for heterotrophs. Our analysis does not imply, however, that such sub-$\mu$m microbes originated in these environments or that they would necessarily manifest this size. Nonetheless, our results indicate at the minimum that information-centric arguments taken in isolation are insufficient to rule out the existence of organisms with these dimensions in the neighbourhood of AHVs. 

One might, in fact, contemplate whether the steep gradients at AHVs, which facilitate the reduction in $\calr_\mathrm{min}$, were accordingly amenable to the evolution of sub-$\mu$m microbes in AHV habitats. Lastly, if the parameters we have adopted in the relevant formulae are roughly characteristic of AHVs on icy moons and planets in the outer Solar system like Enceladus and perhaps Europa \citep{NP16,HSHC,JHC20,Tau20}, it is conceivable that those worlds could also support similar organisms. We caution that small microbes with radii $\gtrsim 0.1$ $\mu$m delineated in the above paragraph might prove to be parasitic, in which case it is easier to search for their larger hosts when conducting life-detection experiments. However, on the basis of the simple mechanistic model developed herein, it is not possible to gauge \emph{a priori} whether the smallest microbes with effective gradient sensing and dispersal abilities will always turn out to be obligate parasites or whether they may end up being symbionts or free-living cells. Further empirical research, allied to theoretical modeling, is needed to resolve this issue and come up with optimal search strategies in hunting for microorganisms on other worlds.

In the second instance, we chose to tackle Titan, largely because of its unique potential for hosting exotic life in non-aqueous solvents. By taking the properties of the new solvent (methane) into account as well as the astrophysical attributes of Titan (e.g., its size and location), we computed the constraints on the cell size and thereby deduced $\calr_\mathrm{min}$. We found that, when all other traits such as the microenvironment and organismal physiology are held fixed, $\calr_\mathrm{min}$ is virtually identical to its equivalent on Earth.In mathematical terms, this result was along expected lines because the various formulae for the microbial size exhibit a fairly weak dependence on the relevant physicochemical variables, thereby engendering some degree of universality, provided that the gradient length scales are held fixed. Hence, taken at face value, this result implies that organisms on Titan with the capacity to sense gradients efficiently may perhaps possess a similar lower bound on their size despite the dissimilarities between this world and present-day Earth.

Our work has a number of practical implications in the search for extraterrestrial life via \emph{in situ} life-detection missions. The presence of sharp gradients drives down $\calr_\mathrm{min}$ and might therefore have aided in the early evolution of life. Thus, the search for biomarkers may benefit from prioritizing environments where such gradients exist today or were prevalent in the past; the Gusev crater on Mars with its opaline silica deposits reminiscent of hot springs on Earth is an intriguing example \citep{RF16,MBG18,RCV20}. On the other hand, organisms dwelling in relatively homogeneous environments could require higher $\calr_\mathrm{min}$ if they are to be effective at gradient sensing because of the joint requirements enforced by (\ref{RchemS})-(\ref{SNR}).

The detection of larger organisms is probably more feasible, as is their likelihood of undergoing fossilisation and possibly being unearthed by field studies. In either event, our framework offers a simple heuristic for predicting what are the smallest organisms with gradient sensing and motility in a given domain, and this tool can be gainfully employed in the selection and design of apposite instrumentation. The latter subject is being vigorously pursued vis-\`a-vis Mars, Venus, and the subsurface ocean worlds of the Solar system \citep{BGLK07,HGD17,VWP17,CHP,WFS18,SBZ19,DUP20,Tau20,CBM20,HLE20}. It is not implausible, therefore, that the discovery of extraterrestrial life might progress \emph{a posse ad esse} in the coming decades.

Naturally, theoretical consequences of some import emerge from this study as well. As noted above, a reduction in $\calr_\mathrm{min}$ might prove amenable to the evolution of ``minimalistic'' lifeforms. Furthermore, due to the lowered constraints on information sensing and motility set by size in this case, the transition to more complex organisms may have encountered fewer hurdles. There are reasons to suppose that chemotaxis and movement are intertwined with the prospects for initiating symbiotic interactions and the emergence of complex spatio-temporal aggregates \citep{BB95,Har03,WA04,DK10,PWA11,RFL19}. 

Among the critical steps, the advent of multicellularity stands out because it is distinguished by sophisticated intercellular signalling, coordination and specialization \citep{Shap98,GS07,LK15}, each of which presumably entailed significant ``motion'' in both the physical and informational realms \citep{BCL00,DK10,ZDS12,GA15}. It is tempting to speculate that environments that are more conducive to information acquisition of the type analysed herein would have relatively higher likelihoods for the evolution of simple and complex multicellularity and the profound ecological and evolutionary changes that accompany these transitions \citep{JMS95,Knoll,LL21}; needless to say, this conjecture remains unproven.

\section*{Acknowledgments}
The author is grateful to Michael Russell, Chris McKay, and Charles Cockell for the valuable comments and references pertaining to certain facets of this work. The insightful and constructive feedback provided by the reviewers is also duly acknowledged. This research was supported by the Florida Institute of Technology and the resources provided by the Harvard Library system were of much use during the course of undertaking this study.

%\bibliographystyle{abbrvnat}
%\bibliography{Motility}

\begin{thebibliography}{315}
\providecommand{\natexlab}[1]{#1}
\providecommand{\url}[1]{\texttt{#1}}
\expandafter\ifx\csname urlstyle\endcsname\relax
  \providecommand{\doi}[1]{doi: #1}\else
  \providecommand{\doi}{doi: \begingroup \urlstyle{rm}\Url}\fi

\bibitem[{Adler}(1966)]{Ad66}
J.~{Adler}.
\newblock {Chemotaxis in Bacteria}.
\newblock \emph{Science}, 153\penalty0 (3737):\penalty0 708--716, Aug. 1966.
\newblock \doi{10.1126/science.153.3737.708}.

\bibitem[{Albers} and {Jarrell}(2018)]{AJ18}
S.-V. {Albers} and K.~F. {Jarrell}.
\newblock {The Archaellum: An Update on the Unique Archaeal Motility
  Structure}.
\newblock \emph{Trends Microbiol.}, 26\penalty0 (4):\penalty0 351--362, 2018.
\newblock \doi{10.1016/j.tim.2018.01.004}.

\bibitem[{Alexandre}(2015)]{GA15}
G.~{Alexandre}.
\newblock {Chemotaxis Control of Transient Cell Aggregation}.
\newblock \emph{J. Bacteriol.}, 197\penalty0 (20):\penalty0 3230--3237, 2015.
\newblock \doi{10.1128/JB.00121-15}.

\bibitem[{Alexandre} et~al.(2004){Alexandre}, {Greer-Phillips}, and
  {Zhulin}]{AGZ}
G.~{Alexandre}, S.~{Greer-Phillips}, and I.~B. {Zhulin}.
\newblock {Ecological role of energy taxis in microorganisms}.
\newblock \emph{FEMS Microbiol. Rev.}, 28\penalty0 (1):\penalty0 113--126,
  2004.
\newblock \doi{10.1016/j.femsre.2003.10.003}.

\bibitem[{Alon} et~al.(1999){Alon}, {Surette}, {Barkai}, and {Leibler}]{ASBL}
U.~{Alon}, M.~G. {Surette}, N.~{Barkai}, and S.~{Leibler}.
\newblock {Robustness in bacterial chemotaxis}.
\newblock \emph{Nature}, 397\penalty0 (6715):\penalty0 168--171, Jan. 1999.
\newblock \doi{10.1038/16483}.

\bibitem[{Altair} et~al.(2020){Altair}, {Sartori}, {Rodrigues}, {de Avellar},
  and {Galante}]{ASR20}
T.~{Altair}, L.~M. {Sartori}, F.~{Rodrigues}, M.~G.~B. {de Avellar}, and
  D.~{Galante}.
\newblock {Natural Radioactive Environments as Sources of Local Disequilibrium
  for the Emergence of Life}.
\newblock \emph{Astrobiology}, 20\penalty0 (12):\penalty0 1489--1497, Dec.
  2020.
\newblock \doi{10.1089/ast.2019.2133}.

\bibitem[{Andersen} et~al.(2016){Andersen}, {Berge}, {Gon{\c{c}}alves},
  {Hartvig}, {Heuschele}, {Hylander}, {Jacobsen}, {Lindemann}, {Martens},
  {Neuheimer}, {Olsson}, {Palacz}, {Prowe}, {Sainmont}, {Traving}, {Visser},
  {Wadhwa}, and {Ki{\o}rboe}]{ABG}
K.~H. {Andersen}, T.~{Berge}, R.~J. {Gon{\c{c}}alves}, M.~{Hartvig},
  J.~{Heuschele}, S.~{Hylander}, N.~S. {Jacobsen}, C.~{Lindemann}, E.~A.
  {Martens}, A.~B. {Neuheimer}, K.~{Olsson}, A.~{Palacz}, A.~E.~F. {Prowe},
  J.~{Sainmont}, S.~J. {Traving}, A.~W. {Visser}, N.~{Wadhwa}, and
  T.~{Ki{\o}rboe}.
\newblock {Characteristic Sizes of Life in the Oceans, from Bacteria to
  Whales}.
\newblock \emph{Annu. Rev. Mar. Sci.}, 8:\penalty0 217--241, Jan. 2016.
\newblock \doi{10.1146/annurev-marine-122414-034144}.

\bibitem[{Angilletta}(2009)]{Ang09}
M.~J. {Angilletta}.
\newblock \emph{{Thermal Adaptation: A Theoretical and Empirical Synthesis}}.
\newblock Oxford: Oxford University Press, 2009.

\bibitem[{Armitage}(1999)]{Ar99}
J.~P. {Armitage}.
\newblock {Bacterial Tactic Responses}.
\newblock \emph{Adv. Microb. Physiol.}, 41:\penalty0 229--289, 1999.
\newblock \doi{10.1016/S0065-2911(08)60168-X}.

\bibitem[{Baaske} et~al.(2007){Baaske}, {Weinert}, {Duhr}, {Lemke}, {Russell},
  and {Braun}]{BWD07}
P.~{Baaske}, F.~M. {Weinert}, S.~{Duhr}, K.~H. {Lemke}, M.~J. {Russell}, and
  D.~{Braun}.
\newblock {Extreme accumulation of nucleotides in simulated hydrothermal pore
  systems}.
\newblock \emph{Proc. Natl. Acad. Sci. USA}, 104\penalty0 (22):\penalty0
  9346--9351, May 2007.
\newblock \doi{10.1073/pnas.0609592104}.

\bibitem[{Bada}(2004)]{Bad04}
J.~L. {Bada}.
\newblock {How life began on Earth: a status report}.
\newblock \emph{Earth Planet. Sci. Lett.}, 226\penalty0 (1-2):\penalty0 1--15,
  Sept. 2004.
\newblock \doi{10.1016/j.epsl.2004.07.036}.

\bibitem[{Bains}(2004)]{Ba04}
W.~{Bains}.
\newblock {Many Chemistries Could Be Used to Build Living Systems}.
\newblock \emph{Astrobiology}, 4\penalty0 (2):\penalty0 137--167, June 2004.
\newblock \doi{10.1089/153110704323175124}.

\bibitem[{Ball} et~al.(2007){Ball}, {Garry}, {Lorenz}, and
  {Kerzhanovich}]{BGLK07}
A.~J. {Ball}, J.~R.~C. {Garry}, R.~D. {Lorenz}, and V.~V. {Kerzhanovich}.
\newblock \emph{{Planetary Landers and Entry Probes}}.
\newblock Cambridge: Cambridge University Press, 2007.

\bibitem[{Ballesteros} et~al.(2019){Ballesteros}, {Fernandez-Soto}, and
  {Mart{\'\i}nez}]{BFM19}
F.~J. {Ballesteros}, A.~{Fernandez-Soto}, and V.~J. {Mart{\'\i}nez}.
\newblock {Diving into Exoplanets: Are Water Seas the Most Common?}
\newblock \emph{Astrobiology}, 19\penalty0 (5):\penalty0 642--654, May 2019.
\newblock \doi{10.1089/ast.2017.1720}.

\bibitem[{Barge} et~al.(2017){Barge}, {Branscomb}, {Brucato}, {Cardoso},
  {Cartwright}, {Danielache}, {Galante}, {Kee}, {Miguel}, {Mojzsis},
  {Robinson}, {Russell}, {Simoncini}, and {Sobron}]{BB17}
L.~M. {Barge}, E.~{Branscomb}, J.~R. {Brucato}, S.~S.~S. {Cardoso}, J.~H.~E.
  {Cartwright}, S.~O. {Danielache}, D.~{Galante}, T.~P. {Kee}, Y.~{Miguel},
  S.~{Mojzsis}, K.~J. {Robinson}, M.~J. {Russell}, E.~{Simoncini}, and
  P.~{Sobron}.
\newblock {Thermodynamics, Disequilibrium, Evolution: Far-From-Equilibrium
  Geological and Chemical Considerations for Origin-Of-Life Research}.
\newblock \emph{Orig. Life Evol. Biosph.}, 47\penalty0 (1):\penalty0 39--56,
  Mar. 2017.
\newblock \doi{10.1007/s11084-016-9508-z}.

\bibitem[{Barge} et~al.(2019){Barge}, {Flores}, {Baum}, {Vand erVelde}, and
  {Russell}]{BFB}
L.~M. {Barge}, E.~{Flores}, M.~M. {Baum}, D.~G. {Vand erVelde}, and M.~J.
  {Russell}.
\newblock {Redox and pH gradients drive amino acid synthesis in iron
  oxyhydroxide mineral systems}.
\newblock \emph{Proc. Natl. Acad. Sci. USA}, 116\penalty0 (11):\penalty0
  4828--4833, Mar. 2019.
\newblock \doi{10.1073/pnas.1812098116}.

\bibitem[{Barge} et~al.(2020{\natexlab{a}}){Barge}, {Flores}, {VanderVelde},
  {Weber}, {Baum}, and {Castonguay}]{BFV20}
L.~M. {Barge}, E.~{Flores}, D.~G. {VanderVelde}, J.~M. {Weber}, M.~M. {Baum},
  and A.~{Castonguay}.
\newblock {Effects of Geochemical and Environmental Parameters on Abiotic
  Organic Chemistry Driven by Iron Hydroxide Minerals}.
\newblock \emph{J. Geophys. Res. Planets}, 125\penalty0 (11):\penalty0 e06423,
  Nov. 2020{\natexlab{a}}.
\newblock \doi{10.1029/2020JE006423}.

\bibitem[{Barge} et~al.(2020{\natexlab{b}}){Barge}, {Jones}, {Pagano},
  {Martinez}, and {Bescup}]{BJP20}
L.~M. {Barge}, J.-P. {Jones}, J.~J. {Pagano}, E.~{Martinez}, and J.~{Bescup}.
\newblock {Three-Dimensional Analysis of a Simulated Prebiotic Hydrothermal
  Chimney}.
\newblock \emph{ACS Earth Space Chem.}, 4\penalty0 (9):\penalty0 1663--1669,
  2020{\natexlab{b}}.
\newblock \doi{10.1021/acsearthspacechem.0c00186}.

\bibitem[{Baross} and {Hoffman}(1985)]{BH85}
J.~A. {Baross} and S.~E. {Hoffman}.
\newblock {Submarine hydrothermal vents and associated gradient environments as
  sites for the origin and evolution of life}.
\newblock \emph{Orig. Life Evol. Biosph.}, 15\penalty0 (4):\penalty0 327--345,
  Dec. 1985.
\newblock \doi{10.1007/BF01808177}.

\bibitem[{Bates} et~al.(2010){Bates}, {Lee}, {Tunnicliffe}, and
  {Lamare}]{BLT10}
A.~E. {Bates}, R.~W. {Lee}, V.~{Tunnicliffe}, and M.~D. {Lamare}.
\newblock {Deep-sea hydrothermal vent animals seek cool fluids in a highly
  variable thermal environment}.
\newblock \emph{Nat. Commun.}, 1\penalty0 (2):\penalty0 14, May 2010.
\newblock \doi{10.1038/ncomms1014}.

\bibitem[{Bazylinski} and {Frankel}(2004)]{BF04}
D.~A. {Bazylinski} and R.~B. {Frankel}.
\newblock {Magnetosome formation in prokaryotes}.
\newblock \emph{Nat. Rev. Microbiol.}, 2\penalty0 (3):\penalty0 217--230, 2004.
\newblock \doi{10.1038/nrmicro842}.

\bibitem[{Beatty} et~al.(2005){Beatty}, {Overmann}, {Lince}, {Manske}, {Lang},
  {Blankenship}, {Van Dover}, {Martinson}, and {Plumley}]{BOL05}
J.~T. {Beatty}, J.~{Overmann}, M.~T. {Lince}, A.~K. {Manske}, A.~S. {Lang},
  R.~E. {Blankenship}, C.~L. {Van Dover}, T.~A. {Martinson}, and F.~G.
  {Plumley}.
\newblock {An obligately photosynthetic bacterial anaerobe from a deep-sea
  hydrothermal vent}.
\newblock \emph{Proc. Natl. Acad. Sci. USA}, 102\penalty0 (26):\penalty0
  9306--9310, June 2005.
\newblock \doi{10.1073/pnas.0503674102}.

\bibitem[{Bechinger} et~al.(2016){Bechinger}, {Di Leonardo}, {L{\"o}wen},
  {Reichhardt}, {Volpe}, and {Volpe}]{BDL16}
C.~{Bechinger}, R.~{Di Leonardo}, H.~{L{\"o}wen}, C.~{Reichhardt}, G.~{Volpe},
  and G.~{Volpe}.
\newblock {Active particles in complex and crowded environments}.
\newblock \emph{Rev. Mod. Phys.}, 88\penalty0 (4):\penalty0 045006, Nov. 2016.
\newblock \doi{10.1103/RevModPhys.88.045006}.

\bibitem[{Beeby} et~al.(2020){Beeby}, {Ferreira}, {Tripp}, {Albers}, and
  {Mitchell}]{BFT20}
M.~{Beeby}, J.~L. {Ferreira}, P.~{Tripp}, S.-V. {Albers}, and D.~R. {Mitchell}.
\newblock {Propulsive nanomachines: the convergent evolution of archaella,
  flagella and cilia}.
\newblock \emph{FEMS Microbiol. Rev.}, 44\penalty0 (3):\penalty0 253--304,
  2020.
\newblock \doi{10.1093/femsre/fuaa006}.

\bibitem[{Ben-Jacob} et~al.(2000){Ben-Jacob}, {Cohen}, and {Levine}]{BCL00}
E.~{Ben-Jacob}, I.~{Cohen}, and H.~{Levine}.
\newblock {Cooperative self-organization of microorganisms}.
\newblock \emph{Adv. Phys.}, 49\penalty0 (4):\penalty0 395--554, June 2000.
\newblock \doi{10.1080/000187300405228}.

\bibitem[{Benner} et~al.(2004){Benner}, {Ricardo}, and {Carrigan}]{BRC04}
S.~A. {Benner}, A.~{Ricardo}, and M.~A. {Carrigan}.
\newblock {Is there a common chemical model for life in the universe?}
\newblock \emph{Current opinion in chemical biology}, 8\penalty0 (6):\penalty0
  672--689, 2004.
\newblock \doi{10.1016/j.cbpa.2004.10.003}.

\bibitem[{Berg}(1993)]{Berg}
H.~C. {Berg}.
\newblock \emph{{Random Walks in Biology}}.
\newblock Princeton: Princeton University Press, 2nd edition, 1993.

\bibitem[{Berg} and {Brown}(1972)]{BB72}
H.~C. {Berg} and D.~A. {Brown}.
\newblock {Chemotaxis in Escherichia coli analysed by Three-dimensional
  Tracking}.
\newblock \emph{Nature}, 239\penalty0 (5374):\penalty0 500--504, Oct. 1972.
\newblock \doi{10.1038/239500a0}.

\bibitem[{Berg} and {Purcell}(1977)]{BP77}
H.~C. {Berg} and E.~M. {Purcell}.
\newblock {Physics of chemoreception}.
\newblock \emph{Biophys. J.}, 20\penalty0 (2):\penalty0 193--219, 1977.
\newblock \doi{10.1016/S0006-3495(77)85544-6}.

\bibitem[{Berg}(2011)]{IB11}
I.~A. {Berg}.
\newblock {Ecological Aspects of the Distribution of Different Autotrophic
  CO$_2$ Fixation Pathways}.
\newblock \emph{Appl. Environ. Microbiol.}, 77\penalty0 (6):\penalty0
  1925--1936, 2011.
\newblock \doi{10.1128/AEM.02473-10}.

\bibitem[{Bi} and {Sourjik}(2018)]{BS18}
S.~{Bi} and V.~{Sourjik}.
\newblock {Stimulus sensing and signal processing in bacterial chemotaxis}.
\newblock \emph{Curr. Opin. Microbiol.}, 45:\penalty0 22--29, 2018.
\newblock \doi{10.1016/j.mib.2018.02.002}.

\bibitem[{Bialek} and {Setayeshgar}(2005)]{BS05}
W.~{Bialek} and S.~{Setayeshgar}.
\newblock {Physical limits to biochemical signaling}.
\newblock \emph{Proc. Natl. Acad. Sci. USA}, 102\penalty0 (29):\penalty0
  10040--10045, July 2005.
\newblock \doi{10.1073/pnas.0504321102}.

\bibitem[{Birge}(1990)]{Bir90}
R.~R. {Birge}.
\newblock {Photophysics and Molecular Electronic Applications of the
  Rhodopsins}.
\newblock \emph{Annu. Rev. Phys. Chem.}, 41:\penalty0 683--733, Oct. 1990.
\newblock \doi{10.1146/annurev.pc.41.100190.003343}.

\bibitem[{Bj{\"o}rn}(2015)]{LOB}
L.~O. {Bj{\"o}rn}.
\newblock \emph{{Photobiology: The Science of Light and Life}}.
\newblock New York: Springer, 3rd edition, 2015.

\bibitem[{Blakemore}(1975)]{RB75}
R.~{Blakemore}.
\newblock {Magnetotactic Bacteria}.
\newblock \emph{Science}, 190\penalty0 (4212):\penalty0 377--379, Oct. 1975.
\newblock \doi{10.1126/science.170679}.

\bibitem[{Blanckenhorn}(2000)]{WB00}
W.~U. {Blanckenhorn}.
\newblock {The Evolution of Body Size: What Keeps Organisms Small?}
\newblock \emph{Q. Rev. Biol.}, 75\penalty0 (4):\penalty0 385--407, 2000.
\newblock \doi{10.1086/393620}.

\bibitem[{Boetius} et~al.(2015){Boetius}, {Anesio}, {Deming}, {Mikucki}, and
  {Rapp}]{BAD15}
A.~{Boetius}, A.~M. {Anesio}, J.~W. {Deming}, J.~A. {Mikucki}, and J.~Z.
  {Rapp}.
\newblock {Microbial ecology of the cryosphere: sea ice and glacial habitats}.
\newblock \emph{Nat. Rev. Microbiol.}, 13\penalty0 (11):\penalty0 677--690,
  2015.
\newblock \doi{10.1038/nrmicro3522}.

\bibitem[{Bonner}(1947)]{Bon47}
J.~T. {Bonner}.
\newblock {Evidence for the formation of cell aggregates by chemotaxis in the
  development of the slime mold Dictyostelium discoideum}.
\newblock \emph{J. Exp. Zool.}, 106\penalty0 (1):\penalty0 1--26, 1947.
\newblock \doi{10.1002/jez.1401060102}.

\bibitem[{Bonner}(2006)]{Bon06}
J.~T. {Bonner}.
\newblock \emph{{Why Size Matters: From Bacteria to Blue Whales}}.
\newblock Princeton: Princeton University Press, 2006.

\bibitem[{Boyer} et~al.(2020){Boyer}, {Schubotz}, {Summons}, {Woods}, and
  {Shock}]{BSS20}
G.~M. {Boyer}, F.~{Schubotz}, R.~E. {Summons}, J.~{Woods}, and E.~L. {Shock}.
\newblock {Carbon Oxidation State in Microbial Polar Lipids Suggests Adaptation
  to Hot Spring Temperature and Redox Gradients}.
\newblock \emph{Front. Microbiol.}, 11:\penalty0 229, 2020.
\newblock \doi{10.3389/fmicb.2020.00229}.

\bibitem[{Brain} et~al.(2016){Brain}, {Bagenal}, {Ma}, {Nilsson}, and {Stenberg
  Wieser}]{BBM16}
D.~A. {Brain}, F.~{Bagenal}, Y.~J. {Ma}, H.~{Nilsson}, and G.~{Stenberg
  Wieser}.
\newblock {Atmospheric escape from unmagnetized bodies}.
\newblock \emph{J. Geophys. Res. Planets}, 121\penalty0 (12):\penalty0
  2364--2385, Dec. 2016.
\newblock \doi{10.1002/2016JE005162}.

\bibitem[{Branscomb} et~al.(2017){Branscomb}, {Biancalani}, {Goldenfeld}, and
  {Russell}]{BBGM}
E.~{Branscomb}, T.~{Biancalani}, N.~{Goldenfeld}, and M.~{Russell}.
\newblock {Escapement mechanisms and the conversion of disequilibria; \emph{the
  engines of creation}}.
\newblock \emph{Phys. Rep.}, 677:\penalty0 1--60, 2017.
\newblock \doi{10.1016/j.physrep.2017.02.001}.

\bibitem[{Brenner} et~al.(1998){Brenner}, {Levitov}, and {Budrene}]{BLB98}
M.~{Brenner}, L.~{Levitov}, and E.~{Budrene}.
\newblock {Physical Mechanisms for Chemotactic Pattern Formation by Bacteria}.
\newblock \emph{Biophys. J.}, 74\penalty0 (4):\penalty0 1677--1693, Apr. 1998.
\newblock \doi{10.1016/S0006-3495(98)77880-4}.

\bibitem[{Brock} and {Brock}(1968)]{BB68}
T.~D. {Brock} and M.~L. {Brock}.
\newblock {Relationship between Environmental Temperature and Optimum
  Temperature of Bacteria along a Hot Spring Thermal Gradient}.
\newblock \emph{J. Appl. Bacteriol.}, 31\penalty0 (1):\penalty0 54--58, 1968.
\newblock \doi{10.1111/j.1365-2672.1968.tb00340.x}.

\bibitem[{Brown} et~al.(2004){Brown}, {Gillooly}, {Allen}, {Savage}, and
  {West}]{BGA04}
J.~H. {Brown}, J.~F. {Gillooly}, A.~P. {Allen}, V.~M. {Savage}, and G.~B.
  {West}.
\newblock Toward a metabolic theory of ecology.
\newblock \emph{Ecology}, 85\penalty0 (7):\penalty0 1771--1789, 2004.
\newblock \doi{10.1890/03-9000}.

\bibitem[{Brown} et~al.(2010){Brown}, {Lebreton}, and {Waite}]{BLW10}
R.~H. {Brown}, J.-P. {Lebreton}, and J.~H. {Waite}, editors.
\newblock \emph{{Titan from Cassini-Huygens}}.
\newblock Dordrecht: Springer, 2010.
\newblock \doi{10.1007/978-1-4020-9215-2}.

\bibitem[{Bryant} and {Frigaard}(2006)]{BF06}
D.~A. {Bryant} and N.-U. {Frigaard}.
\newblock {Prokaryotic photosynthesis and phototrophy illuminated}.
\newblock \emph{Trends Microbiol.}, 14\penalty0 (11):\penalty0 488--496, 2006.
\newblock \doi{10.1016/j.tim.2006.09.001}.

\bibitem[{Budrene} and {Berg}(1995)]{BB95}
E.~O. {Budrene} and H.~C. {Berg}.
\newblock {Dynamics of formation of symmetrical patterns by chemotactic
  bacteria}.
\newblock \emph{Nature}, 376\penalty0 (6535):\penalty0 49--53, July 1995.
\newblock \doi{10.1038/376049a0}.

\bibitem[{Burcar} et~al.(2015){Burcar}, {Barge}, {Trail}, {Watson}, {Russell},
  and {McGown}]{BBT}
B.~T. {Burcar}, L.~M. {Barge}, D.~{Trail}, E.~B. {Watson}, M.~J. {Russell}, and
  L.~B. {McGown}.
\newblock {RNA Oligomerization in Laboratory Analogues of Alkaline Hydrothermal
  Vent Systems}.
\newblock \emph{Astrobiology}, 15\penalty0 (7):\penalty0 509--522, July 2015.
\newblock \doi{10.1089/ast.2014.1280}.

\bibitem[{Bywater} and {Conde-Frieboesk}(2005)]{BC05}
R.~P. {Bywater} and K.~{Conde-Frieboesk}.
\newblock {Did Life Begin on the Beach?}
\newblock \emph{Astrobiology}, 5\penalty0 (4):\penalty0 568--574, Aug. 2005.
\newblock \doi{10.1089/ast.2005.5.568}.

\bibitem[{Cadogan} et~al.(2014){Cadogan}, {Maitland}, and {Trusler}]{CMT14}
S.~P. {Cadogan}, G.~C. {Maitland}, and J.~P.~M. {Trusler}.
\newblock {Diffusion Coefficients of CO$_2$ and N$_2$ in Water at Temperatures
  between 298.15 K and 423.15 K at Pressures up to 45 MPa}.
\newblock \emph{J. Chem. Eng. Data}, 59\penalty0 (2):\penalty0 519--525, 2014.
\newblock \doi{10.1021/je401008s}.

\bibitem[{Caldwell}(1974)]{Cal74}
D.~R. {Caldwell}.
\newblock {Thermal conductivity of sea water}.
\newblock \emph{Deep-Sea Res. Oceanogr. Abstr.}, 21\penalty0 (2):\penalty0
  131--137, Jan. 1974.
\newblock \doi{10.1016/0011-7471(74)90070-9}.

\bibitem[{Camley}(2018)]{Cam18}
B.~A. {Camley}.
\newblock {Collective gradient sensing and chemotaxis: modeling and recent
  developments}.
\newblock \emph{J. Phys. Condens. Matter}, 30\penalty0 (22):\penalty0 223001,
  June 2018.
\newblock \doi{10.1088/1361-648X/aabd9f}.

\bibitem[{Camprub{\'\i}} et~al.(2019){Camprub{\'\i}}, {de Leeuw}, {House},
  {Raulin}, {Russell}, {Spang}, {Tirumalai}, and {Westall}]{CLH19}
E.~{Camprub{\'\i}}, J.~W. {de Leeuw}, C.~H. {House}, F.~{Raulin}, M.~J.
  {Russell}, A.~{Spang}, M.~R. {Tirumalai}, and F.~{Westall}.
\newblock {The Emergence of Life}.
\newblock \emph{Space Sci. Rev.}, 215\penalty0 (8):\penalty0 56, Dec. 2019.
\newblock \doi{10.1007/s11214-019-0624-8}.

\bibitem[{Carrer} et~al.(2020){Carrer}, {Skrbic}, {Bore}, {Milano}, {Cascella},
  and {Giacometti}]{CSB20}
M.~{Carrer}, T.~{Skrbic}, S.~L. {Bore}, G.~{Milano}, M.~{Cascella}, and
  A.~{Giacometti}.
\newblock {Can Polarity-Inverted Surfactants Self-Assemble in Nonpolar
  Solvents?}
\newblock \emph{J. Phys. Chem. B}, 124\penalty0 (29):\penalty0 6448--6458,
  2020.
\newblock \doi{10.1021/acs.jpcb.0c04842}.

\bibitem[{Carrier} et~al.(2020){Carrier}, {Beaty}, {Meyer}, {Blank}, {Chou},
  {DasSarma}, {Des Marais}, {Eigenbrode}, {Grefenstette}, {Lanza}, {Schuerger},
  {Schwendner}, {Smith}, {Stoker}, {Tarnas}, {Webster}, {Bakermans}, {Baxter},
  {Bell}, {Benner}, {Bolivar Torres}, {Boston}, {Bruner}, {Clark}, {DasSarma},
  {Engelhart}, {Gallegos}, {Garvin}, {Gasda}, {Green}, {Harris}, {Hoffman},
  {Kieft}, {Koeppel}, {Lee}, {Li}, {Lynch}, {Mackelprang}, {Mahaffy},
  {Matthies}, {Nellessen}, {Newsom}, {Northup}, {O'Connor}, {Perl}, {Quinn},
  {Rowe}, {Sauterey}, {Schneegurt}, {Schulze-Makuch}, {Scuderi}, {Spilde},
  {Stamenkovi{\'c}}, {Torres Celis}, {Viola}, {Wade}, {Walker}, {Wiens},
  {Williams}, {Williams}, and {Xu}]{CBM20}
B.~L. {Carrier}, D.~W. {Beaty}, M.~A. {Meyer}, J.~G. {Blank}, L.~{Chou},
  S.~{DasSarma}, D.~J. {Des Marais}, J.~L. {Eigenbrode}, N.~{Grefenstette},
  N.~L. {Lanza}, A.~C. {Schuerger}, P.~{Schwendner}, H.~D. {Smith}, C.~R.
  {Stoker}, J.~D. {Tarnas}, K.~D. {Webster}, C.~{Bakermans}, B.~K. {Baxter},
  M.~S. {Bell}, S.~A. {Benner}, H.~H. {Bolivar Torres}, P.~J. {Boston},
  R.~{Bruner}, B.~C. {Clark}, P.~{DasSarma}, A.~E. {Engelhart}, Z.~E.
  {Gallegos}, Z.~K. {Garvin}, P.~J. {Gasda}, J.~H. {Green}, R.~L. {Harris},
  M.~E. {Hoffman}, T.~{Kieft}, A.~H.~D. {Koeppel}, P.~A. {Lee}, X.~{Li}, K.~L.
  {Lynch}, R.~{Mackelprang}, P.~R. {Mahaffy}, L.~H. {Matthies}, M.~A.
  {Nellessen}, H.~E. {Newsom}, D.~E. {Northup}, B.~R.~W. {O'Connor}, S.~M.
  {Perl}, R.~C. {Quinn}, L.~A. {Rowe}, B.~{Sauterey}, M.~A. {Schneegurt},
  D.~{Schulze-Makuch}, L.~A. {Scuderi}, M.~N. {Spilde}, V.~{Stamenkovi{\'c}},
  J.~A. {Torres Celis}, D.~{Viola}, B.~D. {Wade}, C.~J. {Walker}, R.~C.
  {Wiens}, A.~J. {Williams}, J.~M. {Williams}, and J.~{Xu}.
\newblock {Mars Extant Life: What's Next? Conference Report}.
\newblock \emph{Astrobiology}, 20\penalty0 (6):\penalty0 785--814, June 2020.
\newblock \doi{10.1089/ast.2020.2237}.

\bibitem[{Cartwright} and {Russell}(2019)]{CR19}
J.~H.~E. {Cartwright} and M.~J. {Russell}.
\newblock {The origin of life: the submarine alkaline vent theory at 30}.
\newblock \emph{Interface Focus}, 9\penalty0 (6):\penalty0 20190104, 2019.
\newblock \doi{10.1098/rsfs.2019.0104}.

\bibitem[{Celani} and {Vergassola}(2010)]{CV10}
A.~{Celani} and M.~{Vergassola}.
\newblock {Bacterial strategies for chemotaxis response}.
\newblock \emph{Proc. Natl. Acad. Sci. USA}, 107\penalty0 (4):\penalty0
  1391--1396, Jan. 2010.
\newblock \doi{10.1073/pnas.0909673107}.

\bibitem[{Chan} et~al.(2019){Chan}, {Hinman}, {Potter-McIntyre}, {Schubert},
  {Gillams}, {Awramik}, {Boston}, {Bower}, {Des Marais}, {Farmer}, {Jia},
  {King}, {Hazen}, {L{\'e}veill{\'e}}, {Papineau}, {Rempfert},
  {S{\'a}nchez-Rom{\'a}n}, {Spear}, {Southam}, {Stern}, and {Cleaves}]{CHP}
M.~A. {Chan}, N.~W. {Hinman}, S.~L. {Potter-McIntyre}, K.~E. {Schubert}, R.~J.
  {Gillams}, S.~M. {Awramik}, P.~J. {Boston}, D.~M. {Bower}, D.~J. {Des
  Marais}, J.~D. {Farmer}, T.~Z. {Jia}, P.~L. {King}, R.~M. {Hazen}, R.~J.
  {L{\'e}veill{\'e}}, D.~{Papineau}, K.~R. {Rempfert},
  M.~{S{\'a}nchez-Rom{\'a}n}, J.~R. {Spear}, G.~{Southam}, J.~C. {Stern}, and
  H.~J. {Cleaves}.
\newblock {Deciphering Biosignatures in Planetary Contexts}.
\newblock \emph{Astrobiology}, 19\penalty0 (9):\penalty0 1075--1102, Sept.
  2019.
\newblock \doi{10.1089/ast.2018.1903}.

\bibitem[{Cheng} et~al.(2014){Cheng}, {Zeng}, {Guo}, and {Zhu}]{CZG14}
X.~{Cheng}, Y.~{Zeng}, Z.~{Guo}, and L.~{Zhu}.
\newblock {Diffusion of Nitrogen and Phosphorus Across the Sediment-Water
  Interface and In Seawater at Aquaculture Areas of Daya Bay, China}.
\newblock \emph{Int. J. Environ. Res. Public Health}, 11\penalty0 (2):\penalty0
  1557--1572, 2014.
\newblock \doi{10.3390/ijerph110201557}.

\bibitem[{Chiasson}(2016)]{Chi16}
A.~D. {Chiasson}.
\newblock \emph{{Geothermal Heat Pump and Heat Engine Systems: Theory And
  Practice}}.
\newblock Chichester: John Wiley \& Sons, 2016.

\bibitem[{Christensen}(2010)]{URC10}
U.~R. {Christensen}.
\newblock {Dynamo Scaling Laws and Applications to the Planets}.
\newblock \emph{Space Sci. Rev.}, 152\penalty0 (1-4):\penalty0 565--590, May
  2010.
\newblock \doi{10.1007/s11214-009-9553-2}.

\bibitem[{Cleaves} et~al.(2009){Cleaves}, {Aubrey}, and {Bada}]{CAB}
H.~J. {Cleaves}, A.~D. {Aubrey}, and J.~L. {Bada}.
\newblock {An Evaluation of the Critical Parameters for Abiotic Peptide
  Synthesis in Submarine Hydrothermal Systems}.
\newblock \emph{Orig. Life Evol. Biosph.}, 39\penalty0 (2):\penalty0 109--126,
  Apr. 2009.
\newblock \doi{10.1007/s11084-008-9154-1}.

\bibitem[{Cockell}(1999)]{Co99}
C.~S. {Cockell}.
\newblock {Life on Venus}.
\newblock \emph{Planet. Space Sci.}, 47\penalty0 (12):\penalty0 1487--1501,
  Dec. 1999.
\newblock \doi{10.1016/S0032-0633(99)00036-7}.

\bibitem[{Cockell}(2018)]{Co18}
C.~S. {Cockell}.
\newblock \emph{{The Equations of Life: How Physics Shapes Evolution}}.
\newblock New York: Basic Books, 2018.

\bibitem[{Cockell}(2020)]{Co20}
C.~S. {Cockell}.
\newblock \emph{{Astrobiology: Understanding Life in the Universe}}.
\newblock Hoboken: John Wiley \& Sons, 2nd edition, 2020.

\bibitem[{Cockell} et~al.(2021){Cockell}, {McMahon}, and {Biddle}]{CMB21}
C.~S. {Cockell}, S.~{McMahon}, and J.~F. {Biddle}.
\newblock {When Is Life a Viable Hypothesis? The Case of Venusian Phosphine}.
\newblock \emph{Astrobiology}, 21\penalty0 (2), 2021.
\newblock \doi{10.1089/ast.2020.2390}.

\bibitem[{Corliss} et~al.(1981){Corliss}, {Baross}, and {Hoffman}]{CBH}
J.~B. {Corliss}, J.~A. {Baross}, and S.~E. {Hoffman}.
\newblock {An hypothesis concerning the relationship between submarine hot
  springs and the origin of life on Earth}.
\newblock \emph{Oceanol. Acta Sp.}, 4:\penalty0 59--69, 1981.
\newblock URL \url{https://archimer.ifremer.fr/doc/00245/35661/34170.pdf}.

\bibitem[{Cremer} et~al.(2019){Cremer}, {Honda}, {Tang}, {Wong-Ng},
  {Vergassola}, and {Hwa}]{CHY}
J.~{Cremer}, T.~{Honda}, Y.~{Tang}, J.~{Wong-Ng}, M.~{Vergassola}, and
  T.~{Hwa}.
\newblock {Chemotaxis as a navigation strategy to boost range expansion}.
\newblock \emph{Nature}, 575\penalty0 (7784):\penalty0 658--663, Nov. 2019.
\newblock \doi{10.1038/s41586-019-1733-y}.

\bibitem[{Cuppen} et~al.(2017){Cuppen}, {Walsh}, {Lamberts}, {Semenov},
  {Garrod}, {Penteado}, and {Ioppolo}]{CWL17}
H.~M. {Cuppen}, C.~{Walsh}, T.~{Lamberts}, D.~{Semenov}, R.~T. {Garrod}, E.~M.
  {Penteado}, and S.~{Ioppolo}.
\newblock {Grain Surface Models and Data for Astrochemistry}.
\newblock \emph{Space Sci. Rev.}, 212\penalty0 (1-2):\penalty0 1--58, Oct.
  2017.
\newblock \doi{10.1007/s11214-016-0319-3}.

\bibitem[{Dachwald} et~al.(2020){Dachwald}, {Ulamec}, {Postberg}, {Sohl}, {de
  Vera}, {Waldmann}, {Lorenz}, {Zacny}, {Hellard}, {Biele}, and
  {Rettberg}]{DUP20}
B.~{Dachwald}, S.~{Ulamec}, F.~{Postberg}, F.~{Sohl}, J.-P. {de Vera},
  C.~{Waldmann}, R.~D. {Lorenz}, K.~A. {Zacny}, H.~{Hellard}, J.~{Biele}, and
  P.~{Rettberg}.
\newblock {Key Technologies and Instrumentation for Subsurface Exploration of
  Ocean Worlds}.
\newblock \emph{Space Sci. Rev.}, 216\penalty0 (5):\penalty0 83, June 2020.
\newblock \doi{10.1007/s11214-020-00707-5}.

\bibitem[{Damer} and {Deamer}(2020)]{DD20}
B.~{Damer} and D.~{Deamer}.
\newblock {The Hot Spring Hypothesis for an Origin of Life}.
\newblock \emph{Astrobiology}, 20\penalty0 (4):\penalty0 429--452, Apr. 2020.
\newblock \doi{10.1089/ast.2019.2045}.

\bibitem[{DasSarma} and {Schwieterman}(2018)]{DSS18}
S.~{DasSarma} and E.~W. {Schwieterman}.
\newblock {Early evolution of purple retinal pigments on Earth and implications
  for exoplanet biosignatures}.
\newblock \emph{Int. J. Astrobiol.}, pages 1--10, 2018.
\newblock \doi{10.1017/S1473550418000423}.

\bibitem[{de Pater} and {Lissauer}(2015)]{DPL15}
I.~{de Pater} and J.~J. {Lissauer}.
\newblock \emph{{Planetary Sciences}}.
\newblock Cambridge: Cambridge University Press, 2015.

\bibitem[{Deamer} and {Damer}(2017)]{DD17}
D.~{Deamer} and B.~{Damer}.
\newblock {Can Life Begin on Enceladus? A Perspective from Hydrothermal
  Chemistry}.
\newblock \emph{Astrobiology}, 17\penalty0 (9):\penalty0 834--839, Sept. 2017.
\newblock \doi{10.1089/ast.2016.1610}.

\bibitem[{Deamer} et~al.(2019){Deamer}, {Damer}, and {Kompanichenko}]{DDK19}
D.~{Deamer}, B.~{Damer}, and V.~{Kompanichenko}.
\newblock {Hydrothermal Chemistry and the Origin of Cellular Life}.
\newblock \emph{Astrobiology}, 19\penalty0 (12):\penalty0 1523--1537, Dec.
  2019.
\newblock \doi{10.1089/ast.2018.1979}.

\bibitem[{Deamer}(2019)]{Deam19}
D.~W. {Deamer}.
\newblock \emph{{Assembling Life: How Can Life Begin on Earth and Other
  Habitable Planets?}}
\newblock Oxford: Oxford University Press, 2019.

\bibitem[{DeLong} et~al.(2010){DeLong}, {Okie}, {Moses}, {Sibly}, and
  {Brown}]{DOM10}
J.~P. {DeLong}, J.~G. {Okie}, M.~E. {Moses}, R.~M. {Sibly}, and J.~H. {Brown}.
\newblock {Shifts in metabolic scaling, production, and efficiency across major
  evolutionary transitions of life}.
\newblock \emph{Proc. Natl. Acad. Sci. USA}, 107\penalty0 (29):\penalty0
  12941--12945, July 2010.
\newblock \doi{10.1073/pnas.1007783107}.

\bibitem[{Demir} and {Salman}(2012)]{DS12}
M.~{Demir} and H.~{Salman}.
\newblock {Bacterial Thermotaxis by Speed Modulation}.
\newblock \emph{Biophys. J.}, 103\penalty0 (8):\penalty0 1683--1690, Oct. 2012.
\newblock \doi{10.1016/j.bpj.2012.09.005}.

\bibitem[{Denny}(1993)]{Den93}
M.~{Denny}.
\newblock \emph{{Air and Water: The Biology and Physics of Life's Media}}.
\newblock Princeton: Princeton University Press, 1993.

\bibitem[{Des Marais} and {Walter}(2019)]{DMW19}
D.~J. {Des Marais} and M.~R. {Walter}.
\newblock {Terrestrial Hot Spring Systems: Introduction}.
\newblock \emph{Astrobiology}, 19\penalty0 (12):\penalty0 1419--1432, Dec.
  2019.
\newblock \doi{10.1089/ast.2018.1976}.

\bibitem[{Dong} et~al.(2017){Dong}, {Lingam}, {Ma}, and {Cohen}]{DLMC}
C.~{Dong}, M.~{Lingam}, Y.~{Ma}, and O.~{Cohen}.
\newblock {Is Proxima Centauri b Habitable? A Study of Atmospheric Loss}.
\newblock \emph{Astrophys. J. Lett.}, 837\penalty0 (2):\penalty0 L26, Mar.
  2017.
\newblock \doi{10.3847/2041-8213/aa6438}.

\bibitem[{Dong} et~al.(2018){Dong}, {Jin}, {Lingam}, {Airapetian}, {Ma}, and
  {van der Holst}]{DJL18}
C.~{Dong}, M.~{Jin}, M.~{Lingam}, V.~S. {Airapetian}, Y.~{Ma}, and B.~{van der
  Holst}.
\newblock {Atmospheric escape from the TRAPPIST-1 planets and implications for
  habitability}.
\newblock \emph{Proc. Natl. Acad. Sci. USA}, 115\penalty0 (2):\penalty0
  260--265, Jan. 2018.
\newblock \doi{10.1073/pnas.1708010115}.

\bibitem[{Dusenbery}(1997)]{Dus97}
D.~B. {Dusenbery}.
\newblock {Minimum Size Limit for Useful Locomotion by Free-Swimming Microbes}.
\newblock \emph{Proc. Natl. Acad. Sci. USA}, 94\penalty0 (20):\penalty0
  10949--10954, Sept. 1997.
\newblock \doi{10.1073/pnas.94.20.10949}.

\bibitem[{Dusenbery}(2009)]{Dus09}
D.~B. {Dusenbery}.
\newblock \emph{{Living at Micro Scale: The Unexpected Physics of Being
  Small}}.
\newblock Cambridge: Harvard University Press, 2009.

\bibitem[{Dyson}(1999)]{Dyson}
F.~{Dyson}.
\newblock \emph{{Origins of Life}}.
\newblock Cambridge: Cambridge University Press, 1999.

\bibitem[{Engel} et~al.(2015){Engel}, {Schaffer}, {Cuellar}, {Villa},
  {Plitzko}, and {Baumeister}]{ESC15}
B.~D. {Engel}, M.~{Schaffer}, L.~K. {Cuellar}, E.~{Villa}, J.~M. {Plitzko}, and
  W.~{Baumeister}.
\newblock {Native architecture of the \emph{Chlamydomonas} chloroplast revealed
  by in situ cryo-electron tomography}.
\newblock \emph{eLife}, 4:\penalty0 e04889, 2015.
\newblock \doi{10.7554/eLife.04889}.

\bibitem[{Engelmann}(1881)]{Eng81}
T.~W. {Engelmann}.
\newblock {Neue Methode zur Untersuchung der Sauerstoffausscheidung
  pflanzlicher und thierischer Organismen}.
\newblock \emph{Pfl{\"u}ger Arch. Gesamte Physiol. Menschen Tiere.},
  25\penalty0 (1):\penalty0 285--292, 1881.
\newblock \doi{10.1007/BF01661982}.

\bibitem[{Erglis} et~al.(2007){Erglis}, {Wen}, {Ose}, {Zeltins}, {Sharipo},
  {Janmey}, and {Cebers}]{EWO07}
K.~{Erglis}, Q.~{Wen}, V.~{Ose}, A.~{Zeltins}, A.~{Sharipo}, P.~{Janmey}, and
  A.~{Cebers}.
\newblock {Dynamics of Magnetotactic Bacteria in a Rotating Magnetic Field}.
\newblock \emph{Biophys. J.}, 93\penalty0 (4):\penalty0 1402--1412, Aug. 2007.
\newblock \doi{10.1529/biophysj.107.107474}.

\bibitem[{Ernst} et~al.(2014){Ernst}, {Lodowski}, {Elstner}, {Hegemann},
  {Brown}, and {Kandori}]{EL14}
O.~P. {Ernst}, D.~T. {Lodowski}, M.~{Elstner}, P.~{Hegemann}, L.~S. {Brown},
  and H.~{Kandori}.
\newblock {Microbial and Animal Rhodopsins: Structures, Functions, and
  Molecular Mechanisms}.
\newblock \emph{Chem. Rev.}, 114\penalty0 (1):\penalty0 126--163, 2014.
\newblock \doi{10.1021/cr4003769}.

\bibitem[{Faivre} and {Schuler}(2008)]{FS08}
D.~{Faivre} and D.~{Schuler}.
\newblock {Magnetotactic Bacteria and Magnetosomes}.
\newblock \emph{Chem. Rev.}, 108\penalty0 (11):\penalty0 4875--4898, 2008.
\newblock \doi{10.1021/cr078258w}.

\bibitem[{Farley} et~al.(2020){Farley}, {Williford}, {Stack}, {Bhartia},
  {Chen}, {de la Torre}, {Hand}, {Goreva}, {Herd}, {Hueso}, {Liu}, {Maki},
  {Martinez}, {Moeller}, {Nelessen}, {Newman}, {Nunes}, {Ponce}, {Spanovich},
  {Willis}, {Beegle}, {Bell}, {Brown}, {Hamran}, {Hurowitz}, {Maurice},
  {Paige}, {Rodriguez-Manfredi}, {Schulte}, and {Wiens}]{FWS20}
K.~A. {Farley}, K.~H. {Williford}, K.~M. {Stack}, R.~{Bhartia}, A.~{Chen},
  M.~{de la Torre}, K.~{Hand}, Y.~{Goreva}, C.~D.~K. {Herd}, R.~{Hueso},
  Y.~{Liu}, J.~N. {Maki}, G.~{Martinez}, R.~C. {Moeller}, A.~{Nelessen}, C.~E.
  {Newman}, D.~{Nunes}, A.~{Ponce}, N.~{Spanovich}, P.~A. {Willis}, L.~W.
  {Beegle}, J.~F. {Bell}, A.~J. {Brown}, S.-E. {Hamran}, J.~A. {Hurowitz},
  S.~{Maurice}, D.~A. {Paige}, J.~A. {Rodriguez-Manfredi}, M.~{Schulte}, and
  R.~C. {Wiens}.
\newblock {Mars 2020 Mission Overview}.
\newblock \emph{Space Sci. Rev.}, 216\penalty0 (8):\penalty0 142, Dec. 2020.
\newblock \doi{10.1007/s11214-020-00762-y}.

\bibitem[{Fenchel}(2002)]{Fen02}
T.~{Fenchel}.
\newblock {Microbial Behavior in a Heterogeneous World}.
\newblock \emph{Science}, 296\penalty0 (5570):\penalty0 1068--1071, May 2002.
\newblock \doi{10.1126/science.1070118}.

\bibitem[{Figueroa-Morales} et~al.(2020){Figueroa-Morales}, {Soto}, {Junot},
  {Darnige}, {Douarche}, {Martinez}, {Lindner}, and {Cl{\'e}ment}]{FSJ20}
N.~{Figueroa-Morales}, R.~{Soto}, G.~{Junot}, T.~{Darnige}, C.~{Douarche},
  V.~A. {Martinez}, A.~{Lindner}, and {\'E}.~{Cl{\'e}ment}.
\newblock {3D Spatial Exploration by E. coli Echoes Motor Temporal
  Variability}.
\newblock \emph{Phys. Rev. X}, 10\penalty0 (2):\penalty0 021004, Apr. 2020.
\newblock \doi{10.1103/PhysRevX.10.021004}.

\bibitem[{Foster} et~al.(1984){Foster}, {Saranak}, {Patel}, {Zarilli}, {Okabe},
  {Kline}, and {Nakanishi}]{FSP84}
K.~W. {Foster}, J.~{Saranak}, N.~{Patel}, G.~{Zarilli}, M.~{Okabe}, T.~{Kline},
  and K.~{Nakanishi}.
\newblock {A rhodopsin is the functional photoreceptor for phototaxis in the
  unicellular eukaryote Chlamydomonas}.
\newblock \emph{Nature}, 311\penalty0 (5988):\penalty0 756--759, Oct. 1984.
\newblock \doi{10.1038/311756a0}.

\bibitem[{Fry}(2000)]{Fry}
I.~{Fry}.
\newblock \emph{{The Emergence of Life on Earth: A Historical and Scientific
  Overview}}.
\newblock New Brunswick: Rutgers University Press, 2000.

\bibitem[{Fu} et~al.(2018){Fu}, {Kato}, {Long}, {Mattingly}, {He}, {Vural},
  {Zucker}, and {Emonet}]{FKL18}
X.~{Fu}, S.~{Kato}, J.~{Long}, H.~H. {Mattingly}, C.~{He}, D.~C. {Vural}, S.~W.
  {Zucker}, and T.~{Emonet}.
\newblock {Spatial self-organization resolves conflicts between individuality
  and collective migration}.
\newblock \emph{Nat. Commun.}, 9:\penalty0 2177, June 2018.
\newblock \doi{10.1038/s41467-018-04539-4}.

\bibitem[{Fuchs}(2011)]{Fu11}
G.~{Fuchs}.
\newblock {Alternative Pathways of Carbon Dioxide Fixation: Insights into the
  Early Evolution of Life?}
\newblock \emph{Annu. Rev. Microbiol.}, 65:\penalty0 631--658, 2011.
\newblock \doi{10.1146/annurev-micro-090110-102801}.

\bibitem[{Gilliam} and {McKay}(2011)]{GMK11}
A.~E. {Gilliam} and C.~P. {McKay}.
\newblock {Titan under a red dwarf star and as a rogue planet: requirements for
  liquid methane}.
\newblock \emph{Planet. Space Sci.}, 59\penalty0 (9):\penalty0 835--839, July
  2011.
\newblock \doi{10.1016/j.pss.2011.03.012}.

\bibitem[{Ginzberg} et~al.(2015){Ginzberg}, {Kafri}, and {Kirschner}]{GFK15}
M.~B. {Ginzberg}, R.~{Kafri}, and M.~{Kirschner}.
\newblock {On being the right (cell) size}.
\newblock \emph{Science}, 348\penalty0 (6236):\penalty0 1245075, 2015.
\newblock \doi{10.1126/science.1245075}.

\bibitem[{Goffredi} et~al.(1997){Goffredi}, {Childress}, {Desaulniers}, {Lee},
  {Lallier}, and {Hammond}]{GCD97}
S.~K. {Goffredi}, J.~J. {Childress}, N.~T. {Desaulniers}, R.~W. {Lee}, F.~H.
  {Lallier}, and D.~{Hammond}.
\newblock {Inorganic carbon acquisition by the hydrothermal vent tubeworm
  \emph{Riftia pachyptila} depends upon high external P${_\mathrm{CO_2}}$ and
  upon proton-equivalent ion transport by the worm}.
\newblock \emph{J. Exp. Biol.}, 200\penalty0 (5):\penalty0 883--896, 1997.

\bibitem[{G{\'o}mez-Consarnau} et~al.(2019){G{\'o}mez-Consarnau}, {Raven},
  {Levine}, {Cutter}, {Wang}, {Seegers}, {Ar{\'\i}stegui}, {Fuhrman}, {Gasol},
  and {Sa{\~n}udo-Wilhelmy}]{GCR19}
L.~{G{\'o}mez-Consarnau}, J.~A. {Raven}, N.~M. {Levine}, L.~S. {Cutter},
  D.~{Wang}, B.~{Seegers}, J.~{Ar{\'\i}stegui}, J.~A. {Fuhrman}, J.~M. {Gasol},
  and S.~A. {Sa{\~n}udo-Wilhelmy}.
\newblock {Microbial rhodopsins are major contributors to the solar energy
  captured in the sea}.
\newblock \emph{Sci. Adv.}, 5\penalty0 (8):\penalty0 eaaw8855, Aug. 2019.
\newblock \doi{10.1126/sciadv.aaw8855}.

\bibitem[{Gordeliy} et~al.(2002){Gordeliy}, {Labahn}, {Moukhametzianov},
  {Efremov}, {Granzin}, {Schlesinger}, {B{\"u}ldt}, {Savopol}, {Scheidig},
  {Klare}, and {Engelhard}]{GLM02}
V.~I. {Gordeliy}, J.~{Labahn}, R.~{Moukhametzianov}, R.~{Efremov},
  J.~{Granzin}, R.~{Schlesinger}, G.~{B{\"u}ldt}, T.~{Savopol}, A.~J.
  {Scheidig}, J.~P. {Klare}, and M.~{Engelhard}.
\newblock {Molecular basis of transmembrane signalling by sensory rhodopsin
  II-transducer complex}.
\newblock \emph{Nature}, 419\penalty0 (6906):\penalty0 484--487, Oct. 2002.
\newblock \doi{10.1038/nature01109}.

\bibitem[{Gornitz}(2009)]{Gor09}
V.~{Gornitz}, editor.
\newblock \emph{{Encyclopedia of Paleoclimatology and Ancient Environments}}.
\newblock Dordrecht: Springer, 2009.

\bibitem[{Greaves} et~al.(2020){Greaves}, {Richards}, {Bains}, {Rimmer},
  {Sagawa}, {Clements}, {Seager}, {Petkowski}, {Sousa-Silva}, {Ranjan},
  {Drabek-Maunder}, {Fraser}, {Cartwright}, {Mueller-Wodarg}, {Zhan},
  {Friberg}, {Coulson}, {Lee}, and {Hoge}]{GRB20}
J.~S. {Greaves}, A.~M.~S. {Richards}, W.~{Bains}, P.~B. {Rimmer}, H.~{Sagawa},
  D.~L. {Clements}, S.~{Seager}, J.~J. {Petkowski}, C.~{Sousa-Silva},
  S.~{Ranjan}, E.~{Drabek-Maunder}, H.~J. {Fraser}, A.~{Cartwright},
  I.~{Mueller-Wodarg}, Z.~{Zhan}, P.~{Friberg}, I.~{Coulson}, E.~{Lee}, and
  J.~{Hoge}.
\newblock {Phosphine gas in the cloud decks of Venus}.
\newblock \emph{Nat. Astron.}, Sept. 2020.
\newblock \doi{10.1038/s41550-020-1174-4}.

\bibitem[{Grinspoon} and {Bullock}(2007)]{GB07}
D.~H. {Grinspoon} and M.~A. {Bullock}.
\newblock {Astrobiology and Venus exploration}.
\newblock \emph{Geophys. Monogr. Ser.}, 176:\penalty0 191--206, Jan. 2007.
\newblock \doi{10.1029/176GM12}.

\bibitem[{Grosberg} and {Strathmann}(2007)]{GS07}
R.~K. {Grosberg} and R.~R. {Strathmann}.
\newblock {The Evolution of Multicellularity: A Minor Major Transition?}
\newblock \emph{Annu. Rev. Ecol. Evol. Syst.}, 38:\penalty0 621--654, 2007.
\newblock \doi{10.1146/annurev.ecolsys.36.102403.114735}.

\bibitem[{Hakim} and {Silberzan}(2017)]{HS17}
V.~{Hakim} and P.~{Silberzan}.
\newblock {Collective cell migration: a physics perspective}.
\newblock \emph{Rep. Prog. Phys.}, 80\penalty0 (7):\penalty0 076601, July 2017.
\newblock \doi{10.1088/1361-6633/aa65ef}.

\bibitem[{Haldane}(1926)]{Hal26}
J.~B.~S. {Haldane}.
\newblock {On being the right size}.
\newblock \emph{Harper’s Magazine}, 152:\penalty0 424--427, 1926.

\bibitem[{Hand} et~al.(2020){Hand}, {Sotin}, {Hayes}, and {Coustenis}]{HSHC}
K.~P. {Hand}, C.~{Sotin}, A.~{Hayes}, and A.~{Coustenis}.
\newblock {On the Habitability and Future Exploration of Ocean Worlds}.
\newblock \emph{Space Science Reviews}, 216\penalty0 (5):\penalty0 1--24, 2020.
\newblock \doi{10.1007/s11214-020-00713-7}.

\bibitem[{Hansen} et~al.(2007){Hansen}, {Lundholm}, and {Rost}]{HLR}
P.~J. {Hansen}, N.~{Lundholm}, and B.~{Rost}.
\newblock {Growth limitation in marine red-tide dinoflagellates: effects of pH
  versus inorganic carbon availability}.
\newblock \emph{Mar. Ecol. Prog. Ser.}, 334:\penalty0 63--71, Mar. 2007.
\newblock \doi{10.3354/meps334063}.

\bibitem[{Hao} et~al.(2020){Hao}, {Knoll}, {Huang}, {Schieber}, {Hazen}, and
  {Daniel}]{Hao20}
J.~{Hao}, A.~H. {Knoll}, F.~{Huang}, J.~{Schieber}, R.~M. {Hazen}, and
  I.~{Daniel}.
\newblock {Cycling phosphorus on the Archean Earth: Part II. Phosphorus
  limitation on primary production in Archean ecosystems}.
\newblock \emph{Geochim. Cosmochim. Acta}, 280:\penalty0 360--377, July 2020.
\newblock \doi{10.1016/j.gca.2020.04.005}.

\bibitem[{Hargrave}(2001)]{Har01}
P.~A. {Hargrave}.
\newblock {Rhodopsin Structure, Function, and Topography The Friedenwald
  Lecture}.
\newblock \emph{Investig. Ophthalmol. Vis. Sci.}, 42\penalty0 (1):\penalty0
  3--9, 2001.

\bibitem[{Harris} and {Trappeniers}(1980)]{HT80}
K.~R. {Harris} and N.~J. {Trappeniers}.
\newblock {The density dependence of the self-diffusion coefficient of liquid
  methane}.
\newblock \emph{Physica A}, 104\penalty0 (1):\penalty0 262--280, Nov. 1980.
\newblock \doi{10.1016/0378-4371(80)90087-4}.

\bibitem[{Harrison} and {Lane}(2018)]{HL18}
S.~A. {Harrison} and N.~{Lane}.
\newblock {Life as a guide to prebiotic nucleotide synthesis}.
\newblock \emph{Nat. Commun.}, 9:\penalty0 5176, Dec. 2018.
\newblock \doi{10.1038/s41467-018-07220-y}.

\bibitem[{Harshey}(2003)]{Har03}
R.~M. {Harshey}.
\newblock {Bacterial Motility on a Surface: Many Ways to a Common Goal}.
\newblock \emph{Annu. Rev. Microbiol.}, 57\penalty0 (1):\penalty0 249--273,
  2003.
\newblock \doi{10.1146/annurev.micro.57.030502.091014}.

\bibitem[{Harvey}(1924)]{Har24}
R.~B. {Harvey}.
\newblock {Enzymes of Thermal Algae}.
\newblock \emph{Science}, 60\penalty0 (1560):\penalty0 481--482, Nov. 1924.
\newblock \doi{10.1126/science.60.1560.481}.

\bibitem[{Hayes}(2016)]{Hay16}
A.~G. {Hayes}.
\newblock {The Lakes and Seas of Titan}.
\newblock \emph{Annu. Rev. Earth Planet. Sci.}, 44:\penalty0 57--83, June 2016.
\newblock \doi{10.1146/annurev-earth-060115-012247}.

\bibitem[{Hays} et~al.(2017){Hays}, {Graham}, {Des Marais}, {Hausrath},
  {Horgan}, {McCollom}, {Parenteau}, {Potter-McIntyre}, {Williams}, and
  {Lynch}]{HGD17}
L.~E. {Hays}, H.~V. {Graham}, D.~J. {Des Marais}, E.~M. {Hausrath},
  B.~{Horgan}, T.~M. {McCollom}, M.~N. {Parenteau}, S.~L. {Potter-McIntyre},
  A.~J. {Williams}, and K.~L. {Lynch}.
\newblock {Biosignature Preservation and Detection in Mars Analog
  Environments}.
\newblock \emph{Astrobiology}, 17\penalty0 (4):\penalty0 363--400, Apr. 2017.
\newblock \doi{10.1089/ast.2016.1627}.

\bibitem[{Hazelbauer}(2012)]{GLH12}
G.~L. {Hazelbauer}.
\newblock {Bacterial Chemotaxis: The Early Years of Molecular Studies}.
\newblock \emph{Annu. Rev. Microbiol.}, 66:\penalty0 285--303, 2012.
\newblock \doi{10.1146/annurev-micro-092611-150120}.

\bibitem[{Hegemann}(2008)]{Heg08}
P.~{Hegemann}.
\newblock {Algal Sensory Photoreceptors}.
\newblock \emph{Annu. Rev. Plant Biol.}, 59:\penalty0 167--189, 2008.
\newblock \doi{10.1146/annurev.arplant.59.032607.092847}.

\bibitem[{Hein} et~al.(2020){Hein}, {Lingam}, {Marshall Eubanks}, {Hibberd},
  {Fries}, and {Blase}]{HLE20}
A.~M. {Hein}, M.~{Lingam}, T.~{Marshall Eubanks}, A.~{Hibberd}, D.~{Fries}, and
  W.~P. {Blase}.
\newblock {A Precursor Balloon Mission for Venusian Astrobiology}.
\newblock \emph{Astrophys. J. Lett.}, 903\penalty0 (2):\penalty0 L36, Nov.
  2020.
\newblock \doi{10.3847/2041-8213/abc347}.

\bibitem[{Herzog} and {Wirth}(2012)]{HW12}
B.~{Herzog} and R.~{Wirth}.
\newblock {Swimming Behavior of Selected Species of \emph{Archaea}}.
\newblock \emph{Appl. Environ. Microbiol.}, 78\penalty0 (6):\penalty0
  1670--1674, 2012.
\newblock \doi{10.1128/AEM.06723-11}.

\bibitem[{Horner} et~al.(2020){Horner}, {Kane}, {Marshall}, {Dalba}, {Holt},
  {Wood}, {Maynard-Casely}, {Wittenmyer}, {Lykawka}, {Hill}, {Salmeron},
  {Bailey}, {L{\"o}hne}, {Agnew}, {Carter}, and {Tylor}]{HKM20}
J.~{Horner}, S.~R. {Kane}, J.~P. {Marshall}, P.~A. {Dalba}, T.~R. {Holt},
  J.~{Wood}, H.~E. {Maynard-Casely}, R.~{Wittenmyer}, P.~S. {Lykawka},
  M.~{Hill}, R.~{Salmeron}, J.~{Bailey}, T.~{L{\"o}hne}, M.~{Agnew}, B.~D.
  {Carter}, and C.~C.~E. {Tylor}.
\newblock {Solar System Physics for Exoplanet Research}.
\newblock \emph{Publ. Astron. Soc. Pac.}, 132\penalty0 (1016):\penalty0 102001,
  Oct. 2020.
\newblock \doi{10.1088/1538-3873/ab8eb9}.

\bibitem[{H{\"o}rst}(2017)]{Hor17}
S.~M. {H{\"o}rst}.
\newblock {Titan's atmosphere and climate}.
\newblock \emph{J. Geophys. Res. Planets}, 122\penalty0 (3):\penalty0 432--482,
  Mar. 2017.
\newblock \doi{10.1002/2016JE005240}.

\bibitem[{Hu} and {Tu}(2014)]{HT14}
B.~{Hu} and Y.~{Tu}.
\newblock {Behaviors and Strategies of Bacterial Navigation in Chemical and
  Nonchemical Gradients}.
\newblock \emph{PLoS Comput. Biol.}, 10\penalty0 (6):\penalty0 e1003672, 2014.
\newblock \doi{10.1371/journal.pcbi.1003672}.

\bibitem[{Hudson} et~al.(2020){Hudson}, {de Graaf}, {Rodin}, {Ohno}, {Lane},
  {McGlynn}, {Yamada}, {Nakamura}, {Barge}, {Braun}, and {Sojo}]{Hud20}
R.~{Hudson}, R.~{de Graaf}, M.~S. {Rodin}, A.~{Ohno}, N.~{Lane}, S.~E.
  {McGlynn}, Y.~M.~A. {Yamada}, R.~{Nakamura}, L.~M. {Barge}, D.~{Braun}, and
  V.~{Sojo}.
\newblock {CO$_2$ reduction driven by a pH gradient}.
\newblock \emph{Proc. Natl. Acad. Sci. USA}, 117\penalty0 (37):\penalty0
  22873--22879, 2020.
\newblock \doi{10.1073/pnas.2002659117}.

\bibitem[{Irwin} and {Schulze-Makuch}(2020)]{ISM20}
L.~N. {Irwin} and D.~{Schulze-Makuch}.
\newblock {The Astrobiology of Alien Worlds: Known and Unknown Forms of Life}.
\newblock \emph{Universe}, 6\penalty0 (9):\penalty0 130, Aug. 2020.
\newblock \doi{10.3390/universe6090130}.

\bibitem[{Jackson}(2016)]{Ja16}
J.~B. {Jackson}.
\newblock {Natural pH Gradients in Hydrothermal Alkali Vents Were Unlikely to
  Have Played a Role in the Origin of Life}.
\newblock \emph{J. Mol. Evol.}, 83\penalty0 (1-2):\penalty0 1--11, Aug. 2016.
\newblock \doi{10.1007/s00239-016-9756-6}.

\bibitem[{Jebbar} et~al.(2020){Jebbar}, {Hickman-Lewis}, {Cavalazzi},
  {Taubner}, {Rittmann}, and {Antunes}]{JHC20}
M.~{Jebbar}, K.~{Hickman-Lewis}, B.~{Cavalazzi}, R.-S. {Taubner}, S.~K. M.~R.
  {Rittmann}, and A.~{Antunes}.
\newblock {Microbial Diversity and Biosignatures: An Icy Moons Perspective}.
\newblock \emph{Space Sci. Rev.}, 216\penalty0 (1):\penalty0 10, Jan. 2020.
\newblock \doi{10.1007/s11214-019-0620-z}.

\bibitem[{J{\'e}kely}(2009)]{Jek09}
G.~{J{\'e}kely}.
\newblock {Evolution of phototaxis}.
\newblock \emph{Phil. Trans. R. Soc. B}, 364\penalty0 (1531):\penalty0
  2795--2808, 2009.
\newblock \doi{10.1098/rstb.2009.0072}.

\bibitem[{Jennings} et~al.(2016){Jennings}, {Cottini}, {Nixon}, {Achterberg},
  {Flasar}, {Kunde}, {Romani}, {Samuelson}, {Mamoutkine}, {Gorius},
  {Coustenis}, and {Tokano}]{JC16}
D.~E. {Jennings}, V.~{Cottini}, C.~A. {Nixon}, R.~K. {Achterberg}, F.~M.
  {Flasar}, V.~G. {Kunde}, P.~N. {Romani}, R.~E. {Samuelson}, A.~{Mamoutkine},
  N.~J.~P. {Gorius}, A.~{Coustenis}, and T.~{Tokano}.
\newblock {Surface Temperatures on Titan during Northern Winter and Spring}.
\newblock \emph{Astrophys. J. Lett.}, 816\penalty0 (1):\penalty0 L17, Jan.
  2016.
\newblock \doi{10.3847/2041-8205/816/1/L17}.

\bibitem[{Jia} et~al.(2018){Jia}, {Kivelson}, {Khurana}, and {Kurth}]{JKK18}
X.~{Jia}, M.~G. {Kivelson}, K.~K. {Khurana}, and W.~S. {Kurth}.
\newblock {Evidence of a plume on Europa from Galileo magnetic and plasma wave
  signatures}.
\newblock \emph{Nat. Astron.}, 2:\penalty0 459--464, May 2018.
\newblock \doi{10.1038/s41550-018-0450-z}.

\bibitem[{Jones}(1993)]{Jo93}
J.~G. {Jones}.
\newblock \emph{{Advances in Microbial Ecology}}, volume~13.
\newblock New York: Springer, 1993.
\newblock \doi{10.1007/978-1-4615-2858-6}.

\bibitem[{Kearns}(2010)]{DK10}
D.~B. {Kearns}.
\newblock {A field guide to bacterial swarming motility}.
\newblock \emph{Nat. Rev. Microbiol.}, 8\penalty0 (9):\penalty0 634--644, 2010.
\newblock \doi{10.1038/nrmicro2405}.

\bibitem[{Keller} and {Segel}(1971)]{KS71}
E.~F. {Keller} and L.~A. {Segel}.
\newblock {Model for chemotaxis}.
\newblock \emph{J. Theor. Biol.}, 30\penalty0 (2):\penalty0 225--234, 1971.
\newblock \doi{10.1016/0022-5193(71)90050-6}.

\bibitem[{Kelley} et~al.(2002){Kelley}, {Baross}, and {Delaney}]{KBD02}
D.~S. {Kelley}, J.~A. {Baross}, and J.~R. {Delaney}.
\newblock {Volcanoes, Fluids, and Life at Mid-Ocean Ridge Spreading Centers}.
\newblock \emph{Annu. Rev. Earth Planet. Sci.}, 30:\penalty0 385--491, Jan.
  2002.
\newblock \doi{10.1146/annurev.earth.30.091201.141331}.

\bibitem[{Kempes} et~al.(2017){Kempes}, {van Bodegom}, {Wolpert}, {Libby},
  {Amend}, and {Hoehler}]{KVB17}
C.~P. {Kempes}, P.~M. {van Bodegom}, D.~{Wolpert}, E.~{Libby}, J.~{Amend}, and
  T.~{Hoehler}.
\newblock {Drivers of Bacterial Maintenance and Minimal Energy Requirements}.
\newblock \emph{Front. Microbiol.}, 8:\penalty0 31, 2017.
\newblock \doi{10.3389/fmicb.2017.00031}.

\bibitem[{Kempes} et~al.(2019){Kempes}, {Koehl}, and {West}]{Kemp19}
C.~P. {Kempes}, M.~A.~R. {Koehl}, and G.~B. {West}.
\newblock {The Scales That Limit: The Physical Boundaries of Evolution}.
\newblock \emph{Front. Ecol. Evol.}, 7:\penalty0 242, 2019.
\newblock \doi{10.3389/fevo.2019.00242}.

\bibitem[{Ki{\o}rboe}(2008)]{Kio08}
T.~{Ki{\o}rboe}.
\newblock \emph{{A Mechanistic Approach to Plankton Ecology}}.
\newblock Princeton: Princeton University Press, 2008.

\bibitem[{Kirchman}(1994)]{Ki94}
D.~L. {Kirchman}.
\newblock The uptake of inorganic nutrients by heterotrophic bacteria.
\newblock \emph{Microb. Ecol.}, 28\penalty0 (2):\penalty0 255--271, 1994.
\newblock \doi{10.1007/BF00166816}.

\bibitem[{Kirchman}(2018)]{Kir18}
D.~L. {Kirchman}.
\newblock \emph{{Processes in Microbial Ecology}}.
\newblock Oxford: Oxford University Press, 2nd edition, 2018.

\bibitem[{Kitadai} and {Maruyama}(2018)]{KM18}
N.~{Kitadai} and S.~{Maruyama}.
\newblock {Origins of building blocks of life: A review}.
\newblock \emph{Geosci. Front.}, 9\penalty0 (4):\penalty0 1117--1153, 2018.
\newblock \doi{10.1016/j.gsf.2017.07.007}.

\bibitem[{Kitadai} et~al.(2019){Kitadai}, {Nakamura}, {Yamamoto}, {Takai},
  {Yoshida}, and {Oono}]{KNY19}
N.~{Kitadai}, R.~{Nakamura}, M.~{Yamamoto}, K.~{Takai}, N.~{Yoshida}, and
  Y.~{Oono}.
\newblock {Metals likely promoted protometabolism in early ocean alkaline
  hydrothermal systems}.
\newblock \emph{Sci. Adv.}, 5\penalty0 (6):\penalty0 eaav7848, June 2019.
\newblock \doi{10.1126/sciadv.aav7848}.

\bibitem[{Knoll} et~al.(1999){Knoll}, {Osborn}, {Baross}, {Berg}, {Pace}, and
  {Sogin}]{KOB99}
A.~{Knoll}, M.~J. {Osborn}, J.~{Baross}, H.~C. {Berg}, N.~R. {Pace}, and
  M.~{Sogin}.
\newblock \emph{{Size Limits of Very Small Microorganisms: Proceedings of a
  Workshop}}.
\newblock Washington, DC: National Academy Press, 1999.

\bibitem[{Knoll}(2015)]{Knoll}
A.~H. {Knoll}.
\newblock \emph{{Life on a Young Planet: The First Three Billion Years of
  Evolution on Earth}}.
\newblock Princeton Science Library. Princeton: Princeton University Press,
  2015.

\bibitem[{Koch}(1996)]{Koch}
A.~L. {Koch}.
\newblock {What size should a bacterium be? A question of scale}.
\newblock \emph{Annu. Rev. Microbiol.}, 50\penalty0 (1):\penalty0 317--348,
  1996.
\newblock \doi{10.1146/annurev.micro.50.1.317}.

\bibitem[{Kreysing} et~al.(2015){Kreysing}, {Keil}, {Lanzmich}, and
  {Braun}]{KKL15}
M.~{Kreysing}, L.~{Keil}, S.~{Lanzmich}, and D.~{Braun}.
\newblock {Heat flux across an open pore enables the continuous replication and
  selection of oligonucleotides towards increasing length}.
\newblock \emph{Nat. Chem.}, 7\penalty0 (3):\penalty0 203--208, Mar. 2015.
\newblock \doi{10.1038/nchem.2155}.

\bibitem[{Krom} and {Berner}(1980)]{KB80}
M.~D. {Krom} and R.~A. {Berner}.
\newblock {The diffusion coefficients of sulfate, ammonium, and phosphate ions
  in anoxic marine sediments1}.
\newblock \emph{Limnol. Oceanogr.}, 25\penalty0 (2):\penalty0 327--337, Mar.
  1980.
\newblock \doi{10.4319/lo.1980.25.2.0327}.

\bibitem[{Laakso} and {Schrag}(2018)]{LS18}
T.~A. {Laakso} and D.~P. {Schrag}.
\newblock {Limitations on Limitation}.
\newblock \emph{Global Biogeochem. Cy.}, 32\penalty0 (3):\penalty0 486--496,
  Mar. 2018.
\newblock \doi{10.1002/2017GB005832}.

\bibitem[{Lang} and {Brazelton}(2020)]{LB20}
S.~Q. {Lang} and W.~J. {Brazelton}.
\newblock {Habitability of the marine serpentinite subsurface: a case study of
  the Lost City hydrothermal field}.
\newblock \emph{Phil. Trans. R. Soc. A}, 378\penalty0 (2165):\penalty0
  20180429, 2020.
\newblock \doi{10.1098/rsta.2018.0429}.

\bibitem[{LaRowe} and {Amend}(2015)]{LA15}
D.~E. {LaRowe} and J.~P. {Amend}.
\newblock {Power limits for microbial life}.
\newblock \emph{Front. Microbiol.}, 6:\penalty0 718, 2015.
\newblock \doi{10.3389/fmicb.2015.00718}.

\bibitem[{Lathe}(2004)]{La04}
R.~{Lathe}.
\newblock {Fast tidal cycling and the origin of life}.
\newblock \emph{Icarus}, 168\penalty0 (1):\penalty0 18--22, Mar. 2004.
\newblock \doi{10.1016/j.icarus.2003.10.018}.

\bibitem[{Lauga}(2016)]{Lau16}
E.~{Lauga}.
\newblock {Bacterial Hydrodynamics}.
\newblock \emph{Annu. Rev. Fluid Mech.}, 48\penalty0 (1):\penalty0 105--130,
  Jan. 2016.
\newblock \doi{10.1146/annurev-fluid-122414-034606}.

\bibitem[{Lauro} et~al.(2009){Lauro}, {McDougald}, {Thomas}, {Williams},
  {Egan}, {Rice}, {DeMaere}, {Ting}, {Ertan}, {Johnson}, {Ferriera}, {Lapidus},
  {Anderson}, {Kyrpides}, {Munk}, {Detter}, {Han}, {Brown}, {Robb},
  {Kjelleberg}, and {Cavicchioli}]{LMT}
F.~M. {Lauro}, D.~{McDougald}, T.~{Thomas}, T.~J. {Williams}, S.~{Egan},
  S.~{Rice}, M.~Z. {DeMaere}, L.~{Ting}, H.~{Ertan}, J.~{Johnson},
  S.~{Ferriera}, A.~{Lapidus}, I.~{Anderson}, N.~{Kyrpides}, A.~C. {Munk},
  C.~{Detter}, C.~S. {Han}, M.~V. {Brown}, F.~T. {Robb}, S.~{Kjelleberg}, and
  R.~{Cavicchioli}.
\newblock {The genomic basis of trophic strategy in marine bacteria}.
\newblock \emph{Proc. Natl. Acad. Sci. USA}, 106\penalty0 (37):\penalty0
  15527--15533, Sept. 2009.
\newblock \doi{10.1073/pnas.0903507106}.

\bibitem[{Lef{\`e}vre} and {Bazylinski}(2013)]{LB13}
C.~T. {Lef{\`e}vre} and D.~A. {Bazylinski}.
\newblock {Ecology, Diversity, and Evolution of Magnetotactic Bacteria}.
\newblock \emph{Microbiol. Mol. Biol. Rev.}, 77\penalty0 (3):\penalty0
  497--526, 2013.
\newblock \doi{10.1128/MMBR.00021-13}.

\bibitem[{Lever} et~al.(2015){Lever}, {Rogers}, {Lloyd}, {Overmann}, {Schink},
  {Thauer}, {Hoehler}, and {J{\o}rgensen}]{LRL15}
M.~A. {Lever}, K.~L. {Rogers}, K.~G. {Lloyd}, J.~{Overmann}, B.~{Schink}, R.~K.
  {Thauer}, T.~M. {Hoehler}, and B.~B. {J{\o}rgensen}.
\newblock {Life under extreme energy limitation: a synthesis of laboratory- and
  field-based investigations}.
\newblock \emph{FEMS Microbiol. Rev.}, 39\penalty0 (5):\penalty0 688--728,
  2015.
\newblock \doi{10.1093/femsre/fuv020}.

\bibitem[{Li} et~al.(1983){Li}, {Subba Rao}, {Harrison}, {Smith}, {Cullen},
  {Irwin}, and {Platt}]{LSH83}
W.~K.~W. {Li}, D.~V. {Subba Rao}, W.~G. {Harrison}, J.~C. {Smith}, J.~J.
  {Cullen}, B.~{Irwin}, and T.~{Platt}.
\newblock {Autotrophic Picoplankton in the Tropical Ocean}.
\newblock \emph{Science}, 219\penalty0 (4582):\penalty0 292--295, Jan. 1983.
\newblock \doi{10.1126/science.219.4582.292}.

\bibitem[{Limaye} et~al.(2018){Limaye}, {Mogul}, {Smith}, {Ansari},
  {S{\l}owik}, and {Vaishampayan}]{LMS18}
S.~S. {Limaye}, R.~{Mogul}, D.~J. {Smith}, A.~H. {Ansari}, G.~P. {S{\l}owik},
  and P.~{Vaishampayan}.
\newblock {Venus' Spectral Signatures and the Potential for Life in the
  Clouds}.
\newblock \emph{Astrobiology}, 18\penalty0 (9):\penalty0 1181--1198, Sept.
  2018.
\newblock \doi{10.1089/ast.2017.1783}.

\bibitem[{Lingam} and {Loeb}(2018{\natexlab{a}})]{LL18}
M.~{Lingam} and A.~{Loeb}.
\newblock {Implications of Tides for Life on Exoplanets}.
\newblock \emph{Astrobiology}, 18\penalty0 (7):\penalty0 967--982, July
  2018{\natexlab{a}}.
\newblock \doi{10.1089/ast.2017.1718}.

\bibitem[{Lingam} and {Loeb}(2018{\natexlab{b}})]{LiMa18}
M.~{Lingam} and A.~{Loeb}.
\newblock {Is Extraterrestrial Life Suppressed on Subsurface Ocean Worlds due
  to the Paucity of Bioessential Elements?}
\newblock \emph{Astron. J.}, 156\penalty0 (4):\penalty0 151, Oct.
  2018{\natexlab{b}}.
\newblock \doi{10.3847/1538-3881/aada02}.

\bibitem[{Lingam} and {Loeb}(2019{\natexlab{a}})]{MLAL}
M.~{Lingam} and A.~{Loeb}.
\newblock {Dependence of Biological Activity on the Surface Water Fraction of
  Planets}.
\newblock \emph{Astron. J.}, 157\penalty0 (1):\penalty0 25, Jan.
  2019{\natexlab{a}}.
\newblock \doi{10.3847/1538-3881/aaf420}.

\bibitem[{Lingam} and {Loeb}(2019{\natexlab{b}})]{MLi19}
M.~{Lingam} and A.~{Loeb}.
\newblock {Colloquium: Physical constraints for the evolution of life on
  exoplanets}.
\newblock \emph{Rev. Mod. Phys.}, 91\penalty0 (2):\penalty0 021002, Apr.
  2019{\natexlab{b}}.
\newblock \doi{10.1103/RevModPhys.91.021002}.

\bibitem[{Lingam} and {Loeb}(2019{\natexlab{c}})]{MaLi19}
M.~{Lingam} and A.~{Loeb}.
\newblock {Subsurface exolife}.
\newblock \emph{Int. J. Astrobiol.}, 18\penalty0 (2):\penalty0 112--141, Apr.
  2019{\natexlab{c}}.
\newblock \doi{10.1017/S1473550418000083}.

\bibitem[{Lingam} and {Loeb}(2020{\natexlab{a}})]{LL20}
M.~{Lingam} and A.~{Loeb}.
\newblock {Constraints on Aquatic Photosynthesis for Terrestrial Planets around
  Other Stars}.
\newblock \emph{Astrophys. J. Lett.}, 889\penalty0 (1):\penalty0 L15, Jan.
  2020{\natexlab{a}}.
\newblock \doi{10.3847/2041-8213/ab6a14}.

\bibitem[{Lingam} and {Loeb}(2020{\natexlab{b}})]{ML20}
M.~{Lingam} and A.~{Loeb}.
\newblock {On the Habitable Lifetime of Terrestrial Worlds with High
  Radionuclide Abundances}.
\newblock \emph{Astrophys. J. Lett.}, 889\penalty0 (1):\penalty0 L20, Jan.
  2020{\natexlab{b}}.
\newblock \doi{10.3847/2041-8213/ab68e5}.

\bibitem[{Lingam} and {Loeb}(2020{\natexlab{c}})]{ML21}
M.~{Lingam} and A.~{Loeb}.
\newblock {On The Biomass Required To Produce Phosphine Detected In The Cloud
  Decks Of Venus}.
\newblock \emph{Int. J. Astrobiol.}, art. arXiv:2009.07835, Sept.
  2020{\natexlab{c}}.

\bibitem[{Lingam} and {Loeb}(2021{\natexlab{a}})]{LL21}
M.~{Lingam} and A.~{Loeb}.
\newblock \emph{{Life in the Cosmos: From Biosignatures to Technosignatures}}.
\newblock Cambridge: Harvard University Press, 2021{\natexlab{a}}.
\newblock URL \url{https://www.hup.harvard.edu/catalog.php?isbn=9780674987579}.

\bibitem[{Lingam} and {Loeb}(2021{\natexlab{b}})]{MaLi21}
M.~{Lingam} and A.~{Loeb}.
\newblock {Physical Constraints on Motility with Applications to Possible Life
  on Mars and Enceladus}.
\newblock \emph{Planet. Sci. J.}, art. arXiv:2101.06876, Jan.
  2021{\natexlab{b}}.

\bibitem[{Lopatkin} and {Collins}(2020)]{LC20}
A.~J. {Lopatkin} and J.~J. {Collins}.
\newblock {Predictive biology: modelling, understanding and harnessing
  microbial complexity}.
\newblock \emph{Nat. Rev. Microbiol.}, 18:\penalty0 507--520, 2020.
\newblock \doi{10.1038/s41579-020-0372-5}.

\bibitem[{Lorenz}(2019)]{Lor19}
R.~D. {Lorenz}.
\newblock \emph{Exploring Planetary Climate: A History of Scientific Discovery
  on Earth, Mars, Venus and Titan}.
\newblock Cambridge: Cambridge University Press, 2019.

\bibitem[{Lozano} et~al.(2016){Lozano}, {Ten Hagen}, {L{\"o}wen}, and
  {Bechinger}]{LTL16}
C.~{Lozano}, B.~{Ten Hagen}, H.~{L{\"o}wen}, and C.~{Bechinger}.
\newblock {Phototaxis of synthetic microswimmers in optical landscapes}.
\newblock \emph{Nat. Commun.}, 7:\penalty0 12828, Sept. 2016.
\newblock \doi{10.1038/ncomms12828}.

\bibitem[{Luef} et~al.(2015){Luef}, {Frischkorn}, {Wrighton}, {Holman},
  {Birarda}, {Thomas}, {Singh}, {Williams}, {Siegerist}, {Tringe}, {Downing},
  {Comolli}, and {Banfield}]{LFW15}
B.~{Luef}, K.~R. {Frischkorn}, K.~C. {Wrighton}, H.-Y.~N. {Holman},
  G.~{Birarda}, B.~C. {Thomas}, A.~{Singh}, K.~H. {Williams}, C.~E.
  {Siegerist}, S.~G. {Tringe}, K.~H. {Downing}, L.~R. {Comolli}, and J.~F.
  {Banfield}.
\newblock {Diverse uncultivated ultra-small bacterial cells in groundwater}.
\newblock \emph{Nat. Commun.}, 6:\penalty0 6372, Feb. 2015.
\newblock \doi{10.1038/ncomms7372}.

\bibitem[{Lunine}(2009)]{Lun09}
J.~I. {Lunine}.
\newblock {Saturn's Titan: A Strict Test for Life's Cosmic Ubiquity}.
\newblock \emph{Proc. Am. Philos. Soc.}, 153\penalty0 (4):\penalty0 403--418,
  2009.
\newblock \doi{10.2307/20721510}.

\bibitem[{Lyons} and {Kolter}(2015)]{LK15}
N.~A. {Lyons} and R.~{Kolter}.
\newblock {On the evolution of bacterial multicellularity}.
\newblock \emph{Curr. Opin. Microbiol.}, 24:\penalty0 21--28, 2015.
\newblock \doi{10.1016/j.mib.2014.12.007}.

\bibitem[{Makarieva} et~al.(2008){Makarieva}, {Gorshkov}, {Li}, {Chown},
  {Reich}, and {Gavrilov}]{MGL08}
A.~M. {Makarieva}, V.~G. {Gorshkov}, B.-L. {Li}, S.~L. {Chown}, P.~B. {Reich},
  and V.~M. {Gavrilov}.
\newblock {Mean mass-specific metabolic rates are strikingly similar across
  life's major domains: Evidence for life's metabolic optimum}.
\newblock \emph{Proc. Natl. Acad. Sci. USA}, 105\penalty0 (44):\penalty0
  16994--16999, Nov. 2008.
\newblock \doi{10.1073/pnas.0802148105}.

\bibitem[{Malmstrom} et~al.(2004){Malmstrom}, {Kiene}, {Cottrell}, and
  {Kirchman}]{MKC04}
R.~R. {Malmstrom}, R.~P. {Kiene}, M.~T. {Cottrell}, and D.~L. {Kirchman}.
\newblock {Contribution of SAR11 Bacteria to Dissolved
  Dimethylsulfoniopropionate and Amino Acid Uptake in the North Atlantic
  Ocean}.
\newblock \emph{Appl. Environ. Microbiol.}, 70\penalty0 (7):\penalty0
  4129--4135, 2004.
\newblock \doi{10.1128/AEM.70.7.4129-4135.2004}.

\bibitem[{Maniloff} et~al.(1997){Maniloff}, {Nealson}, {Psenner}, {Loferer},
  and {Folk}]{MNP97}
J.~{Maniloff}, K.~H. {Nealson}, R.~{Psenner}, M.~{Loferer}, and R.~L. {Folk}.
\newblock Nannobacteria: size limits and evidence.
\newblock \emph{Science}, 276\penalty0 (5320):\penalty0 1773--1776, 1997.
\newblock \doi{10.1126/science.276.5320.1773e}.

\bibitem[{Mann} and {Lazier}(2006)]{ML06}
K.~H. {Mann} and J.~R.~N. {Lazier}.
\newblock \emph{Dynamics of Marine Ecosystems: Biological-Physical Interactions
  in the Oceans}.
\newblock Oxford: Blackwell Publishing, 3rd edition, 2006.

\bibitem[{Manske} et~al.(2005){Manske}, {Glaeser}, {Kuypers}, and
  {Overmann}]{MGK05}
A.~K. {Manske}, J.~{Glaeser}, M.~M.~M. {Kuypers}, and J.~{Overmann}.
\newblock {Physiology and Phylogeny of Green Sulfur Bacteria Forming a
  Monospecific Phototrophic Assemblage at a Depth of 100 Meters in the Black
  Sea}.
\newblock \emph{Appl. Environ. Microbiol.}, 71\penalty0 (12):\penalty0
  8049--8060, 2005.
\newblock \doi{10.1128/AEM.71.12.8049-8060.2005}.

\bibitem[{Martens} et~al.(2015){Martens}, {Wadhwa}, {Jacobsen}, {Lindemann},
  {Andersen}, and {Visser}]{MWJ15}
E.~A. {Martens}, N.~{Wadhwa}, N.~S. {Jacobsen}, C.~{Lindemann}, K.~H.
  {Andersen}, and A.~{Visser}.
\newblock {Size structures sensory hierarchy in ocean life}.
\newblock \emph{Proc. R. Soc. B}, 282\penalty0 (1815):\penalty0 20151346, 2015.
\newblock \doi{10.1098/rspb.2015.1346}.

\bibitem[{Martin} and {McMinn}(2018)]{MM18}
A.~{Martin} and A.~{McMinn}.
\newblock {Sea ice, extremophiles and life on extra-terrestrial ocean worlds}.
\newblock \emph{Int. J. Astrobiol.}, 17\penalty0 (1):\penalty0 1--16, Jan.
  2018.
\newblock \doi{10.1017/S1473550416000483}.

\bibitem[{Martin} et~al.(2008){Martin}, {Baross}, {Kelley}, and
  {Russell}]{MB08}
W.~{Martin}, J.~{Baross}, D.~{Kelley}, and M.~J. {Russell}.
\newblock {Hydrothermal vents and the origin of life}.
\newblock \emph{Nat. Rev. Microbiol.}, 6\penalty0 (11):\penalty0 805--814,
  2008.
\newblock \doi{10.1038/nrmicro1991}.

\bibitem[{Mast} et~al.(2013){Mast}, {Schink}, {Gerland}, and {Braun}]{MSG13}
C.~B. {Mast}, S.~{Schink}, U.~{Gerland}, and D.~{Braun}.
\newblock {Escalation of polymerization in a thermal gradient}.
\newblock \emph{Proc. Natl. Acad. Sci. USA}, 110\penalty0 (20):\penalty0
  8030--8035, May 2013.
\newblock \doi{10.1073/pnas.1303222110}.

\bibitem[{McIntyre} et~al.(2019){McIntyre}, {Lineweaver}, and {Ireland}]{SM19}
S.~R.~N. {McIntyre}, C.~H. {Lineweaver}, and M.~J. {Ireland}.
\newblock {Planetary magnetism as a parameter in exoplanet habitability}.
\newblock \emph{Mon. Not. R. Astron. Soc.}, 485\penalty0 (3):\penalty0
  3999--4012, May 2019.
\newblock \doi{10.1093/mnras/stz667}.

\bibitem[{McKay}(2014)]{McK14}
C.~P. {McKay}.
\newblock {Requirements and limits for life in the context of exoplanets}.
\newblock \emph{Proc. Natl. Acad. Sci. USA}, 111\penalty0 (35):\penalty0
  12628--12633, Sept. 2014.
\newblock \doi{10.1073/pnas.1304212111}.

\bibitem[{McKay}(2016)]{McK16}
C.~P. {McKay}.
\newblock {Titan as the Abode of Life}.
\newblock \emph{Life}, 6\penalty0 (1):\penalty0 8, Feb. 2016.
\newblock \doi{10.3390/life6010008}.

\bibitem[{McKay} and {Smith}(2005)]{MS05}
C.~P. {McKay} and H.~D. {Smith}.
\newblock {Possibilities for methanogenic life in liquid methane on the surface
  of Titan}.
\newblock \emph{Icarus}, 178\penalty0 (1):\penalty0 274--276, Nov. 2005.
\newblock \doi{10.1016/j.icarus.2005.05.018}.

\bibitem[{McMahon} et~al.(2018){McMahon}, {Bosak}, {Grotzinger}, {Milliken},
  {Summons}, {Daye}, {Newman}, {Fraeman}, {Williford}, and {Briggs}]{MBG18}
S.~{McMahon}, T.~{Bosak}, J.~P. {Grotzinger}, R.~E. {Milliken}, R.~E.
  {Summons}, M.~{Daye}, S.~A. {Newman}, A.~{Fraeman}, K.~H. {Williford}, and
  D.~E.~G. {Briggs}.
\newblock {A Field Guide to Finding Fossils on Mars}.
\newblock \emph{J. Geophys. Res. Planets}, 123\penalty0 (5):\penalty0
  1012--1040, May 2018.
\newblock \doi{10.1029/2017JE005478}.

\bibitem[{Meadows} et~al.(2020){Meadows}, {Arney}, {Schmidt}, and {Des
  Marais}]{MASD}
V.~S. {Meadows}, G.~N. {Arney}, B.~E. {Schmidt}, and D.~J. {Des Marais},
  editors.
\newblock \emph{{Planetary Astrobiology}}.
\newblock Space Science Series. Tucson: University of Arizona Press, 2020.

\bibitem[{M{\'e}nez} et~al.(2018){M{\'e}nez}, {Pisapia}, {Andreani}, {Jamme},
  {Vanbellingen}, {Brunelle}, {Richard}, {Dumas}, and
  {R{\'e}fr{\'e}giers}]{MPA18}
B.~{M{\'e}nez}, C.~{Pisapia}, M.~{Andreani}, F.~{Jamme}, Q.~P. {Vanbellingen},
  A.~{Brunelle}, L.~{Richard}, P.~{Dumas}, and M.~{R{\'e}fr{\'e}giers}.
\newblock {Abiotic synthesis of amino acids in the recesses of the oceanic
  lithosphere}.
\newblock \emph{Nature}, 564\penalty0 (7734):\penalty0 59--63, Nov. 2018.
\newblock \doi{10.1038/s41586-018-0684-z}.

\bibitem[{Meyer-Vernet} and {Rospars}(2016)]{MR16}
N.~{Meyer-Vernet} and J.-P. {Rospars}.
\newblock {Maximum relative speeds of living organisms: Why do bacteria perform
  as fast as ostriches?}
\newblock \emph{Phys. Biol.}, 13\penalty0 (6):\penalty0 066006, Dec. 2016.
\newblock \doi{10.1088/1478-3975/13/6/066006}.

\bibitem[{Micali} and {Endres}(2016)]{ME16}
G.~{Micali} and R.~G. {Endres}.
\newblock {Bacterial chemotaxis: information processing, thermodynamics, and
  behavior}.
\newblock \emph{Curr. Opin. Microbiol.}, 30:\penalty0 8--15, 2016.
\newblock \doi{10.1016/j.mib.2015.12.001}.

\bibitem[{Michalski} et~al.(2018){Michalski}, {Onstott}, {Mojzsis}, {Mustard},
  {Chan}, {Niles}, and {Johnson}]{MOM18}
J.~R. {Michalski}, T.~C. {Onstott}, S.~J. {Mojzsis}, J.~{Mustard}, Q.~H.~S.
  {Chan}, P.~B. {Niles}, and S.~S. {Johnson}.
\newblock {The Martian subsurface as a potential window into the origin of
  life}.
\newblock \emph{Nat. Geosci.}, 11\penalty0 (1):\penalty0 21--26, Dec. 2018.
\newblock \doi{10.1038/s41561-017-0015-2}.

\bibitem[{Milo} and {Phillips}(2016)]{MP16}
R.~{Milo} and R.~{Phillips}.
\newblock \emph{{Cell Biology by the Numbers}}.
\newblock New York: Garland Science, 2016.

\bibitem[{Minic} and {Thongbam}(2011)]{Mi11}
Z.~{Minic} and P.~D. {Thongbam}.
\newblock {The Biological Deep Sea Hydrothermal Vent as a Model to Study Carbon
  Dioxide Capturing Enzymes}.
\newblock \emph{Mar. Drugs}, 9\penalty0 (5):\penalty0 719--738, 2011.
\newblock \doi{10.3390/md9050719}.

\bibitem[{Mispelaer} et~al.(2013){Mispelaer}, {Theul{\'e}}, {Aouididi},
  {Noble}, {Duvernay}, {Danger}, {Roubin}, {Morata}, {Hasegawa}, and
  {Chiavassa}]{MTA13}
F.~{Mispelaer}, P.~{Theul{\'e}}, H.~{Aouididi}, J.~{Noble}, F.~{Duvernay},
  G.~{Danger}, P.~{Roubin}, O.~{Morata}, T.~{Hasegawa}, and T.~{Chiavassa}.
\newblock {Diffusion measurements of CO, HNCO, H$_{2}$CO, and NH$_{3}$ in
  amorphous water ice}.
\newblock \emph{Astron. Astrophys.}, 555:\penalty0 A13, July 2013.
\newblock \doi{10.1051/0004-6361/201220691}.

\bibitem[{M{\"o}glich} et~al.(2010){M{\"o}glich}, {Yang}, {Ayers}, and
  {Moffat}]{MYA10}
A.~{M{\"o}glich}, X.~{Yang}, R.~A. {Ayers}, and K.~{Moffat}.
\newblock {Structure and Function of Plant Photoreceptors}.
\newblock \emph{Annu. Rev. Plant Biol.}, 61:\penalty0 21--47, 2010.
\newblock \doi{10.1146/annurev-arplant-042809-112259}.

\bibitem[{M{\"o}ller} et~al.(2017){M{\"o}ller}, {Kriegel}, {Kie{\ss}}, {Sojo},
  and {Braun}]{MKK17}
F.~M. {M{\"o}ller}, F.~{Kriegel}, M.~{Kie{\ss}}, V.~{Sojo}, and D.~{Braun}.
\newblock {Steep pH Gradients and Directed Colloid Transport in a Microfluidic
  Alkaline Hydrothermal Pore}.
\newblock \emph{Angew. Chem. Int. Ed.}, 56\penalty0 (9):\penalty0 2340--2344,
  2017.
\newblock \doi{10.1002/anie.201610781}.

\bibitem[{Mori}(1999)]{Mo99}
I.~{Mori}.
\newblock {Genetics of Chemotaxis and Thermotaxis in the Nematode
  \emph{Caenorhabditis Elegans}}.
\newblock \emph{Annu. Rev. Genet.}, 33\penalty0 (1):\penalty0 399--422, 1999.
\newblock \doi{10.1146/annurev.genet.33.1.399}.

\bibitem[{Morowitz} and {Sagan}(1967)]{MS67}
H.~{Morowitz} and C.~{Sagan}.
\newblock {Life in the Clouds of Venus?}
\newblock \emph{Nature}, 215\penalty0 (5107):\penalty0 1259--1260, Sept. 1967.
\newblock \doi{10.1038/2151259a0}.

\bibitem[{Morowitz}(1967)]{Mor67}
H.~J. {Morowitz}.
\newblock Biological self-replicating systems.
\newblock \emph{Prog. Theoret. Biol.}, 1:\penalty0 35--58, 1967.

\bibitem[{Muchowska} et~al.(2020){Muchowska}, {Varma}, and {Moran}]{MVM20}
K.~B. {Muchowska}, S.~J. {Varma}, and J.~{Moran}.
\newblock {Nonenzymatic Metabolic Reactions and Life's Origins}.
\newblock \emph{Chem. Rev.}, 120\penalty0 (15):\penalty0 7708--7744, 2020.
\newblock \doi{10.1021/acs.chemrev.0c00191}.

\bibitem[{Mulkidjanian} et~al.(2012){Mulkidjanian}, {Bychkov}, {Dibrova},
  {Galperin}, and {Koonin}]{MB12}
A.~Y. {Mulkidjanian}, A.~Y. {Bychkov}, D.~V. {Dibrova}, M.~Y. {Galperin}, and
  E.~V. {Koonin}.
\newblock {PNAS Plus: Origin of first cells at terrestrial, anoxic geothermal
  fields}.
\newblock \emph{Proc. Natl. Acad. Sci. USA}, 109\penalty0 (14):\penalty0
  E821--E830, Apr. 2012.
\newblock \doi{10.1073/pnas.1117774109}.

\bibitem[{Nadeau} et~al.(2016){Nadeau}, {Lindensmith}, {Deming}, {Fernandez},
  and {Stocker}]{NLD16}
J.~{Nadeau}, C.~{Lindensmith}, J.~W. {Deming}, V.~I. {Fernandez}, and
  R.~{Stocker}.
\newblock {Microbial Morphology and Motility as Biosignatures for Outer Planet
  Missions}.
\newblock \emph{Astrobiology}, 16\penalty0 (10):\penalty0 755--774, Oct. 2016.
\newblock \doi{10.1089/ast.2015.1376}.

\bibitem[{Nadeau} et~al.(2018){Nadeau}, {Bedrossian}, and {Lindensmith}]{NBL18}
J.~L. {Nadeau}, M.~{Bedrossian}, and C.~A. {Lindensmith}.
\newblock {Imaging technologies and strategies for detection of extant
  extraterrestrial microorganisms}.
\newblock \emph{Adv. Phys. X}, 3\penalty0 (1):\penalty0 1424032, 2018.
\newblock \doi{10.1080/23746149.2018.1424032}.

\bibitem[{Neveu} et~al.(2018){Neveu}, {Hays}, {Voytek}, {New}, and
  {Schulte}]{NHV18}
M.~{Neveu}, L.~E. {Hays}, M.~A. {Voytek}, M.~H. {New}, and M.~D. {Schulte}.
\newblock {The Ladder of Life Detection}.
\newblock \emph{Astrobiology}, 18\penalty0 (11):\penalty0 1375--1402, Nov.
  2018.
\newblock \doi{10.1089/ast.2017.1773}.

\bibitem[{Neveu} et~al.(2020){Neveu}, {Anbar}, {Davila}, {Glavin}, {Mackenzie},
  {Phillips-Lander}, {Sherwood}, {Takano}, {Williams}, and {Yano}]{NAD20}
M.~{Neveu}, A.~{Anbar}, A.~F. {Davila}, D.~P. {Glavin}, S.~M. {Mackenzie},
  C.~{Phillips-Lander}, B.~{Sherwood}, Y.~{Takano}, P.~{Williams}, and
  H.~{Yano}.
\newblock {Returning Samples From Enceladus for Life Detection}.
\newblock \emph{Front. Astron. Space Sci.}, 7:\penalty0 26, 2020.
\newblock \doi{10.3389/fspas.2020.00026}.

\bibitem[{Niether} et~al.(2016){Niether}, {Afanasenkau}, and {Dhont}]{NAD16}
D.~{Niether}, D.~{Afanasenkau}, and J.~K.~G. {Dhont}.
\newblock {Accumulation of formamide in hydrothermal pores to form prebiotic
  nucleobases}.
\newblock \emph{Proc. Natl. Acad. Sci. USA}, 113\penalty0 (16):\penalty0
  4272--4277, Apr. 2016.
\newblock \doi{10.1073/pnas.1600275113}.

\bibitem[{Nimmo} and {Pappalardo}(2016)]{NP16}
F.~{Nimmo} and R.~T. {Pappalardo}.
\newblock {Ocean worlds in the outer solar system}.
\newblock \emph{J. Geophys. Res. Planets}, 121\penalty0 (8):\penalty0
  1378--1399, Aug. 2016.
\newblock \doi{10.1002/2016JE005081}.

\bibitem[{Nisbet} and {Fowler}(1996)]{NF96}
E.~G. {Nisbet} and C.~M.~R. {Fowler}.
\newblock {The hydrothermal imprint on life: did heat-shock proteins,
  metalloproteins and photosynthesis begin around hydrothermal vents?}
\newblock \emph{Geol. Soc. London Spec. Publ.}, 118\penalty0 (1):\penalty0
  239--251, Jan. 1996.
\newblock \doi{10.1144/GSL.SP.1996.118.01.15}.

\bibitem[{Nisbet} et~al.(1995){Nisbet}, {Cann}, and {Van Dover}]{NCV95}
E.~G. {Nisbet}, J.~R. {Cann}, and C.~L. {Van Dover}.
\newblock {Origins of photosynthesis}.
\newblock \emph{Nature}, 373\penalty0 (6514):\penalty0 479--480, Feb. 1995.
\newblock \doi{10.1038/373479a0}.

\bibitem[{Okubo} and {Levin}(2001)]{OL01}
A.~{Okubo} and S.~A. {Levin}.
\newblock \emph{{Diffusion and Ecological Problems: Modern Perspectives}},
  volume~14 of \emph{Interdisciplinary Applied Mathematics}.
\newblock New York: Springer-Verlag, 2nd edition, 2001.

\bibitem[{Olson} et~al.(2020){Olson}, {Jansen}, and {Abbot}]{OJA20}
S.~L. {Olson}, M.~{Jansen}, and D.~S. {Abbot}.
\newblock {Oceanographic Considerations for Exoplanet Life Detection}.
\newblock \emph{Astrophys. J.}, 895\penalty0 (1):\penalty0 19, May 2020.
\newblock \doi{10.3847/1538-4357/ab88c9}.

\bibitem[{Orgel}(2008)]{Org08}
L.~E. {Orgel}.
\newblock {The Implausibility of Metabolic Cycles on the Prebiotic Earth}.
\newblock \emph{PLoS Biol}, 6\penalty0 (1):\penalty0 e18, 2008.
\newblock \doi{10.1371/journal.pbio.0060018}.

\bibitem[{Or{\'o}} and {Lazcano}(1984)]{OL84}
J.~{Or{\'o}} and A.~{Lazcano}.
\newblock {A minimal living system and the origin of a protocell}.
\newblock \emph{Adv. Space Res}, 4\penalty0 (12):\penalty0 167--176, Jan. 1984.
\newblock \doi{10.1016/0273-1177(84)90559-3}.

\bibitem[{Osinski} et~al.(2013){Osinski}, {Tornabene}, {Banerjee}, {Cockell},
  {Flemming}, {Izawa}, {McCutcheon}, {Parnell}, {Preston}, {Pickersgill},
  {Pontefract}, {Sapers}, and {Southam}]{OTB13}
G.~R. {Osinski}, L.~L. {Tornabene}, N.~R. {Banerjee}, C.~S. {Cockell},
  R.~{Flemming}, M.~R.~M. {Izawa}, J.~{McCutcheon}, J.~{Parnell}, L.~J.
  {Preston}, A.~E. {Pickersgill}, A.~{Pontefract}, H.~M. {Sapers}, and
  G.~{Southam}.
\newblock {Impact-generated hydrothermal systems on Earth and Mars}.
\newblock \emph{Icarus}, 224\penalty0 (2):\penalty0 347--363, June 2013.
\newblock \doi{10.1016/j.icarus.2012.08.030}.

\bibitem[{Painter}(2019)]{Pa19}
K.~J. {Painter}.
\newblock {Mathematical models for chemotaxis and their applications in
  self-organisation phenomena}.
\newblock \emph{J. Theor. Biol.}, 481:\penalty0 162--182, 2019.
\newblock \doi{10.1016/j.jtbi.2018.06.019}.

\bibitem[{Palczewski}(2006)]{Pal06}
K.~{Palczewski}.
\newblock G protein--coupled receptor rhodopsin.
\newblock \emph{Annu. Rev. Biochem.}, 75:\penalty0 743--767, 2006.
\newblock \doi{10.1146/annurev.biochem.75.103004.142743}.

\bibitem[{Paytan} and {McLaughlin}(2007)]{PM07}
A.~{Paytan} and K.~{McLaughlin}.
\newblock {The Oceanic Phosphorus Cycle}.
\newblock \emph{Chem. Rev.}, 107\penalty0 (2):\penalty0 563--576, 2007.
\newblock \doi{10.1021/cr0503613}.

\bibitem[{Pfeffer}(1884)]{Pfe84}
W.~{Pfeffer}.
\newblock {Locomotorische Richtungsbewegungen durch chemische Reize}.
\newblock \emph{Untersuch. Botan. Inst. T{\"u}bingen}, 1:\penalty0 363--482,
  1884.

\bibitem[{Phillips} et~al.(2013){Phillips}, {Kondev}, {Theriot}, and
  {Garcia}]{PKTG}
R.~{Phillips}, J.~{Kondev}, J.~{Theriot}, and H.~G. {Garcia}.
\newblock \emph{{Physical Biology of the Cell}}.
\newblock New York: Garland Science, 2nd edition, 2013.

\bibitem[{Porter} et~al.(2011){Porter}, {Wadhams}, and {Armitage}]{PWA11}
S.~L. {Porter}, G.~H. {Wadhams}, and J.~P. {Armitage}.
\newblock {Signal processing in complex chemotaxis pathways}.
\newblock \emph{Nat. Rev. Microbiol.}, 9\penalty0 (3):\penalty0 153--165, 2011.
\newblock \doi{10.1038/nrmicro2505}.

\bibitem[{Postberg} et~al.(2018){Postberg}, {Khawaja}, {Abel}, {Choblet},
  {Glein}, {Gudipati}, {Henderson}, {Hsu}, {Kempf}, {Klenner},
  {Moragas-Klostermeyer}, {Magee}, {N{\"o}lle}, {Perry}, {Reviol}, {Schmidt},
  {Srama}, {Stolz}, {Tobie}, {Trieloff}, and {Waite}]{PKN18}
F.~{Postberg}, N.~{Khawaja}, B.~{Abel}, G.~{Choblet}, C.~R. {Glein}, M.~S.
  {Gudipati}, B.~L. {Henderson}, H.-W. {Hsu}, S.~{Kempf}, F.~{Klenner},
  G.~{Moragas-Klostermeyer}, B.~{Magee}, L.~{N{\"o}lle}, M.~{Perry},
  R.~{Reviol}, J.~{Schmidt}, R.~{Srama}, F.~{Stolz}, G.~{Tobie}, M.~{Trieloff},
  and J.~H. {Waite}.
\newblock {Macromolecular organic compound s from the depths of Enceladus}.
\newblock \emph{Nature}, 558\penalty0 (7711):\penalty0 564--568, June 2018.
\newblock \doi{10.1038/s41586-018-0246-4}.

\bibitem[{Preiner} et~al.(2020){Preiner}, {Igarashi}, {Muchowska}, {Yu},
  {Varma}, {Kleinermanns}, {Nobu}, {Kamagata}, {T{\"u}ys{\"u}z}, {Moran}, and
  {Martin}]{PIM20}
M.~{Preiner}, K.~{Igarashi}, K.~B. {Muchowska}, M.~{Yu}, S.~J. {Varma},
  K.~{Kleinermanns}, M.~K. {Nobu}, Y.~{Kamagata}, H.~{T{\"u}ys{\"u}z},
  J.~{Moran}, and W.~F. {Martin}.
\newblock {A hydrogen-dependent geochemical analogue of primordial carbon and
  energy metabolism}.
\newblock \emph{Nat. Ecol. Evol.}, 4:\penalty0 534--542, 2020.
\newblock \doi{10.1038/s41559-020-1125-6}.

\bibitem[{Price}(2007)]{PBP07}
P.~B. {Price}.
\newblock {Microbial life in glacial ice and implications for a cold origin of
  life}.
\newblock \emph{FEMS Microbiol. Ecol.}, 59\penalty0 (2):\penalty0 217--231,
  2007.
\newblock \doi{10.1111/j.1574-6941.2006.00234.x}.

\bibitem[{Prigogine} and {Nicolis}(1971)]{PN71}
I.~{Prigogine} and G.~{Nicolis}.
\newblock {Biological order, structure and instabilities}.
\newblock \emph{Q. Rev. Biophys.}, 4\penalty0 (2-3):\penalty0 107--148, 1971.
\newblock \doi{10.1017/S0033583500000615}.

\bibitem[{Priye} et~al.(2017){Priye}, {Yu}, {Hassan}, and {Ugaz}]{PYH17}
A.~{Priye}, Y.~{Yu}, Y.~A. {Hassan}, and V.~M. {Ugaz}.
\newblock {Synchronized chaotic targeting and acceleration of surface chemistry
  in prebiotic hydrothermal microenvironments}.
\newblock \emph{Proc. Natl. Acad. Sci. USA}, 114\penalty0 (6):\penalty0
  1275--1280, Feb. 2017.
\newblock \doi{10.1073/pnas.1612924114}.

\bibitem[{Purcell}(1977)]{Pur77}
E.~M. {Purcell}.
\newblock {Life at low Reynolds number}.
\newblock \emph{Am. J. Phys.}, 45\penalty0 (1):\penalty0 3--11, Jan. 1977.
\newblock \doi{10.1119/1.10903}.

\bibitem[{Raina} et~al.(2019){Raina}, {Fernandez}, {Lambert}, {Stocker}, and
  {Seymour}]{RFL19}
J.-B. {Raina}, V.~{Fernandez}, B.~{Lambert}, R.~{Stocker}, and J.~R. {Seymour}.
\newblock {The role of microbial motility and chemotaxis in symbiosis}.
\newblock \emph{Nat. Rev. Microbiol.}, 17\penalty0 (5):\penalty0 284--294,
  2019.
\newblock \doi{10.1038/s41579-019-0182-9}.

\bibitem[{Rapp{\'e}} et~al.(2002){Rapp{\'e}}, {Connon}, {Vergin}, and
  {Giovannoni}]{RCV02}
M.~S. {Rapp{\'e}}, S.~A. {Connon}, K.~L. {Vergin}, and S.~J. {Giovannoni}.
\newblock {Cultivation of the ubiquitous SAR11 marine bacterioplankton clade}.
\newblock \emph{Nature}, 418\penalty0 (6898):\penalty0 630--633, Aug. 2002.
\newblock \doi{10.1038/nature00917}.

\bibitem[{Raulin} et~al.(2012){Raulin}, {Brasse}, {Poch}, and {Coll}]{RBPC}
F.~{Raulin}, C.~{Brasse}, O.~{Poch}, and P.~{Coll}.
\newblock {Prebiotic-like chemistry on Titan}.
\newblock \emph{Chem. Soc. Rev.}, 41\penalty0 (16):\penalty0 5380--5393, 2012.
\newblock \doi{10.1039/C2CS35014A}.

\bibitem[{Raven}(1994)]{Rav94}
J.~A. {Raven}.
\newblock {Why are there no picoplanktonic O$_2$ evolvers with volumes less
  than $10^{-19}$ m$^3$?}
\newblock \emph{J. Plankton Res.}, 16\penalty0 (5):\penalty0 565--580, 1994.
\newblock \doi{10.1093/plankt/16.5.565}.

\bibitem[{Raven} and {Donnelly}(2013)]{RaD13}
J.~A. {Raven} and S.~{Donnelly}.
\newblock {Brown Dwarfs and Black Smokers: The Potential for Photosynthesis
  Using Radiation from Low-Temperature Black Bodies}.
\newblock In J.-P. {de Vera} and J.~{Seckbach}, editors, \emph{{Habitability of
  Other Planets and Satellites}}, pages 267--284. Springer, 2013.
\newblock \doi{10.1007/978-94-007-6546-7_15}.

\bibitem[{Raven} et~al.(2000){Raven}, {K{\"u}bler}, and {Beardall}]{RKB00}
J.~A. {Raven}, J.~E. {K{\"u}bler}, and J.~{Beardall}.
\newblock {Put out the light, and then put out the light}.
\newblock \emph{J. Mar. Biol. Assoc. UK}, 80\penalty0 (1):\penalty0 1--25,
  2000.
\newblock \doi{10.1017/s0025315499001526}.

\bibitem[{Raven} et~al.(2013){Raven}, {Beardall}, {Larkum}, and
  {S{\'a}nchez-Baracaldo}]{RBL13}
J.~A. {Raven}, J.~{Beardall}, A.~W.~D. {Larkum}, and
  P.~{S{\'a}nchez-Baracaldo}.
\newblock {Interactions of photosynthesis with genome size and function}.
\newblock \emph{Phil. Trans. R. Soc. B}, 368\penalty0 (1622):\penalty0
  20120264, 2013.
\newblock \doi{10.1098/rstb.2012.0264}.

\bibitem[{Riekeles} et~al.(2021){Riekeles}, {Schirmack}, and
  {Schulze-Makuch}]{RSS21}
M.~{Riekeles}, J.~{Schirmack}, and D.~{Schulze-Makuch}.
\newblock {Machine Learning Algorithms Applied to Identify Microbial Species by
  Their Motility}.
\newblock \emph{Life}, 11\penalty0 (1):\penalty0 44, 2021.
\newblock \doi{10.3390/life11010044}.

\bibitem[{Ruff} and {Farmer}(2016)]{RF16}
S.~W. {Ruff} and J.~D. {Farmer}.
\newblock {Silica deposits on Mars with features resembling hot spring
  biosignatures at El Tatio in Chile}.
\newblock \emph{Nat. Commun.}, 7:\penalty0 13554, Nov. 2016.
\newblock \doi{10.1038/ncomms13554}.

\bibitem[{Ruff} et~al.(2020){Ruff}, {Campbell}, {Van Kranendonk}, {Rice}, and
  {Farmer}]{RCV20}
S.~W. {Ruff}, K.~A. {Campbell}, M.~J. {Van Kranendonk}, M.~S. {Rice}, and J.~D.
  {Farmer}.
\newblock {The Case for Ancient Hot Springs in Gusev Crater, Mars}.
\newblock \emph{Astrobiology}, 20\penalty0 (4):\penalty0 475--499, Apr. 2020.
\newblock \doi{10.1089/ast.2019.2044}.

\bibitem[{Russell} and {Hall}(1997)]{RH97}
M.~J. {Russell} and A.~J. {Hall}.
\newblock {The emergence of life from iron monosulphide bubbles at a submarine
  hydrothermal redox and pH front}.
\newblock \emph{J. Geol. Soc.}, 154\penalty0 (3):\penalty0 377--402, 1997.
\newblock \doi{10.1144/gsjgs.154.3.0377}.

\bibitem[{Russell} and {Martin}(2004)]{RM04}
M.~J. {Russell} and W.~{Martin}.
\newblock {The rocky roots of the acetyl-CoA pathway}.
\newblock \emph{Trends Biochem. Sci.}, 29\penalty0 (7):\penalty0 358--363,
  2004.
\newblock \doi{10.1016/j.tibs.2004.05.007}.

\bibitem[{Russell} and {Ponce}(2020)]{RP20}
M.~J. {Russell} and A.~{Ponce}.
\newblock Six `must-have' minerals for life’s emergence: Olivine, pyrrhotite,
  bridgmanite, serpentine, fougerite and mackinawite.
\newblock \emph{Life}, 10\penalty0 (11):\penalty0 291, 2020.
\newblock \doi{10.3390/life10110291}.

\bibitem[{Russell} et~al.(1994){Russell}, {Daniel}, {Hall}, and
  {Sherringham}]{RDHS}
M.~J. {Russell}, R.~M. {Daniel}, A.~J. {Hall}, and J.~A. {Sherringham}.
\newblock {A hydrothermally precipitated catalytic iron sulphide membrane as a
  first step toward life}.
\newblock \emph{J. Mol. Evol.}, 39\penalty0 (3):\penalty0 231--243, Sept. 1994.
\newblock \doi{10.1007/BF00160147}.

\bibitem[{Russell} et~al.(2014){Russell}, {Barge}, {Bhartia}, {Bocanegra},
  {Bracher}, {Branscomb}, {Kidd}, {McGlynn}, {Meier}, {Nitschke}, {Shibuya},
  {Vance}, {White}, and {Kanik}]{RBB14}
M.~J. {Russell}, L.~M. {Barge}, R.~{Bhartia}, D.~{Bocanegra}, P.~J. {Bracher},
  E.~{Branscomb}, R.~{Kidd}, S.~{McGlynn}, D.~H. {Meier}, W.~{Nitschke},
  T.~{Shibuya}, S.~{Vance}, L.~{White}, and I.~{Kanik}.
\newblock {The Drive to Life on Wet and Icy Worlds}.
\newblock \emph{Astrobiology}, 14\penalty0 (4):\penalty0 308--343, Apr. 2014.
\newblock \doi{10.1089/ast.2013.1110}.

\bibitem[{Sagan}(1994)]{Sag94}
C.~{Sagan}.
\newblock {The Search for Extraterrestrial Life}.
\newblock \emph{Sci. Am.}, 271\penalty0 (4):\penalty0 92--99, Oct. 1994.
\newblock \doi{10.1038/scientificamerican1094-92}.

\bibitem[{Sagan} et~al.(1992){Sagan}, {Thompson}, and {Khare}]{STK92}
C.~{Sagan}, W.~R. {Thompson}, and B.~N. {Khare}.
\newblock {Titan: a laboratory for prebiological organic chemistry}.
\newblock \emph{Acc. Chem. Res.}, 25\penalty0 (7):\penalty0 286--292, 1992.
\newblock \doi{10.1021/ar00019a003}.

\bibitem[{Salditt} et~al.(2020){Salditt}, {Keil}, {Horning}, {Mast}, {Joyce},
  and {Braun}]{SKH20}
A.~{Salditt}, L.~M.~R. {Keil}, D.~P. {Horning}, C.~B. {Mast}, G.~F. {Joyce},
  and D.~{Braun}.
\newblock {Thermal Habitat for RNA Amplification and Accumulation}.
\newblock \emph{Phys. Rev. Lett.}, 125\penalty0 (4):\penalty0 048104, July
  2020.
\newblock \doi{10.1103/PhysRevLett.125.048104}.

\bibitem[{Sandstr{\"o}m} and {Rahm}(2020)]{SR20}
H.~{Sandstr{\"o}m} and M.~{Rahm}.
\newblock {Can polarity-inverted membranes self-assemble on Titan?}
\newblock \emph{Sci. Adv.}, 6\penalty0 (4):\penalty0 eaax0272, Jan. 2020.
\newblock \doi{10.1126/sciadv.aax0272}.

\bibitem[{Sarmiento} and {Gruber}(2006)]{SG06}
J.~L. {Sarmiento} and N.~{Gruber}.
\newblock \emph{{Ocean Biogeochemical Dynamics}}.
\newblock Princeton: Princeton University Press, 2006.

\bibitem[{Schlesinger} and {Bernhardt}(2013)]{SB13}
W.~H. {Schlesinger} and E.~S. {Bernhardt}.
\newblock \emph{{Biogeochemistry: An Analysis of Global Change}}.
\newblock Waltham: Academic Press, 3rd edition, 2013.

\bibitem[{Schmelzer} et~al.(2005){Schmelzer}, {Zanotto}, and {Fokin}]{SZF05}
J.~W.~P. {Schmelzer}, E.~D. {Zanotto}, and V.~M. {Fokin}.
\newblock {Pressure dependence of viscosity}.
\newblock \emph{J. Chem. Phys.}, 122\penalty0 (7):\penalty0 074511--074511,
  Feb. 2005.
\newblock \doi{10.1063/1.1851510}.

\bibitem[{Schmidt} et~al.(1991){Schmidt}, {DeLong}, and {Pace}]{SDP91}
T.~M. {Schmidt}, E.~F. {DeLong}, and N.~R. {Pace}.
\newblock {Analysis of a marine picoplankton community by 16S rRNA gene cloning
  and sequencing}.
\newblock \emph{J. Bacteriol.}, 173\penalty0 (14):\penalty0 4371--4378, 1991.
\newblock \doi{10.1128/jb.173.14.4371-4378.1991}.

\bibitem[{Schr{\"o}dinger}(1944)]{Sch44}
E.~{Schr{\"o}dinger}.
\newblock \emph{{What Is Life? The Physical Aspect of the Living Cell}}.
\newblock Cambridge: Cambridge University Press, 1944.

\bibitem[{Schuech} et~al.(2019){Schuech}, {Hoehfurtner}, {Smith}, and
  {Humphries}]{SHS19}
R.~{Schuech}, T.~{Hoehfurtner}, D.~J. {Smith}, and S.~{Humphries}.
\newblock {Motile curved bacteria are Pareto-optimal}.
\newblock \emph{Proc. Natl. Acad. Sci. USA}, 116\penalty0 (29):\penalty0
  14440--14447, 2019.
\newblock \doi{10.1073/pnas.1818997116}.

\bibitem[{Sch{\"u}ler}(2007)]{Sch07}
D.~{Sch{\"u}ler}, editor.
\newblock \emph{{Magnetoreception and Magnetosomes in Bacteria}}, volume~3 of
  \emph{Microbiology Monographs}.
\newblock Berlin: Springer-Verlag, 2007.
\newblock \doi{10.1007/11741862}.

\bibitem[{Schulze-Makuch} and {Grinspoon}(2005)]{SG05}
D.~{Schulze-Makuch} and D.~H. {Grinspoon}.
\newblock {Biologically Enhanced Energy and Carbon Cycling on Titan?}
\newblock \emph{Astrobiology}, 5\penalty0 (4):\penalty0 560--567, Aug. 2005.
\newblock \doi{10.1089/ast.2005.5.560}.

\bibitem[{Schulze-Makuch} and {Irwin}(2018)]{SMI18}
D.~{Schulze-Makuch} and L.~N. {Irwin}.
\newblock \emph{{Life in the Universe: Expectations and Constraints}}.
\newblock Cham: Springer, 3rd edition, 2018.

\bibitem[{Schulze-Makuch} et~al.(2004){Schulze-Makuch}, {Grinspoon}, {Abbas},
  {Irwin}, and {Bullock}]{SGA04}
D.~{Schulze-Makuch}, D.~H. {Grinspoon}, O.~{Abbas}, L.~N. {Irwin}, and M.~A.
  {Bullock}.
\newblock {A Sulfur-Based Survival Strategy for Putative Phototrophic Life in
  the Venusian Atmosphere}.
\newblock \emph{Astrobiology}, 4\penalty0 (1):\penalty0 11--18, Mar. 2004.
\newblock \doi{10.1089/153110704773600203}.

\bibitem[{Seager} et~al.(2021){Seager}, {Petkowski}, {Gao}, {Bains}, {Bryan},
  {Ranjan}, and {Greaves}]{SPG20}
S.~{Seager}, J.~J. {Petkowski}, P.~{Gao}, W.~{Bains}, N.~C. {Bryan},
  S.~{Ranjan}, and J.~{Greaves}.
\newblock {The Venusian Lower Atmosphere Haze as a Depot for Desiccated
  Microbial Life: A Proposed Life Cycle for Persistence of the Venusian Aerial
  Biosphere}.
\newblock \emph{Astrobiology}, 21\penalty0 (2):\penalty0 arXiv:2009.06474, Feb.
  2021.
\newblock \doi{10.1089/ast.2020.2244}.

\bibitem[{Shapiro}(1998)]{Shap98}
J.~A. {Shapiro}.
\newblock {Thinking about Bacterial Populations as Multicellular Organisms}.
\newblock \emph{Annu. Rev. Microbiol.}, 52\penalty0 (1):\penalty0 81--104,
  1998.
\newblock \doi{10.1146/annurev.micro.52.1.81}.

\bibitem[{Sleep}(2018)]{Sle18}
N.~H. {Sleep}.
\newblock {Geological and Geochemical Constraints on the Origin and Evolution
  of Life}.
\newblock \emph{Astrobiology}, 18\penalty0 (9):\penalty0 1199--1219, Sept.
  2018.
\newblock \doi{10.1089/ast.2017.1778}.

\bibitem[{Smith} and {Morowitz}(2016)]{SM16}
E.~{Smith} and H.~J. {Morowitz}.
\newblock \emph{{The Origin and Nature of Life on Earth}}.
\newblock Cambridge: Cambridge University Press, 2016.

\bibitem[{Smith} and {Szathm{\'a}ry}(1995)]{JMS95}
J.~M. {Smith} and E.~{Szathm{\'a}ry}.
\newblock \emph{{The Major Transitions in Evolution}}.
\newblock Oxford: Oxford University Press, 1995.

\bibitem[{Soderblom} et~al.(1990){Soderblom}, {Kieffer}, {Becker}, {Brown},
  {Cook}, {Hansen}, {Johnson}, {Kirk}, and {Shoemaker}]{SKB90}
L.~A. {Soderblom}, S.~W. {Kieffer}, T.~L. {Becker}, R.~H. {Brown}, I.~{Cook},
  A.~F., C.~J. {Hansen}, T.~V. {Johnson}, R.~L. {Kirk}, and E.~M. {Shoemaker}.
\newblock {Triton's Geyser-Like Plumes: Discovery and Basic Characterization}.
\newblock \emph{Science}, 250\penalty0 (4979):\penalty0 410--415, Oct. 1990.
\newblock \doi{10.1126/science.250.4979.410}.

\bibitem[{Sojo} et~al.(2016){Sojo}, {Herschy}, {Whicher}, {Camprub{\'\i}}, and
  {Lane}]{SHW16}
V.~{Sojo}, B.~{Herschy}, A.~{Whicher}, E.~{Camprub{\'\i}}, and N.~{Lane}.
\newblock {The Origin of Life in Alkaline Hydrothermal Vents}.
\newblock \emph{Astrobiology}, 16\penalty0 (2):\penalty0 181--197, Feb. 2016.
\newblock \doi{10.1089/ast.2015.1406}.

\bibitem[{Sojo} et~al.(2019){Sojo}, {Ohno}, {McGlynn}, {Yamada}, and
  {Nakamura}]{SOM19}
V.~{Sojo}, A.~{Ohno}, S.~E. {McGlynn}, Y.~{Yamada}, and R.~{Nakamura}.
\newblock {Microfluidic Reactors for Carbon Fixation under Ambient-Pressure
  Alkaline-Hydrothermal-Vent Conditions}.
\newblock \emph{Life}, 9\penalty0 (1):\penalty0 16, 2019.
\newblock \doi{10.3390/life9010016}.

\bibitem[{Sourjik} and {Wingreen}(2012)]{SW12}
V.~{Sourjik} and N.~S. {Wingreen}.
\newblock {Responding to chemical gradients: bacterial chemotaxis}.
\newblock \emph{Curr. Opin. Cell Biol.}, 24\penalty0 (2):\penalty0 262--268,
  2012.
\newblock \doi{10.1016/j.ceb.2011.11.008}.

\bibitem[{Sparks} et~al.(2017){Sparks}, {Schmidt}, {McGrath}, {Hand},
  {Spencer}, {Cracraft}, and {E Deustua}]{SSM17}
W.~B. {Sparks}, B.~E. {Schmidt}, M.~A. {McGrath}, K.~P. {Hand}, J.~R.
  {Spencer}, M.~{Cracraft}, and S.~{E Deustua}.
\newblock {Active Cryovolcanism on Europa?}
\newblock \emph{Astrophys. J. Lett.}, 839\penalty0 (2):\penalty0 L18, Apr.
  2017.
\newblock \doi{10.3847/2041-8213/aa67f8}.

\bibitem[{Spitzer}(2021)]{Spitz}
J.~{Spitzer}.
\newblock \emph{{How Molecular Forces and Rotating Planets Create Life: The
  Emergence and Evolution of Prokaryotic Cells}}.
\newblock Cambridge: The MIT Press, 2021.

\bibitem[{Stamenkovi{\'c}} et~al.(2019){Stamenkovi{\'c}}, {Beegle}, {Zacny},
  {Arumugam}, {Baglioni}, {Barba}, {Baross}, {Bell}, {Bhartia}, {Blank},
  {Boston}, {Breuer}, {Brinckerhoff}, {Burgin}, {Cooper}, {Cormarkovic},
  {Davila}, {Davis}, {Edwards}, {Etiope}, {Fischer}, {Glavin}, {Grimm},
  {Inagaki}, {Kirschvink}, {Kobayashi}, {Komarek}, {Malaska}, {Michalski},
  {M{\'e}nez}, {Mischna}, {Moser}, {Mustard}, {Onstott}, {Orphan}, {Osburn},
  {Plaut}, {Plesa}, {Putzig}, {Rogers}, {Rothschild}, {Russell}, {Sapers},
  {Lollar}, {Spohn}, {Tarnas}, {Tuite}, {Viola}, {Ward}, {Wilcox}, and
  {Woolley}]{SBZ19}
V.~{Stamenkovi{\'c}}, L.~W. {Beegle}, K.~{Zacny}, D.~D. {Arumugam},
  P.~{Baglioni}, N.~{Barba}, J.~{Baross}, M.~S. {Bell}, R.~{Bhartia}, J.~G.
  {Blank}, P.~J. {Boston}, D.~{Breuer}, W.~{Brinckerhoff}, M.~S. {Burgin},
  I.~{Cooper}, V.~{Cormarkovic}, A.~{Davila}, R.~M. {Davis}, C.~{Edwards},
  G.~{Etiope}, W.~W. {Fischer}, D.~P. {Glavin}, R.~E. {Grimm}, F.~{Inagaki},
  J.~L. {Kirschvink}, A.~{Kobayashi}, T.~{Komarek}, M.~{Malaska},
  J.~{Michalski}, B.~{M{\'e}nez}, M.~{Mischna}, D.~{Moser}, J.~{Mustard}, T.~C.
  {Onstott}, V.~J. {Orphan}, M.~R. {Osburn}, J.~{Plaut}, A.~C. {Plesa},
  N.~{Putzig}, K.~L. {Rogers}, L.~{Rothschild}, M.~{Russell}, H.~{Sapers},
  B.~S. {Lollar}, T.~{Spohn}, J.~D. {Tarnas}, M.~{Tuite}, D.~{Viola}, L.~M.
  {Ward}, B.~{Wilcox}, and R.~{Woolley}.
\newblock {The next frontier for planetary and human exploration}.
\newblock \emph{Nat. Astron.}, 3:\penalty0 116--120, Jan. 2019.
\newblock \doi{10.1038/s41550-018-0676-9}.

\bibitem[{Stevenson}(2010)]{DJS10}
D.~J. {Stevenson}.
\newblock {Planetary Magnetic Fields: Achievements and Prospects}.
\newblock \emph{Space Sci. Rev.}, 152\penalty0 (1-4):\penalty0 651--664, May
  2010.
\newblock \doi{10.1007/s11214-009-9572-z}.

\bibitem[{Stevenson} et~al.(2015){Stevenson}, {Lunine}, and {Clancy}]{SLC15}
J.~{Stevenson}, J.~{Lunine}, and P.~{Clancy}.
\newblock {Membrane alternatives in worlds without oxygen: Creation of an
  azotosome}.
\newblock \emph{Sci. Adv.}, 1\penalty0 (1):\penalty0 1400067, Feb. 2015.
\newblock \doi{10.1126/sciadv.1400067}.

\bibitem[{Stocker}(2012)]{Sto12}
R.~{Stocker}.
\newblock {Marine Microbes See a Sea of Gradients}.
\newblock \emph{Science}, 338\penalty0 (6107):\penalty0 628, Nov. 2012.
\newblock \doi{10.1126/science.1208929}.

\bibitem[{St{\"u}eken} et~al.(2013){St{\"u}eken}, {Anderson}, {Bowman},
  {Brazelton}, {Colangelo-Lillis}, {Goldman}, {Som}, and {Baross}]{SAB13}
E.~E. {St{\"u}eken}, R.~E. {Anderson}, J.~S. {Bowman}, W.~J. {Brazelton},
  J.~{Colangelo-Lillis}, A.~D. {Goldman}, S.~M. {Som}, and J.~A. {Baross}.
\newblock {Did life originate from a global chemical reactor?}
\newblock \emph{Geobiology}, 11\penalty0 (2):\penalty0 101--126, 2013.
\newblock \doi{10.1111/gbi.12025}.

\bibitem[{Sutherland}(2017)]{Suth17}
J.~D. {Sutherland}.
\newblock {Studies on the origin of life -- the end of the beginning}.
\newblock \emph{Nat. Rev. Chem.}, 1:\penalty0 0012, 2017.
\newblock \doi{10.1038/s41570-016-0012}.

\bibitem[{Svenstrup} et~al.(2003){Svenstrup}, {Fedder}, {Abraham-Peskir},
  {Birkelund}, and {Christiansen}]{SFA03}
H.~F. {Svenstrup}, J.~{Fedder}, J.~{Abraham-Peskir}, S.~{Birkelund}, and
  G.~{Christiansen}.
\newblock {\emph{Mycoplasma genitalium} attaches to human spermatozoa}.
\newblock \emph{Hum. Reprod.}, 18\penalty0 (10):\penalty0 2103--2109, 2003.
\newblock \doi{10.1093/humrep/deg392}.

\bibitem[{Swingley} et~al.(2012){Swingley}, {Meyer-Dombard}, {Shock}, {Alsop},
  {Falenski}, {Havig}, and {Raymond}]{SDS12}
W.~D. {Swingley}, D.~R. {Meyer-Dombard}, E.~L. {Shock}, E.~B. {Alsop}, H.~D.
  {Falenski}, J.~R. {Havig}, and J.~{Raymond}.
\newblock {Coordinating Environmental Genomics and Geochemistry Reveals
  Metabolic Transitions in a Hot Spring Ecosystem}.
\newblock \emph{PLoS ONE}, 7\penalty0 (6):\penalty0 e38108, 2012.
\newblock \doi{10.1371/journal.pone.0038108}.

\bibitem[{Taktikos} et~al.(2013){Taktikos}, {Stark}, and {Zaburdaev}]{TSZ13}
J.~{Taktikos}, H.~{Stark}, and V.~{Zaburdaev}.
\newblock {How the Motility Pattern of Bacteria Affects Their Dispersal and
  Chemotaxis}.
\newblock \emph{PLoS ONE}, 8\penalty0 (12):\penalty0 e81936, 2013.
\newblock \doi{10.1371/journal.pone.0081936}.

\bibitem[{Taubner} et~al.(2020){Taubner}, {Olsson-Francis}, {Vance},
  {Ramkissoon}, {Postberg}, {de Vera}, {Antunes}, {Camprubi Casas}, {Sekine},
  {Noack}, {Barge}, {Goodman}, {Jebbar}, {Journaux}, {Karatekin}, {Klenner},
  {Rabbow}, {Rettberg}, {R{\"u}ckriemen-Bez}, {Saur}, {Shibuya}, and
  {Soderlund}]{Tau20}
R.-S. {Taubner}, K.~{Olsson-Francis}, S.~D. {Vance}, N.~K. {Ramkissoon},
  F.~{Postberg}, J.-P. {de Vera}, A.~{Antunes}, E.~{Camprubi Casas},
  Y.~{Sekine}, L.~{Noack}, L.~{Barge}, J.~{Goodman}, M.~{Jebbar},
  B.~{Journaux}, {\"O}.~{Karatekin}, F.~{Klenner}, E.~{Rabbow}, P.~{Rettberg},
  T.~{R{\"u}ckriemen-Bez}, J.~{Saur}, T.~{Shibuya}, and K.~M. {Soderlund}.
\newblock {Experimental and Simulation Efforts in the Astrobiological
  Exploration of Exooceans}.
\newblock \emph{Space Sci. Rev.}, 216\penalty0 (1):\penalty0 9, Jan. 2020.
\newblock \doi{10.1007/s11214-020-0635-5}.

\bibitem[{Tu}(2013)]{Tu13}
Y.~{Tu}.
\newblock {Quantitative Modeling of Bacterial Chemotaxis: Signal Amplification
  and Accurate Adaptation}.
\newblock \emph{Annu. Rev. Biophys.}, 42:\penalty0 337--359, 2013.
\newblock \doi{10.1146/annurev-biophys-083012-130358}.

\bibitem[{Tyrrell}(1999)]{Ty99}
T.~{Tyrrell}.
\newblock {The relative influences of nitrogen and phosphorus on oceanic
  primary production}.
\newblock \emph{Nature}, 400\penalty0 (6744):\penalty0 525--531, Aug. 1999.
\newblock \doi{10.1038/22941}.

\bibitem[{Uebe} and {Sch{\"u}ler}(2016)]{US16}
R.~{Uebe} and D.~{Sch{\"u}ler}.
\newblock Magnetosome biogenesis in magnetotactic bacteria.
\newblock \emph{Nat. Rev. Microbiol.}, 14\penalty0 (10):\penalty0 621--637,
  2016.
\newblock \doi{10.1038/nrmicro.2016.99}.

\bibitem[{Vago} et~al.(2017){Vago}, {Westall}, {Pasteur Instrument Team},
  {Pasteur Landing Team}, {Coates}, {Jaumann}, {Korablev}, {Ciarletti},
  {Mitrofanov}, {Josset}, {De Sanctis}, {Bibring}, {Rull}, {Goesmann},
  {Steininger}, {Goetz}, {Brinckerhoff}, {Szopa}, {Raulin}, {Westall},
  {Edwards}, {Whyte}, {Fair{\'e}n}, {Bibring}, {Bridges}, {Hauber}, {Ori},
  {Werner}, {Loizeau}, {Kuzmin}, {Williams}, {Flahaut}, {Forget}, {Vago},
  {Rodionov}, {Korablev}, {Svedhem}, {Sefton-Nash}, {Kminek}, {Lorenzoni},
  {Joudrier}, {Mikhailov}, {Zashchirinskiy}, {Alexashkin}, {Calantropio},
  {Merlo}, {Poulakis}, {Witasse}, {Bayle}, {Bay{\'o}n}, {Meierhenrich},
  {Carter}, {Garc{\'\i}a-Ruiz}, {Baglioni}, {Haldemann}, {Ball}, {Debus},
  {Lindner}, {Haessig}, {Monteiro}, {Trautner}, {Voland}, {Rebeyre}, {Goulty},
  {Didot}, {Durrant}, {Zekri}, {Koschny}, {Toni}, {Visentin}, {Zwick}, {van
  Winnendael}, {Azkarate}, {Carreau}, and {ExoMars Project Team}]{VWP17}
J.~L. {Vago}, F.~{Westall}, {Pasteur Instrument Team}, {Pasteur Landing Team},
  A.~J. {Coates}, R.~{Jaumann}, O.~{Korablev}, V.~{Ciarletti}, I.~{Mitrofanov},
  J.-L. {Josset}, M.~C. {De Sanctis}, J.-P. {Bibring}, F.~{Rull},
  F.~{Goesmann}, H.~{Steininger}, W.~{Goetz}, W.~{Brinckerhoff}, C.~{Szopa},
  F.~{Raulin}, F.~{Westall}, H.~G.~M. {Edwards}, L.~G. {Whyte}, A.~G.
  {Fair{\'e}n}, J.-P. {Bibring}, J.~{Bridges}, E.~{Hauber}, G.~G. {Ori},
  S.~{Werner}, D.~{Loizeau}, R.~O. {Kuzmin}, R.~M.~E. {Williams}, J.~{Flahaut},
  F.~{Forget}, J.~L. {Vago}, D.~{Rodionov}, O.~{Korablev}, H.~{Svedhem},
  E.~{Sefton-Nash}, G.~{Kminek}, L.~{Lorenzoni}, L.~{Joudrier}, V.~{Mikhailov},
  A.~{Zashchirinskiy}, S.~{Alexashkin}, F.~{Calantropio}, A.~{Merlo},
  P.~{Poulakis}, O.~{Witasse}, O.~{Bayle}, S.~{Bay{\'o}n}, U.~{Meierhenrich},
  J.~{Carter}, J.~M. {Garc{\'\i}a-Ruiz}, P.~{Baglioni}, A.~{Haldemann}, A.~J.
  {Ball}, A.~{Debus}, R.~{Lindner}, F.~{Haessig}, D.~{Monteiro}, R.~{Trautner},
  C.~{Voland}, P.~{Rebeyre}, D.~{Goulty}, F.~{Didot}, S.~{Durrant}, E.~{Zekri},
  D.~{Koschny}, A.~{Toni}, G.~{Visentin}, M.~{Zwick}, M.~{van Winnendael},
  M.~{Azkarate}, C.~{Carreau}, and {ExoMars Project Team}.
\newblock {Habitability on Early Mars and the Search for Biosignatures with the
  ExoMars Rover}.
\newblock \emph{Astrobiology}, 17\penalty0 (6-7):\penalty0 471--510, July 2017.
\newblock \doi{10.1089/ast.2016.1533}.

\bibitem[{Van Dover} et~al.(1996){Van Dover}, {Reynolds}, {Chave}, and
  {Tyson}]{VRC96}
C.~L. {Van Dover}, G.~T. {Reynolds}, A.~D. {Chave}, and J.~A. {Tyson}.
\newblock {Light at deep-sea hydrothermal vents}.
\newblock \emph{Geophys. Res. Lett.}, 23\penalty0 (16):\penalty0 2049--2052,
  Jan. 1996.
\newblock \doi{10.1029/96GL02151}.

\bibitem[{Van Loef}(1978)]{VL78}
J.~J. {Van Loef}.
\newblock {Temperature and density dependence of the self-diffusion coefficient
  in compressed liquid methane}.
\newblock \emph{Physica B+C}, 94\penalty0 (1):\penalty0 105--107, Apr. 1978.
\newblock \doi{10.1016/0378-4363(78)90081-5}.

\bibitem[{Vance} et~al.(2007){Vance}, {Harnmeijer}, {Kimura}, {Hussmann},
  {deMartin}, and {Brown}]{VHK07}
S.~{Vance}, J.~{Harnmeijer}, J.~{Kimura}, H.~{Hussmann}, B.~{deMartin}, and
  J.~M. {Brown}.
\newblock {Hydrothermal Systems in Small Ocean Planets}.
\newblock \emph{Astrobiology}, 7\penalty0 (6):\penalty0 987--1005, Dec. 2007.
\newblock \doi{10.1089/ast.2007.0075}.

\bibitem[{Varennes} et~al.(2016){Varennes}, {Han}, and {Mugler}]{VHM16}
J.~{Varennes}, B.~{Han}, and A.~{Mugler}.
\newblock {Collective Chemotaxis through Noisy Multicellular Gradient Sensing}.
\newblock \emph{Biophys. J.}, 111\penalty0 (3):\penalty0 640--649, Aug. 2016.
\newblock \doi{10.1016/j.bpj.2016.06.040}.

\bibitem[{Vargaftik} et~al.(1994){Vargaftik}, {Filippov}, {Tarzimanov}, and
  {Totskii}]{VFTT}
N.~B. {Vargaftik}, L.~P. {Filippov}, A.~A. {Tarzimanov}, and E.~E. {Totskii}.
\newblock \emph{{Handbook of Thermal Conductivity of Liquids and Gases}}.
\newblock Boca Raton: CRC Press, 1994.

\bibitem[{Velimirov}(2001)]{Vel01}
B.~{Velimirov}.
\newblock {Nanobacteria, Ultramicrobacteria and Starvation Forms: A Search for
  the Smallest Metabolizing Bacterium}.
\newblock \emph{Microbes Environ.}, 16\penalty0 (2):\penalty0 67--77, 2001.
\newblock \doi{10.1264/jsme2.2001.67}.

\bibitem[{Vicsek} and {Zafeiris}(2012)]{VZ12}
T.~{Vicsek} and A.~{Zafeiris}.
\newblock {Collective motion}.
\newblock \emph{Phys. Rep.}, 517\penalty0 (3-4):\penalty0 71--140, Aug. 2012.
\newblock \doi{10.1016/j.physrep.2012.03.004}.

\bibitem[{Vogel}(2008)]{Vog08}
S.~{Vogel}.
\newblock {Modes and scaling in aquatic locomotion}.
\newblock \emph{Integr. Comp. Biol.}, 48\penalty0 (6):\penalty0 702--712, 2008.
\newblock \doi{10.1093/icb/icn014}.

\bibitem[{Wadhams} and {Armitage}(2004)]{WA04}
G.~H. {Wadhams} and J.~P. {Armitage}.
\newblock {Making sense of it all: bacterial chemotaxis}.
\newblock \emph{Nat. Rev. Mol. Cell Biol.}, 5\penalty0 (12):\penalty0
  1024--1037, 2004.
\newblock \doi{10.1038/nrm1524}.

\bibitem[{Waite} et~al.(2018){Waite}, {Frankel}, and {Emonet}]{WFE}
A.~J. {Waite}, N.~W. {Frankel}, and T.~{Emonet}.
\newblock {Behavioral Variability and Phenotypic Diversity in Bacterial
  Chemotaxis}.
\newblock \emph{Annu. Rev. Biophys.}, 47:\penalty0 595--616, 2018.
\newblock \doi{10.1016/j.mib.2014.12.007}.

\bibitem[{Waite} et~al.(2017){Waite}, {Glein}, {Perryman}, {Teolis}, {Magee},
  {Miller}, {Grimes}, {Perry}, {Miller}, {Bouquet}, {Lunine}, {Brockwell}, and
  {Bolton}]{WGP17}
J.~H. {Waite}, C.~R. {Glein}, R.~S. {Perryman}, B.~D. {Teolis}, B.~A. {Magee},
  G.~{Miller}, J.~{Grimes}, M.~E. {Perry}, K.~E. {Miller}, A.~{Bouquet}, J.~I.
  {Lunine}, T.~{Brockwell}, and S.~J. {Bolton}.
\newblock {Cassini finds molecular hydrogen in the Enceladus plume: Evidence
  for hydrothermal processes}.
\newblock \emph{Science}, 356\penalty0 (6334):\penalty0 155--159, Apr. 2017.
\newblock \doi{10.1126/science.aai8703}.

\bibitem[{Wan} and {J{\'e}kely}(2021)]{WJ20}
K.~Y. {Wan} and G.~{J{\'e}kely}.
\newblock {Origins of eukaryotic excitability}.
\newblock \emph{Phil. Trans. R. Soc. B}, 376\penalty0 (1820):\penalty0
  20190758, Mar. 2021.
\newblock \doi{10.1098/rstb.2019.0758}.

\bibitem[{Wang} and {Steinbock}(2020)]{WS20}
Q.~{Wang} and O.~{Steinbock}.
\newblock {Materials Synthesis and Catalysis in Microfluidic Devices: Prebiotic
  Chemistry in Mineral Membranes}.
\newblock \emph{ChemCatChem}, 12\penalty0 (1):\penalty0 63--74, 2020.
\newblock \doi{10.1002/cctc.201901495}.

\bibitem[{Ward} and {Shih}(2019)]{WS19}
L.~M. {Ward} and P.~M. {Shih}.
\newblock {The evolution and productivity of carbon fixation pathways in
  response to changes in oxygen concentration over geological time}.
\newblock \emph{Free Radic. Biol. Med.}, 140:\penalty0 188--199, 2019.
\newblock \doi{10.1016/j.freeradbiomed.2019.01.049}.

\bibitem[{Warrant} and {Nilsson}(1998)]{WN98}
E.~J. {Warrant} and D.-E. {Nilsson}.
\newblock {Absorption of white light in photoreceptors}.
\newblock \emph{Vision Res.}, 38\penalty0 (2):\penalty0 195--207, 1998.
\newblock \doi{10.1016/S0042-6989(97)00151-X}.

\bibitem[{Webster} and {Weissburg}(2009)]{WW09}
D.~R. {Webster} and M.~J. {Weissburg}.
\newblock {The Hydrodynamics of Chemical Cues Among Aquatic Organisms}.
\newblock \emph{Annu. Rev. Fluid Mech.}, 41\penalty0 (1):\penalty0 73--90, Jan.
  2009.
\newblock \doi{10.1146/annurev.fluid.010908.165240}.

\bibitem[{Weiss} et~al.(2018){Weiss}, {Preiner}, {Xavier}, {Zimorski}, and
  {Martin}]{WPX18}
M.~C. {Weiss}, M.~{Preiner}, J.~C. {Xavier}, V.~{Zimorski}, and W.~F. {Martin}.
\newblock {The last universal common ancestor between ancient Earth chemistry
  and the onset of genetics}.
\newblock \emph{PLoS Genet.}, 14\penalty0 (8):\penalty0 e1007518, 2018.
\newblock \doi{10.1371/journal.pgen.1007518}.

\bibitem[{Weissburg}(2000)]{Wei00}
M.~J. {Weissburg}.
\newblock {The fluid dynamical context of chemosensory behavior}.
\newblock \emph{Biol. Bull.}, 198\penalty0 (2):\penalty0 188--202, 2000.
\newblock \doi{10.2307/1542523}.

\bibitem[{Went}(1968)]{Went}
F.~W. {Went}.
\newblock {The size of man}.
\newblock \emph{Am. Sci.}, 56\penalty0 (4):\penalty0 400--413, 1968.

\bibitem[{Westall} et~al.(2015){Westall}, {Foucher}, {Bost}, {Bertrand},
  {Loizeau}, {Vago}, {Kminek}, {Gaboyer}, {Campbell}, {Br{\'e}h{\'e}ret},
  {Gautret}, and {Cockell}]{WFB15}
F.~{Westall}, F.~{Foucher}, N.~{Bost}, M.~{Bertrand}, D.~{Loizeau}, J.~L.
  {Vago}, G.~{Kminek}, F.~{Gaboyer}, K.~A. {Campbell}, J.-G.
  {Br{\'e}h{\'e}ret}, P.~{Gautret}, and C.~S. {Cockell}.
\newblock {Biosignatures on Mars: What, Where, and How? Implications for the
  Search for Martian Life}.
\newblock \emph{Astrobiology}, 15\penalty0 (11):\penalty0 998--1029, Nov. 2015.
\newblock \doi{10.1089/ast.2015.1374}.

\bibitem[{Westall} et~al.(2018){Westall}, {Hickman-Lewis}, {Hinman}, {Gautret},
  {Campbell}, {Br{\'e}h{\'e}ret}, {Foucher}, {Hubert}, {Sorieul}, {Dass},
  {Kee}, {Georgelin}, and {Brack}]{WHH18}
F.~{Westall}, K.~{Hickman-Lewis}, N.~{Hinman}, P.~{Gautret}, K.~A. {Campbell},
  J.~G. {Br{\'e}h{\'e}ret}, F.~{Foucher}, A.~{Hubert}, S.~{Sorieul}, A.~V.
  {Dass}, T.~P. {Kee}, T.~{Georgelin}, and A.~{Brack}.
\newblock {A Hydrothermal-Sedimentary Context for the Origin of Life}.
\newblock \emph{Astrobiology}, 18\penalty0 (3):\penalty0 259--293, Mar. 2018.
\newblock \doi{10.1089/ast.2017.1680}.

\bibitem[{Wheat} et~al.(1996){Wheat}, {Feely}, and {Mottl}]{WFM96}
C.~G. {Wheat}, R.~A. {Feely}, and M.~J. {Mottl}.
\newblock {Phosphate removal by oceanic hydrothermal processes: An update of
  the phosphorus budget in the oceans}.
\newblock \emph{Geochim. Cosmochim. Acta}, 60\penalty0 (19):\penalty0
  3593--3608, Oct. 1996.
\newblock \doi{10.1016/0016-7037(96)00189-5}.

\bibitem[{White} et~al.(2000){White}, {Chave}, {Reynolds}, {Gaidos}, {Tyson},
  and {Van Dover}]{WCR00}
S.~N. {White}, A.~D. {Chave}, G.~T. {Reynolds}, E.~J. {Gaidos}, J.~A. {Tyson},
  and C.~L. {Van Dover}.
\newblock {Variations in ambient light emission from black smokers and flange
  pools on the Juan De Fuca Ridge}.
\newblock \emph{Geophys. Res. Lett.}, 27\penalty0 (8):\penalty0 1151--1154,
  Apr. 2000.
\newblock \doi{10.1029/1999GL011074}.

\bibitem[{White} et~al.(2002){White}, {Chave}, and {Reynolds}]{WCR02}
S.~N. {White}, A.~D. {Chave}, and G.~T. {Reynolds}.
\newblock {Investigations of ambient light emission at deep-sea hydrothermal
  vents}.
\newblock \emph{J. Geophys. Res. Solid Earth}, 107\penalty0 (B1):\penalty0
  1.1--1.13, Jan. 2002.
\newblock \doi{10.1029/2000JB000015}.

\bibitem[{Wilde} and {Mullineaux}(2017)]{WW17}
A.~{Wilde} and C.~W. {Mullineaux}.
\newblock {Light-controlled motility in prokaryotes and the problem of
  directional light perception}.
\newblock \emph{FEMS Microbiol. Rev.}, 41\penalty0 (6):\penalty0 900--922,
  2017.
\newblock \doi{10.1093/femsre/fux045}.

\bibitem[{Williford} et~al.(2018){Williford}, {Farley}, {Stack}, {Allwood},
  {Beaty}, {Beegle}, {Bhartia}, {Brown}, {de la Torre Juarez}, {Hamran},
  {Hecht}, {Hurowitz}, {Rodriguez-Manfredi}, {Maurice}, {Milkovich}, and
  {Wiens}]{WFS18}
K.~H. {Williford}, K.~A. {Farley}, K.~M. {Stack}, A.~C. {Allwood}, D.~{Beaty},
  L.~W. {Beegle}, R.~{Bhartia}, A.~J. {Brown}, M.~{de la Torre Juarez}, S.-E.
  {Hamran}, M.~H. {Hecht}, J.~A. {Hurowitz}, J.~A. {Rodriguez-Manfredi},
  S.~{Maurice}, S.~{Milkovich}, and R.~C. {Wiens}.
\newblock \emph{{The NASA Mars 2020 Rover Mission and the Search for
  Extraterrestrial Life}}, pages 275--308.
\newblock Amsterdam: Elsevier, 2018.
\newblock \doi{10.1016/B978-0-12-809935-3.00010-4}.

\bibitem[{Wong-Ng} et~al.(2018){Wong-Ng}, {Celani}, and {Vergassola}]{WCV18}
J.~{Wong-Ng}, A.~{Celani}, and M.~{Vergassola}.
\newblock Exploring the function of bacterial chemotaxis.
\newblock \emph{{Curr. Opin. Microbiol.}}, 45:\penalty0 16--21, 2018.
\newblock \doi{10.1016/j.mib.2018.01.010}.

\bibitem[{Wordsworth} and {Pierrehumbert}(2013)]{WP13}
R.~D. {Wordsworth} and R.~T. {Pierrehumbert}.
\newblock {Water Loss from Terrestrial Planets with CO$_{2}$-rich Atmospheres}.
\newblock \emph{Astrophys. J.}, 778\penalty0 (2):\penalty0 154, Dec 2013.
\newblock \doi{10.1088/0004-637X/778/2/154}.

\bibitem[{Wurch} et~al.(2016){Wurch}, {Giannone}, {Belisle}, {Swift},
  {Utturkar}, {Hettich}, {Reysenbach}, and {Podar}]{WGB16}
L.~{Wurch}, R.~J. {Giannone}, B.~S. {Belisle}, C.~{Swift}, S.~{Utturkar}, R.~L.
  {Hettich}, A.-L. {Reysenbach}, and M.~{Podar}.
\newblock {Genomics-informed isolation and characterization of a symbiotic
  Nanoarchaeota system from a terrestrial geothermal environment}.
\newblock \emph{Nat. Commun.}, 7:\penalty0 12115, July 2016.
\newblock \doi{10.1038/ncomms12115}.

\bibitem[{Yang} et~al.(2018){Yang}, {Lam}, {Adomako}, {Simkovsky}, {Jakob},
  {Rockwell}, {Cohen}, {Taton}, {Wang}, {Lagarias}, {Wilde}, {Nobles}, {Brand},
  and {Golden}]{YRC}
Y.~{Yang}, V.~{Lam}, M.~{Adomako}, R.~{Simkovsky}, A.~{Jakob}, N.~C.
  {Rockwell}, S.~E. {Cohen}, A.~{Taton}, J.~{Wang}, J.~C. {Lagarias},
  A.~{Wilde}, D.~R. {Nobles}, J.~J. {Brand}, and S.~S. {Golden}.
\newblock {Phototaxis in a wild isolate of the cyanobacterium
  \emph{Synechococcus elongatus}}.
\newblock \emph{Proc. Natl. Acad. Sci. USA}, 115\penalty0 (52):\penalty0
  E12378--E12387, 2018.
\newblock \doi{10.1126/science.1070118}.

\bibitem[{Younglove}(1974)]{Yo74}
B.~A. {Younglove}.
\newblock {The Specific Heats, C$_\sigma$, and C$_V$, of Compressed and
  Liquefied Methane}.
\newblock \emph{J. Res. Natl. Bur. Stand. A Phys. Chem.}, 78\penalty0
  (3):\penalty0 401, 1974.
\newblock \doi{10.6028/jres.078a.023}.

\bibitem[{Zhang} et~al.(2012){Zhang}, {Ducret}, {Shaevitz}, and
  {Mignot}]{ZDS12}
Y.~{Zhang}, A.~{Ducret}, J.~{Shaevitz}, and T.~{Mignot}.
\newblock {From individual cell motility to collective behaviors: insights from
  a prokaryote, \emph{Myxococcus xanthus}}.
\newblock \emph{FEMS Microbiol. Rev.}, 36\penalty0 (1):\penalty0 149--164,
  2012.
\newblock \doi{10.1111/j.1574-6976.2011.00307.x}.

\end{thebibliography}

\end{document}